\documentclass[a4paper,11pt]{article}

\usepackage{jheppub} 

\usepackage[T1]{fontenc} 

\usepackage{amsmath}
\usepackage{amsthm}
\usepackage{physics}
\usepackage[compat=1.1.0]{tikz-feynman}
\usepackage[T1]{fontenc} 
\usepackage{tensor}
\usepackage{graphicx}
\usepackage[export]{adjustbox}
\usepackage{slashed}
\usepackage{stackrel,amssymb}
\usepackage{tikz}
\usepackage{mathtools}
\usetikzlibrary{arrows.meta}
\usepackage{comment}
\tikzset{>={Latex[scale=1.1]}}
\usetikzlibrary{arrows.meta,
	bending,
	decorations.markings, decorations.text}
\newcommand{\ee}{\mathrm{e}}
\newcommand{\ii}{\mathrm{i}}
\newcommand{\vvev}[1]{{\left\langle #1 \right\rangle}}
\makeatletter
\newcommand*{\letterdef@}{}
\newcommand*{\letterdef}[3]{%
	\def\letterdef@##1{\expandafter\newcommand\csname #1\endcsname{#2{##1}}}%
	\@tfor\@tempa :=#3\do{\expandafter\letterdef@\expandafter{\@tempa}}}
\makeatother
\letterdef{c#1}{\mathcal}{ABCDEFGHIJKLMNOPQRSTUVWXYZ} 
\letterdef{rm#1}{\mathrm} {dDeimM} 

\title{1/2 BPS Wilson loops in non-conformal $\mathcal{N}=2$ gauge theories and localization: a three-loop analysis}

\author[a]{M. Bill\`o,}
\author[b]{L. Griguolo,}
\author[b]{A. Testa}

\affiliation[a]{Universit\`a di Torino, Dipartimento di Fisica and INFN, Sezione di Torino,\\
	Via P. Giuria 1, I-10125 Torino, Italy}
\affiliation[b]{Dipartimento SMFI, Universit\`a di Parma and INFN, Gruppo Collegato di Parma,
	\\ Viale G.P. Usberti 7/A, 43100 Parma, Italy}

\emailAdd{marco.billo@unito.it}
\emailAdd{luca.griguolo@unipr.it}
\emailAdd{alessandro.testa@unipr.it
}

\abstract{We study the 1/2 BPS circular Wilson loop in four-dimensional SU$(N)$ $\cN = 2$
	SYM theories with massless hypermultiplets and non-vanishing $\beta$-function. Using supersymmetric localization  on $\mathbb{S}^4$, we map the path-integral associated with this observable onto an
	interacting matrix model. Despite the breaking of conformal symmetry at the quantum level,
	we show that, within a specific regime, the matrix model predictions remain consistent with the perturbative results in flat space up to order $g^6$. At this order, our analysis
	reveals that the reorganization of Feynman diagrams based on the matrix model interaction potential,  widely tested in (super)conformal models, also applies to these non-conformal set-ups and is realized by interference mechanisms.}

\begin{document}
	\maketitle
	\flushbottom
	
	\section{Introduction and main results}
	Supersymmetric gauge theories provide a powerful theoretical  laboratory for controlling the dynamics of fields at the quantum level. In four dimensions, these models exhibit interesting dynamics, including confinement without chiral symmetry breaking and the emergence of gapless gauge bosons in the infrared \cite{Seiberg:1994bz,Seiberg:1994pq}. Moreover, through advanced techniques,  such as  dualities \cite{Seiberg:1994rs,Seiberg:1994pq,Gaiotto:2009we} and gauge-gravity correspondences \cite{Maldacena:1997re,Aharony:1999ti}, it has been possible to probe the non-perturbative properties of these models, confirming the presence of  mechanisms that also are expected in physical theories like QCD \cite{Seiberg:1994aj}. 
	
	Recently, extended supersymmetry has allowed to develop new analytical approaches, such as  supersymmetric localization \cite{Pestun:2007rz,Kapustin:2009kz}. Unlike integrability \cite{Minahan:2002ve,Beisert:2006ez}, resurgence \cite{Aniceto:2018bis} and bootstrap approaches \cite{Rattazzi:2008pe,Poland:2010wg}, supersymmetric localization provides a direct technique for computing path integrals. Under suitable conditions, partition functions and classes of local and non-local observables for the theory defined on a compact space-time manifold, such as $\mathbb{S}^4$, can be calculated exactly in terms of matrix models. These are typically characterized by complex interaction potentials that  encode both the conventional perturbative series and non-perturbative contributions. The latter are often associated with semiclassical configurations, such as instantons \cite{Nekrasov:2002qd}, monopoles \cite{Gomis:2011pf} and fluxes \cite{Benini:2012ui,Doroud:2012xw}. Localization thus offers an alternative technique for testing methods that provide informations only in particular regimes and for refining techniques that require external inputs or data\footnote{Localization data have been often used in superconformal bootstrap to refine bounds on anomalous dimensions and OPE coefficients, see for example \cite{Liendo:2018ukf} }. Furthermore, the matrix models generated by supersymmetric localization also offer new insights on the  perturbative techniques, suggesting a convenient reorganization of Feynman diagrams and predicting their large-order behaviours. In four dimensions,  these features have been extensively studied in (super)conformal models, where the computations on compact spaces naturally extend to the Euclidean configurations. Less attention has been given to non-conformal cases.
	
	In this paper, we continue the analysis initiated in \cite{Billo:2023igr} regarding the localization approaches in non-conformal four-dimensional $\mathcal{N}=2$ supersymmetric theories. More precisely,  we will consider SU($N$) $\mathcal{N}=2$ super-Yang-Mills theories (SYM) with massless hypermultiplets in an arbitrary representation $\cR$. In these set-ups, classical conformal symmetry is broken at the quantum level by the  (one-loop exact) $\beta$-function  \cite{Jones:1974mm,Howe:1984xq}
	\begin{equation}
		\label{eq:beta0} \beta(g) =- \epsilon g+	\beta_0 g^3 \ , \quad \text{where} \quad \beta_0 =\dfrac{i_{\mathcal{R}}-N}{8\pi^2}~.
	\end{equation}
	In the previous expression, the first term is the classical contribution in 
	\begin{align}
		\label{epsi}
		d = 4 - 2\epsilon
	\end{align}
	dimensions, while $i_\cR$ denotes the Dynkin index of the representation $\cR$.  In the following, we will focus on asymptotically free theories, where $i_\cR < N$ and  we fix $i_F = 1/2$ for the fundamental representation. Compactifying these theories on the four-sphere $\mathbb{S}^4$, we can employ supersymmetric localization \cite{Pestun:2007rz} to reduce the path-integral associated with the partition function and with the expectation values of protected operators into a matrix model. 
	
	When the theory remains conformal at the quantum level, i.e. when $i_\cR = N$ and the $\beta$-function vanishes, localization results on $\mathbb{S}^4$ naturally extend to flat-space observables. For instance, in  $\mathcal{N}=4$ SYM theories, supersymmetric localization  was employed to derive the analytical expression of the 1/2 BPS Wilson loop \cite{Pestun:2007rz}, originally conjectured in \cite{Erickson:2000af,Drukker:2000rr}.   Moreover, the same technique also applies to supersymmetric  Wilson loops which preserve fewer global supercharges than the circular configuration \cite{Drukker:2007qr}  and families of BPS local operators \cite{Giombi:2009ds}. In these cases, the matrix model generated by  localization is connected to Yang-Mills theories in two-dimensions \cite{Pestun:2009nn,Bassetto:1998sr,Giombi:2009ms} and successfully  captures the  perturbative results  based on standard Feynman diagrams.
	
	Unlike the $\mathcal{N}=4$ theory, where the matrix model generated by localization on $\mathbb{S}^4$ is purely Gaussian, $\mathcal{N}=2$ theories involve  non-trivial interaction potentials. Standard perturbative techniques in flat Euclidean space perfectly reproduce the localization predictions for several protected observables, including supersymmetric Wilson loops \cite{Andree:2010na,Billo:2019fbi,Fiol:2020bhf,Fiol:2020ojn,Galvagno:2021bbj}, chiral operators \cite{Billo:2017glv,Billo:2018oog,Galvagno:2020cgq,Beccaria:2020hgy,Fiol:2021icm} and  Bremsstrahlung functions \cite{Correa:2012at,Bonini:2015fng,Fiol:2015spa,Mitev:2015oty,Gomez:2018usu}. These results show that the perturbative computations in flat space are encoded by a one-loop effective action on $\mathbb{S}^4$ \cite{Pestun:2007rz}, which provides an elegant reorganization of Feynman diagrams. 
	
	However, when the theory involves dimensionful parameters, such as a mass term in the $\mathcal{N}=2^*$ theories or a scale generated by dimensional transmutation, the short and long distance properties of the model are different and it is expected that the flat-space calculations do not coincide with those on the sphere. In particular, when a mass term is present, observables on $\mathbb{S}^4$ naturally depend on  the mass scale and on the radius of the sphere by their product. The dependence on this dimensionless parameter of the observables on $\mathbb{S}^4$ usually differs from the flat-space counterpart. This scenario was analysed in \cite{Belitsky:2020hzs}, where the authors studied  the 1/2 BPS Wilson loop in $\mathcal{N}=2^*$ SYM  and showed that the two-loop perturbative computations of the observable on $\mathbb{S}^4$ coincide with the matrix model predictions, while the analogous flat-space calculation exhibits a different behaviour.
	
	While a mass deformation breaks explicitly conformal symmetry, in theories with massless matter and a non-vanishing $\beta$-function the violation of conformal symmetry occurs  at the quantum level. Compactifying these set-ups on the four-dimensional sphere, we can still apply supersymmetric localization to map the expectation value of specific protected operator into matrix models.  However, when the matter representation $\cR$ is associated with  a non-vanishing $\beta$-function, the one-loop determinants generated by localization requires a regularization based on additional massive supermultiplets of mass $M$ (see in particular Section 4 of \cite{Pestun:2007rz} and Section 2 of \cite{Billo:2023igr} for more details). These are properly introduced in order to make the $\beta$-function vanish and  the one-loop determinants expressible via well-defined products of   $H$-functions\footnote{Ref. \cite{Pestun:2007rz} discusses the case of pure $\mathcal{N}=2$ SYM, while in Section 2 of \cite{Billo:2023igr} the authors describe in detail non-conformal $\mathcal{N}=2$ SQCD and generalize the procedure to the theories under examination.} (see eq. (\ref{eq:H functions})). In the limit $M\to\infty$,  the massive degrees of freedom  decouple and we remain with a well-defined matrix model for  $\mathcal{N}=2$ SYM with massless hypermultiplets in an arbitrary representation $\cR$. This regularization leads to a matrix model which depends on the one-loop exact running coupling 
	\begin{equation}
		\label{eq:classical}
		\frac{1}{{g}^2} = \frac{1}{g_*^2} +\beta_0  \log M^2R^2~ ,
	\end{equation}where  $g_*$ is the renormalized coupling evaluated at the scale  $M$ which, from the perspective of the massless theory,  plays the role of a UV cut-off, while $R$ is the radius of the sphere; it is also the radius of the BPS Wilson loop on $\mathbb{S}^4$. Eq. (\ref{eq:classical}) also describes the running coupling constant of the flat-space theory evaluated at the energy scale $1/R$, with $R$ being the radius of the circular Wilson loop.
	
	The dependence of the matrix model on the running coupling $g$ is obviously expected and analogous to the  flat-space computations. It is therefore important to investigate whether the conventional perturbative series in  Euclidean space, when expressed in terms
	of the running coupling, is encoded in the localization effective action or to understand which part of this series (if any)
	is univocally determined by the localization approach.
	This question was  addressed in \cite{Billo:2019job} for the correlators of chiral primary operators. The analysis revealed  that the flat-space calculation matches the  localization prediction at order $g^4$, while at order $g^6$ the agreement occurs only for dimensionless ratios of correlators.
	A similar analysis is presented in \cite{Billo:2023igr}, where it was showed that the calculation of the 1/2 BPS Wilson loop in flat space matches the localization predictions up to order $g^4$. 
	
	In the present work, which is a detailed  version of a recent short letter \cite{Billo:2024hvf}, we extend the results presented in \cite{Billo:2023igr} up to order $g^6$. In an asymptotically free  $\mathcal{N}=2$ theory with massless hypermultiplets in an representation $\cR$ of SU($N$), the perturbative prediction of the matrix model  for the 1/2 BPS Wilson loops takes the following form
	\begin{equation}
		\label{eq:matrix model predictions introduction}
		\begin{split}
			W(g) = W_0(g) +g^6 \dfrac{3\zeta(3)}{2^8\pi^4N}\cK_4^{\prime} +g^6 \dfrac{\zeta(3)C_FN\beta_0}{16\pi^2} +\cO(g^8)\ .
		\end{split}
	\end{equation} In the previous expression,  $W_0(g)$ is the expectation value of the operator in the Gaussian matrix model, while $\cK^\prime_{4}$ is a colour factor which depends on the representation $\cR$ (see eq. (\ref{eq:def K})). The previous expression is \emph{valid} only in the range of scales \begin{equation}
		\label{eq:range of energies and lambda di strong couplign }
		\frac{1}{\Lambda}\gg R\gg\frac{1}{M} \ , \quad \text{where} \quad \Lambda = M \ee^{\frac{1}{2\beta_0 g_*^2}}
	\end{equation} 
	is the infrared strong coupling scale generated by dimensional transmutation. In this work, we will show that perturbation theory in flat space exactly  reproduces eq. (\ref{eq:matrix model diagrams}) within the regime (\ref{eq:range of energies and lambda di strong couplign }) where the running coupling $g$, defined in eq. (\ref{eq:classical}), is small. Conversely,  for $\Lambda R\sim 1$ the running coupling $g$ grows so that  a resummation of the perturbative series would be needed in order to include in the observables  non-perturbative power-like corrections\footnote{In special multicolour models, such $\mathcal{N}=2^*$ SYM or the massive deformation of superconformal $\mathcal{N}=2$ SQCD, the coefficients $C_n$ can be calculated  on the four-sphere by matrix model generated via supersymmetric localization \cite{Russo:2013kea}. Moreover, also instantons, which we neglected in our analysis, could contribute to the calculation of the observables with power-like corrections.} of the form $C_n(R\Lambda)^n$. 
	
	On general grounds, we expect that the functional dependence of the observable on $R\Lambda$ suffers from a conformal anomaly and differs between the sphere and flat space. Similarly, when $MR\sim 1$, the massive degrees of freedom become relevant and the nature  of the theory changes. As a result, the observables acquire a further dependence on $RM$ which is not purely logarithmic. 
		
	In the following, we will show that  standard perturbation in flat Euclidean space perfectly reproduces eq. (\ref{eq:matrix model predictions introduction}) within the range of validity (\ref{eq:range of energies and lambda di strong couplign }). In particular, the  two $\zeta(3)$-like corrections in eq. (\ref{eq:matrix model predictions introduction}) have a different origin: the contribution proportional to $\cK_{4}^\prime$ is also present in superconformal set-ups \cite{Andree:2010na,Billo:2019fbi}
	and arises from a Feynman integral which retains the same form in flat space and on the sphere, while that involving the coefficient $\beta_0$, emerges by
	interference effects between evanescent terms and the UV
	divergence of the bare coupling constant. Our analysis highlights how the localization matrix model organizes in a compact and elegant way different and complicated diagrammatic contributions, encoding efficiently ultraviolet cancellations and subtle effects resulting from regularizing and renormalizing the flat-space perturbative series.  
	
	\paragraph{Field theory set-up}
	In flat space, we consider  $\mathrm{SU}(N)$ $\mathcal{N}=2$ SYM theories with massless hypermultiplets in an arbitrary representation $\mathcal{R}$ such that the   $\beta$-function is non-vanishing. The explicit expression of the actions is reported in Appendix \ref{sec:actions in flat space}.
	
	The 1/2 BPS Wilson loop operator in the fundamental representation is defined by 
	\begin{equation}
		\label{eq:1/2 BPS supersymmetric Wilson loop}
		\widehat{\cW} = \dfrac{1}{N} \text{tr} \ \mathcal{P} \exp \bigg\{ g_B \int_C \dd{\tau} \bigg[ \mathrm{i} A^{\mu}(x(\tau)) \dot{x}_{\mu}(\tau) + \dfrac{R}{\sqrt{2}} \Big( \Bar{\phi}(x(\tau)) +\phi(x(\tau)) \Big)\bigg]
		\bigg\} \ ,
	\end{equation} where $g_B$ is the bare coupling constant, while $\cP$ denotes the path-ordering operator. In the previous expression, the gauge field $A_\mu(x(\tau))$ and the vector-multiplet  scalar $\phi(x(\tau))$ are integrated over a circle $C$ of radius $R$ and canonically parametrized by \begin{equation}
		\label{eq:parametrization}
		x^{\mu}(\tau)=R(\cos{ \tau }, \sin{ \tau }, 0 ,0) \ ,  \quad \text{with} \quad 0\le\tau<2\pi \ .
	\end{equation}

	The vacuum expectation value of (\ref{eq:1/2 BPS supersymmetric Wilson loop}) contains ultraviolet divergent diagrams. To regularize the singular corrections and preserve the extended supersymmetry we  dimensionally reduce the theory from four to $d = 4 - 2\epsilon$
	dimensions  \cite{Erickson:2000af}. In this scheme, the gauge field $A_\mu$ is a $d$-dimensional vector, while the real scalars generated by the reduction are denoted with $A_i$, with $i=1,\ldots, 2\epsilon$. Since the bare coupling is dimensionless only when $d=4$, this regularization scheme breaks classical conformal symmetry. As a result, the dimensionally regularized observable can only depends on the combination $\hat{g}_B=R^{\epsilon}g_B$. Perturbatively, we expand the expectation value as follows
	\begin{equation}
		\label{eq:dimensionally vev}
		\big<\widehat{\cW}\big> \equiv \cW= 
		1+ \cW_2 + \cW_4 + {\cW}_6 + \cO(\hat{g}_B^8) \ , 
	\end{equation} 
	where the quantities $\cW_{2k}$ are proportional to $\hat g_B^{2k}$. Throughout this work, unless stated otherwise,  the Feynman gauge will be always understood.
	
	\paragraph{Structure of the paper}This paper is organized as follows. In Section \ref{sec:loc}, we present the structure of Pestun's matrix model in general massless $\mathcal{N}=2$ theories with matter representation associated with a non-vanishing $\beta$-function. Subsequently, we consider the insertion of the 1/2 BPS Wilson loops and derive
	the explicit prediction of localization for its perturbative expansion up to order $g^6$. In Section \ref{sec:ft_m}, we present the field theory analysis in flat space. We will first review the two-loop results obtained in \cite{Billo:2023igr} and explain the non-trivial role of additional evanescent terms which result from the integration over the Wilson loop contour. Upon renormalization, these contributions produce finite three-loop corrections which combine with the diagrams presented in subsection \ref{subsec:3lcorr}. Finally, in Section \ref{eq:renormalization}, we discuss the renormalization of the Wilson loop operator. We show that the structure of the divergences respects the usual renormalization properties expected for this operator and that within the specific range of energy scales (\ref{eq:range of energies and lambda di strong couplign }), the perturbative series in flat space coincides with the prediction of the matrix model. Finally, in Section \ref{sec:conclusions}, we draw our conclusions and present some possible future directions. Calculation details of the three-loop diagrams involves several intermediate steps, mainly related to intricate  path-ordered integration over the Wilson loop contour which, as far as we know, have not been performed in the current literature. These computations are presented in detail in five different appendices.
	
	\section{Predictions from localization}
	\label{sec:loc}
	In this work, we consider $\mathcal{N}=2$ theories with SU($N$) gauge group and massless hypermultiplets in an arbitrary representation $\mathcal{R}$ such that the  $\beta$-function is non-vanishing. When these theories are compactified on $\mathbb{S}^4$,  supersymmetric localization enables to reduce the  path integral to an interacting matrix model. However,  the one-loop fluctuation  determinants require a regularization  which involves additional degrees of freedom of mass $M$ \cite{Pestun:2007rz,Billo:2023igr}. The purpose of this section is to introduce the (regularized) matrix model\footnote{In particular, we refer to Section 1 of \cite{Billo:2023igr} for the technical details.} which describes the vacuum expectation value of the 1/2 BPS Wilson loop on $\mathbb{S}^4$ for this class of theories and present the three-loop prediction for this observable.
	
	\subsection{The $\mathbb{S}^4$ partition function} 
	\label{sec:mm}
	Compactifying a  generic SU($N$) $\mathcal{N}=2$ SYM theory on a four-sphere $\mathbb{S}^4$ of radius $R$, localization \cite{Pestun:2007rz} maps the partition function into a matrix model, i.e.
	\begin{equation}
		\label{eq:partition function}
		\mathcal{Z} = \int \mathcal{D}a \  \left|Z(\mathrm{i}a,{g},{R})\right|^2 \ .
	\end{equation} 
	In the previous expression, $a$ is an $N\times N$ Hermitian traceless matrix  whose eigenvalues $a_u$ parametrize the Coulomb moduli space and the integration measure is given by
	\begin{equation}
		\label{eq:matrix model measure}
		\mathcal{D}a=\prod_{u=1}^{N}\dd{a}_u\Delta(a)\delta\bigg(\sum_{v=1}^{N}a_v\bigg)~, \quad \quad \text{with} \quad \quad  \Delta(a)= \prod^N_{u<v}(a_u-a_v)^2 \ .
	\end{equation} 
	denoting the Vandermonde determinant. This quantity represents the Jacobian of the transformation  which connects the integration over a Lie algebra $\mathfrak{g}$ to its Cartan subalgebra $\mathfrak{h}$. This means that $\mathcal{D}a$ is equivalent to the flat integration measure  \begin{equation}
		\label{eq:flat measure}
		\dd{a}=\prod^{N^2-1}_{b=1}\dd a_b \ , \quad \quad \text{where} \quad \quad a=a_bt^b \  .
	\end{equation} In the previous expression, we denoted with $t_n$  the $n$-th hermitian traceless generator of $\mathfrak{su}(n)$ in the fundamental representation where\footnote{The normalization of eq. (\ref{eq:normalization of the generators in the fundamental representation}) fixes the Dynkin index of the fundamental representation to $i_F=1/2$. } 
	\begin{equation}
		\label{eq:normalization of the generators in the fundamental representation}
		\tr t_a t_b =\dfrac{\delta_{ab}}{2}~,
	\end{equation}

	In the localized partition function  (\ref{eq:partition function}), the integrand consists of three different factors
	\begin{equation}
		\label{eq:integrand partition function}
		Z =  Z^\cR_{\mathrm{1-loop}}\, Z_{\mathrm{inst} \ }Z_{\mathrm{cl}}~.
	\end{equation} In the previous expression, $Z_{\rm inst}$ describes the instanton contribution, which can be discarded since we will primarily work in perturbation theory, while  $ Z_{\mathrm{cl}}$ and $Z^\cR_{\mathrm{1-loop}}$ denote, respectively, the classical term   of the matrix model and its interaction potential, which depends on the representation $\mathcal{R}$. These quantities are defined as follows \cite{Billo:2023igr} 
	\begin{equation}
		\label{eq:one-loop in general massless theory}
		\begin{split}
			\left|Z_{\mathrm{cl}}(\mathrm{i}a,{g})\right|^2
			=\mathrm{e}^{-\frac{8\pi^2 {R}^2}{{g}^2}\tr a^2}~, \quad \quad \quad \quad 
			\left|Z^\mathcal{R}_{\text{1-loop}}\right|^2 = \dfrac{\prod_{\mathbf{w}_{\rm Adj}} H( R\, \mathbf{w}_{\rm Adj}\cdot \mathbf{a})}{\prod_{\mathbf{w}_{\mathcal{R}}} H( R\, \mathbf{w}_\mathcal{R}\cdot \mathbf{a})} \ .
		\end{split}
	\end{equation} 
	In the previous expression,  $g$ is the running coupling defined in eq. (\ref{eq:classical}), $\mathbf{a}$ denotes an $N$-dimensional vector containing the eigenvalues of the matrix $a$, while $\mathbf{w}_\mathcal{R}$ and $\mathbf{w}_{\rm Adj}$ are the weight-vectors of the representation $\mathcal{R}$ and of the adjoint one respectively. Moreover, $H(x)$ is defined through the product of   Barnes' G-function as follows \cite{Russo:2013kea}
	\begin{equation}
		\label{eq:H functions}
		\begin{split}
			H(x)&= G(1+\ii x) G(1-\ii x)\, \ee^{-(1+\gamma)x^2}=  \prod_{n=1}^{\infty}\left(1+\frac{x^2}{n^2}\right)^n \mathrm{e}^{-\frac{x^2}{n}} \ ,
		\end{split}
	\end{equation}
	where $\gamma$ is the Euler's constant. Using the properties of the $G$-function, it is straightforward to show that for small values of the argument we have \begin{equation}
		\label{Hsmallx}
		\log H(z) = 
		-\sum_{m=2}^{\infty}(-1)^m \dfrac{\zeta(2m-1)z^{2m}}{m}\ .
	\end{equation} The contribution of the one-loop determinants in eq. (\ref{eq:one-loop in general massless theory}) can be exponentiated and interpreted as an interaction potential for the matrix model, i.e.   \begin{align}
		\label{eq:interaction action}
		S_{\text{int}}(a)  \equiv - \log \big| Z^\mathcal{R}_{\text{1-loop}}\big|^2
		=\left(\Tr_{\mathcal{R}}-\Tr_{\rm Adj}\right)H(Ra) \ .
	\end{align} Combining together the relations of this subsection and rescaling the integration variable according to $a \to \left(\frac{g^2}{8\pi^2 R^2}\right)^{\frac{1}{2}}a $,  we can write the localized partition function with the contribution of the instanton suppressed as follows\footnote{In eq. (\ref{eq:ZN=2generic}), we did not include the Jacobian of the transformation $a \to \left(\frac{g^2}{8\pi^2 R^2}\right)^{\frac{1}{2}}a $ since it introduces a multiplicative constant which  disappears in properly normalized expectation values.} \cite{Billo:2019fbi,Billo:2023igr} \begin{align}
		\label{eq:ZN=2generic}
		%
		\mathcal{Z}&= 
		\int \dd a \   \ee^{- \tr a^2 - S_{\text{int}}(a,g)}\ .
	\end{align} In the previous expression, the measure $\dd{a}$ is defined in eq. (\ref{eq:flat measure}) and is normalized in such a way that $\int \dd a\ \ee^{-\tr a^2 }=1$, while  the interaction potential of eq. (\ref{eq:interaction action}) acquires a dependence on $g$ and can be expended as a power series by eq. (\ref{Hsmallx}), i.e. 
	\begin{align}
		\label{Sintbis}
		\begin{split}
			S_{\text{int}}(a,g) &= - \sum_{m=2}^{\infty}\left(-\frac{g^2}{8\pi^2}\right)^m \dfrac{\zeta(2m-1)}{m} 
			\Tr_\mathcal{R}^\prime a^{2m}~,
		\end{split}
	\end{align}
	where we introduced the primed trace $ \Tr^\prime_\mathcal{R} = \left(\Tr_{\mathcal{R}}-\Tr_{\rm Adj}\right) $. Note that this combination of traces only vanishes in  $\mathcal{N}=4$ SYM theories\footnote{Let us recall that $\mathcal{N}=4$ SYM can be seen as a $\mathcal{N}=2$ vector multiplet coupled to a single adjoint hypermultiplet, i.e. $\cR=\rm Adj$. As a result, the theory is superconformal and $\Tr_{\mathcal{R}}^\prime =0$.}. For general set-ups, the primed trace is non-vanishing and precisely describes the matter sector of the \textit{difference theory}, which arises when we subtract  the field content of $\mathcal{N}=4$ SYM from that associated with $\mathcal{N}=2$ theories  with hypermultiplets in the representation $\cR$. From the perturbative field theory point of view,  the matrix model suggests to construct the interaction contributions by considering the diagrams characterized by internal lines in the representation $\cR$ and by subtracting identical terms in which $\cR=\rm Adj$. For instance, the expected correspondence between a contribution in the matrix model which arises from the quartic vertex  $\Tr_{\mathcal{R}}^\prime a^4$ and the usual Feynman diagrams is    \begin{equation}
		\label{eq:correspondence}
		\Tr_{\mathcal{R}}^\prime a^4 = \mathord{
			\begin{tikzpicture}[radius=2.cm, baseline=-0.65ex,scale=0.5]
				\begin{feynman}
					\vertex (A) at (1,0);
					\vertex (B) at (-1,0);
					\vertex (C) at (0,1);
					\vertex (D) at (0,-1);
					\vertex (O) at (0,0);
					\diagram*{
						(A) --[ plain,thick] (O),
						(D) --[ plain,thick] (O),
						(B)--[plain,thick] (O),
						(C)--[plain,thick] (O),
					};
				\end{feynman}
				\node at (0,0)[circle,fill,inner sep=2pt]{};
		\end{tikzpicture} }  \ \quad \leftrightarrow \quad \mathord{
			\begin{tikzpicture}[radius=2.cm, baseline=-0.65ex,scale=0.55]
				\filldraw[color=white!80, fill=white!15](0,0) circle (1);	
				\draw [black, thick]   (0,0) circle [radius=1.];
				\draw [black, thick, dashed]   (0,0) circle [radius=0.9];
				\begin{feynman}
					\vertex (A) at (0,2.);
					\vertex (C) at (0, 1.);
					\vertex (D) at (0, -1.);
					\vertex (d) at (0.1,0) ;
					\vertex (B) at (0,-2.);
					\vertex (a) at (-1,0);
					\vertex (a1) at (-2,0);
					\vertex (b) at (1,0);
					\vertex (b1) at (2,0);
					\diagram*{
						(A) -- [photon] (C),
						(A) --[ plain] (C),
						(D) -- [photon] (B),
						(D) --[ plain] (B),
						(a1)--[photon] (a),
						(a1)--[plain] (a),
						(b)--[photon] (b1),
						(b)--[plain] (b1),
					};
				\end{feynman}
		\end{tikzpicture} } \ .
	\end{equation} In the previous expression, we used a double dashed/continuos line to denote the propagation of matter in the difference theory approach, while the wiggly/straight lines are associated with vector-multiplet fields. In $\mathcal{N}=2$ superconformal set-ups, the correspondence between matrix model vertices and matter loops  was tested at high orders in perturbation theory for different observables \cite{Andree:2010na,Billo:2017glv,Billo:2019fbi}. However, in non-conformal models, it is no longer obvious whether this connection persists due to the conformal symmetry breaking. 
	
	\subsection{Supersymmetric Wilson loop}
	\label{sec:WL in matrix model}
	In this section, we study the 1/2 BPS circular Wilson loop in the fundamental representation. According to  \cite{Pestun:2007rz}, the vacuum expectation value of this operator can be evaluated via the following matrix model  \cite{Billo:2023igr}
	\begin{align}
		\label{vevWbis}
		W(g)=   \frac{1}{\cZ} \int \dd a\, \ee^{- \tr a^2 - S_{\text{int}}(a,g)}\, \cW(a,g)	~,  
	\end{align} where the matrix operator $\cW(a,g)$ is defined as follows \begin{equation}
		\label{eq:wilson loop matrix operator}
		\cW(a,g)=\dfrac{1}{N} \tr\exp(\frac{ a g }{\sqrt{2}}) = 1 + \dfrac{g^2}{4N}\tr a^2 +\cO(g^2) \ .
	\end{equation}
	
	The matrix model in eq. (\ref{vevWbis}) formally coincides with that considered in \cite{Billo:2019fbi} for the expectation value of the supersymmetric Wilson loop  in  generic superconformal $\mathcal{N}=2$ theories. In the range of energies (\ref{eq:range of energies and lambda di strong couplign }), the running coupling $g$, defined in eq. (\ref{eq:classical}), goes to zero and we can expand  the interaction action via (\ref{Sintbis}). As a result, we find that\footnote{The subscript $0,c$ denotes the connected correlator  in the Gaussian matrix model, i.e. $\vvev{f(a)\,g(a)}_{0,c}=\vvev{f(a)\,g(a)}_0 -\vvev{f(a)}_0 \vvev{g(a)}_0$ with  $f(a)$ and $g(a)$ being arbitrary functions of $a$.}
	\begin{align}
		\label{WexpSint}
		W(g) = W_0(g) 
		+ \left(\frac{g^2}{8\pi^2}\right)^2 \frac{\zeta(3)}{2} \vvev{\cW(a,g)\,\Tr_\mathcal{R}^\prime a^4}_{0,c}	
		+ \cO(g^8) \ .
	\end{align}
	The first term on the right-hand side of the previous expression denotes  the expectation value of the BPS Wilson loop in the Gaussian matrix model, i.e.  \cite{Billo:2019fbi,Billo:2023igr}
	\begin{align}
		\label{expW0}
		W_0&  =\dfrac{1}{N}L^1_{N-1}\left(-\frac{g^2}{4}\right)\,\exp(\frac{{g}^2}{8}\Big(1-\frac{1}{N}\Big)) \notag\\[0.4em]
		&= 1 + \frac{g^2C_F }{4} +   \frac{g^4C_F(2N^2-3)}{192N}
		+  \frac{g^6C_F(N^4-3N^2+3)}{4608 N^2} +\ldots \ , 
	\end{align}
	where $C_F=(N^2-2)/2N$ is the fundamental Casimir, while $L_m^n(x)$ denotes the $n$-th generalized Laguerre polynomial of degree $m$. In $\mathcal{N}=4$ SYM, where the matrix model is Gaussian and $g$ is a pure parameter, the observable is precisely given by the previous expression which, from a diagrammatic point of view, encodes the resummation of the ladder-like corrections   \cite{Erickson:2000af,Drukker:2000rr}. 
	
	Turning our attention to the effects of the interaction action (\ref{WexpSint}), we note that these become evident only at three-loop accuracy. In particular, expanding the Wilson loop operator via eq. (\ref{eq:wilson loop matrix operator}),  we find that the lowest order contribution takes the form
	\begin{align}
		\label{g6corr}
		\left(\frac{g^2}{8\pi^2}\right)^2 \frac{\zeta(3)}{2} \vvev{\cW(a,g)\,\Tr_\mathcal{R}^\prime a^4}_{0,c} = 
		\left(\frac{g^2}{8\pi^2}\right)^2 \frac{\zeta(3)}{2} \frac{g^2}{4N} \vvev{\tr a^2\,\Tr_\mathcal{R}^\prime a^4}_{0,c} + \mathcal{O}(g^8)~.
	\end{align}
	To evaluate the connected correlator for an arbitrary $\cR$ we can introduce the free contraction $\big<{a^a a^b}\big>_0 =\delta^{ab}$ and apply Wick theorem. By considering the legitimate  contractions, it is straightforward to show that 
	\begin{equation}
		\label{eq:conneted correlator}
		\left(\frac{g^2}{8\pi^2}\right)^2 \frac{\zeta(3)}{2} \frac{g^2}{4N} \vvev{\tr a^2\,\Tr_\mathcal{R}^\prime a^4}_{0,c} = \dfrac{g^63\zeta(3)}{2^8\pi^4N}\cK_{4}^\prime + \dfrac{g^6\zeta(3)C_FN\beta_0}{16\pi^2} \ .
	\end{equation} In the previous expression, $\beta_0$ is the one-loop coefficient of the $\beta$-function, defined in eq. (\ref{eq:beta0}), and we  introduced the SU($N$)-invariant quantity \begin{align}
		\label{eq:def K}
		\cK_4^\prime =\Tr_\mathcal{R}^\prime T_aT_eT^aT^e =2N C_F\left(C_\cR i_\cR-\dfrac{N i_\cR}{2}-\dfrac{N^2}{2}\right)  \ .
	\end{align}

	The two interaction contributions in eq. (\ref{eq:conneted correlator}) correspond to the two inequivalent contractions of matrix model quartic vertex \begin{equation}
		\label{eq:matrix model diagrams}
		\mathord{
			\begin{tikzpicture}[baseline=-0.65ex,scale=0.5]
				\draw [black] (0,0) circle [radius=1.5cm];
				\begin{feynman}
					\vertex (A) at (0,1.5);
					\vertex (B) at (0,-1.5);
					\vertex (C) at (0,0);
					\vertex (D) at (-1,0);
					\diagram*{
						(A) -- [plain, thick] (B),
						(C)--[plain,thick,half right] (D),
						(C)--[plain,thick,half left] (D),	
					};
				\end{feynman}
				\node at (0,0)[circle,fill,inner sep=2pt]{};
			\end{tikzpicture} 
		} \ , \quad \quad \mathord{
			\begin{tikzpicture}[baseline=-0.65ex,scale=0.5]
				\draw [black] (0,0) circle [radius=1.5cm];
				\begin{feynman}
					\vertex (A) at (0,1.5);
					\vertex (B) at (0,-1.5);
					\vertex (C) at (0,-0.5);
					\vertex (D) at (0,0.5);
					\diagram*{
						(A) -- [plain, thick] (B),
						(D)-- [plain,half right,thick] (C),
						(D)-- [plain,half left,thick] (C),
					};
				\end{feynman}
				\node at (0,0.5)[circle,fill,inner sep=2pt]{};
			\end{tikzpicture}  
		} \ .
	\end{equation} The correspondence between the matrix model vertices and matter loops (\ref{eq:correspondence}) suggests that these interaction contributions proportional to $\zeta(3)$ should emerge in perturbation theory from two inequivalent  single-exchange diagrams. As we already stressed in the previous section, this correspondence was originally tested in \cite{Andree:2010na,Billo:2019fbi} for generic superconformal set-ups, where only the correction proportional to $\cK_4^\prime$ is present. In non-conformal models, the prediction of the matrix model also includes an additional term proportional to   $\beta_0$. In the following sections, we will show that this novel contribution emerges in perturbative field theory by interference effects between the (UV) poles of the bare coupling and evanescent factors associated with special parts of diagrams which behave as single exchange correction.

	Finally, combining together the relations we derived in this subsection, we  obtain a simple expression for the three-loop prediction, i.e. 
	\begin{equation}
		\label{eq:prediction of localization}
		\begin{split}
			W(g) = W_0(g) + g^6\dfrac{3\zeta(3)}{2^8\pi^4N}\cK_4^{\prime} +g^6 \dfrac{\zeta(3)C_FN\beta_0}{16\pi^2} +\cO(g^8)\ ,
		\end{split}
	\end{equation} where we recall that $W_0(g)$ is given by (\ref{expW0}). Let us stress again that the previous expression is valid within the range  (\ref{eq:range of energies and lambda di strong couplign }). Relaxing this condition, we expect that the observable receives non-perturbative \textit{infrared} corrections (see comments after eq. (\ref{eq:range of energies and lambda di strong couplign })) which make the result on the sphere different  from the flat-space counterpart.
	
	\section{Field theory in flat space}
	\label{sec:ft_m}
	Let us begin with observing that at any perturbative order $\hat g_B^{2k}$,  we can organize the quantities $\cW_{2k}$ of eq. (\ref{eq:dimensionally vev}) as follows:
	\begin{equation}
		\label{W2kis}
		\cW_{2k} = \cW_{2k}^{\rm ladder} + \cW_{2k}^{\rm v.m.} + \cW_{2k}^{\cR}~. 
	\end{equation}	
	The first two contributions capture, respectively,  the ladder-like diagrams, in which the gauge field $A_\mu$ and the scalar field $\phi$ are exchanged at tree-level, and the interaction corrections with internal vertices and lines of the vector multiplet only. These contributions are in common with the $\cN=4$ theory. By $\cW_{2k}^{\cR}$ we denote, instead, the diagrams that contain internal lines associated with the matter hypermultiplets in the representation $\cR$.
	
	It is well known that in the $\cN=4$ theory, where matter transforms in the adjoint representation,  only the ladder-like diagrams contribute to the expectation value of the Wilson loop in the limit  $d\to 4$. This means that, in general, we can write  
	\begin{equation}
		\label{vmisad}
		\cW_{2k}^{\rm v.m.} = - \cW_{2k}^{\rm Adj} + \delta\cW_{2k}^{\rm v.m.}~,  
	\end{equation}
	where $\delta\cW_{2k}^{\rm v.m.}$ is an evanescent corrections: it vanishes for $d=4$ and can be  expanded in power series of $\epsilon = (4-d)/2$. As we will discuss in Section \ref{eq:renormalization}, upon renormalization,  the ultraviolet poles of the bare coupling constant $\hat{g}_B$ interfere with the evanescent terms and produce finite corrections at higher orders in perturbation theory. This means that the renormalized expectation value at three loops, also receives non-trivial contributions from the two-loop evanescent corrections $\delta\cW^{\rm v.m.}_4$ which we will compute explicitly in the following subsection.
	
	Substituting eq. (\ref{vmisad}) into eq. (\ref{W2kis}), we have
	\begin{align}
		\label{W2kisdiff}
		\cW_{2k} = \cW_{2k}^{\rm ladder} + \cW_{2k}^\prime + \delta\cW^{\rm v.m.}_{2k}~, \quad \text{where} \quad \cW_{2k}^\prime \equiv \cW_{2k}^{\cR} - \cW_{2k}^{\rm Adj} \ .
	\end{align}
	Thus, besides the ladder-like diagram and the corrections $\delta \cW_{2k}$, at any perturbative order  the interaction contributions are constructed by subtracting from $\cW_{2k}^{\cR}$ exactly the same diagrams in which the internal matter lines are in the adjoint representation. This combination of contributions, denoted by $\cW_{2k}^\prime$,   precisely encodes the \textit{difference theory} diagrams predicted by the interaction action of the matrix model, see eq. (\ref{eq:interaction action}). 
	
	For $d=4$ the ladder-like contributions $\cW_{2k}^{\rm ladder}$ are known for every $k$ and are captured by a Gaussian matrix model through eq. (\ref{expW0}). 
	Thus, for $d\to 4$, we can write \begin{equation}
		\label{ladderd}
		\cW_{2k}^{\rm ladder} = \left.\cW_{2k}^{\rm ladder}\right|_{d=4} + \delta \cW_{2k}^{\rm ladder}~.
	\end{equation} 
	The evanescent corrections $\delta \cW_{2k}^{\rm ladder}$ can contribute, upon renormalization, to higher perturbative orders. For our purposes, we will have to compute $\delta \cW_{4}^{\rm ladder}$. 
	
	\subsection{One-loop corrections}
	At order $\hat g_B^2$, the Wilson loop expectation value receives contributions from a single class of ladder-like diagrams, i.e. 
	\begin{equation}
		\label{eq:ladder g2 pic}
		\cW_2 =	\mathord{
			\begin{tikzpicture}[baseline=-0.65ex,scale=0.5]
				\draw [black] (0,0) circle [radius=2cm];
				\begin{feynman}
					\vertex (A) at (0,2);
					\vertex (B) at (0,-2);
					\diagram*{
						(A) -- [photon] (B),
					};
				\end{feynman}
			\end{tikzpicture} 
		}+	\mathord{
			\begin{tikzpicture}[baseline=-0.65ex,scale=0.5]
				\draw [black] (0,0) circle [radius=2cm];
				\begin{feynman}
					\vertex (A) at (0,2);
					\vertex (B) at (0,-2);
					\diagram*{
						(A) --[ fermion] (B),
					};
				\end{feynman}
			\end{tikzpicture} 
		} 
		\equiv 	\mathord{
			\begin{tikzpicture}[baseline=-0.65ex,scale=0.5]
				\draw [black] (0,0) circle [radius=2cm];
				\begin{feynman}
					\vertex (A) at (0,2);
					\vertex (B) at (0,-2);
					\diagram*{
						(A) -- [photon] (B),
						(A) --[ fermion] (B),
					};
				\end{feynman}
			\end{tikzpicture} 
		} \  .
	\end{equation} 
	In the previous expression, we employed the double  straight/wiggly line of eq. (\ref{eq:correspondence}) to depict the tree level propagators of the adjoint scalar and of the gauge-field. 
	In the $d$ dimensional Euclidean space, their expression is given by  
	\begin{align}
		\label{eq:tree-level prop}
		\big<\phi^a(x_1)\bar{\phi}^b(x_2)\big>_0&=\delta^{ab}\Delta(x_{12})\notag \ , \\[0.4em] 
		\big<A_\mu^a(x_1)A^b_\nu(x_2)\big>_0	&=\delta_{\mu \nu}\delta^{ab}\Delta(x_{12})  \ ,
	\end{align} 
	where we introduced the notation $x_{12}\equiv x_1-x_2$, while the function $\Delta(x_{12})$ is given by%
	\footnote{This corresponds to the case $s=1$ in eq. (\ref{eq:Fourier transform for massless propagators}), since the in momentum space the tree-level propagator is simply $1/p^2$. }
	\begin{equation}
		\begin{split}
			\label{eq:Deltad}
			\Delta(x_{12})  = \cD(x_{12},1) = \dfrac{\Gamma(1-\epsilon)}{4\pi^{2-\epsilon}\left(x_{12}^2\right)^{1-\epsilon}} \ .
		\end{split}
	\end{equation}

	Expanding the Wilson
	loop (\ref{eq:1/2 BPS supersymmetric Wilson loop}) at order $g_B^2$, and employing the free Wick contractions (\ref{eq:tree-level prop}), we obtain  the following representation for the diagrams in eq.  (\ref{eq:ladder g2 pic}): 
	\begin{equation}
		\label{eq:ladder g2}
		\cW_2 = 
		\mathord{
			\begin{tikzpicture}[baseline=-0.65ex,scale=0.5]
				\draw [black] (0,0) circle [radius=2cm];
				\begin{feynman}
					\vertex (A) at (0,2);
					\vertex (B) at (0,-2);
					\diagram*{
						(A) -- [photon] (B),
						(A) --[ fermion] (B),
					};
				\end{feynman}
			\end{tikzpicture} 
		} =  \dfrac{g_B^2C_F}{2}\oint\dd^2\tau \left(R^2-\dot{x}_1\cdot\dot{x}_2\right) \Delta(x_{12}) \ . 
	\end{equation}
	The two terms above are, respectively, associated with the propagation of the adjoint scalar and of the gauge field inside the Wilson loop. In particular, as will see in the following sections, this combination enters all the diagrams contributing to the BPS Wilson loop (\ref{eq:1/2 BPS supersymmetric Wilson loop}).  Consequently, it is convenient to introduce the following \textit{effective (tree-level) propagator} on the Wilson loop:
	\begin{align}
		\label{defDeltahat}
		\widehat{\Delta}(x_{12})=(R^2-\dot{x}_1\cdot\dot{x}_2)\Delta(x_{12})=  \dfrac{\Gamma(1-\epsilon)}{8\pi^{2-\epsilon}} \left(4 R^2 \sin^2(\frac{\tau_{12}}{2})\right)^{\epsilon}~,
	\end{align}
	where in the second step we used the parametrization (\ref{eq:parametrization}). Substituting the previous expression in eq. (\ref{eq:ladder g2}), we observe that the integration over the contour reduces to a single integral of the form considered in  eq. (\ref{eq:Fourier coefficients}), namely
	\begin{equation}
		\label{a0is}
		a_0(\alpha) \equiv \frac 1\pi \oint \dd\tau \frac{1}{\left(4 \sin^2(\frac{\tau}{2})\right)^{\alpha}} = \frac{\sec(\pi \alpha)\Gamma(\alpha)}{\Gamma(1-\alpha)\Gamma(2\alpha)}~.
	\end{equation}
	As a result, it is straightforward to express the one-loop correction $\cW_2=\cW_{2}^{\rm ladder}$ in a closed form  which is valid for any $d$: 
	\begin{equation}
		\label{eq:ladder g^2 def and definition a_0}
		\cW_2^{\rm ladder} = 
		\hat g_B^2\,C_F \frac{\Gamma(1-\epsilon)}{8\pi^{\epsilon}} a_0(-\epsilon)\\[0.4em]
		\equiv 
		\hat g_B^2\,C_F B_1(\epsilon)~.
	\end{equation}	
	In the previous expression, we introduced, for future convenience,  the  set of functions
	\begin{equation}
		\label{defBn}	
		B_n(\epsilon) =
		\dfrac{\Gamma^n(1-\epsilon)}{8\pi^{n\epsilon}} a_0(-n\epsilon)~,
	\end{equation}  
	which are regular and independent of $n$  for $\epsilon\to 0$. As we will see in the following,  single-exchange contributions, dressed with the $(n-1)$-th loop corrections to the propagators, are expressed in terms of the function $B_n(\epsilon)$. 
	
	Expanding eq. (\ref{eq:ladder g^2 def and definition a_0}) about $\epsilon\to 0$ we can construct explicitly the two terms of eq. (\ref{ladderd}) at one loop. To do this properly we have, however, to re-express the bare coupling in terms of the renormalized one; we will do this in Section \ref{eq:renormalization}. 
	
	\subsection{Two loop corrections}
	\label{subsec:two-loop-corr}
	The two-loop corrections to the expectation value of Wilson loop were analysed in great details in \cite{Billo:2023igr}. We devote this subsection to review the results at order $\hat{g}_B^2$ and determine the relevant evanescent corrections we will employ for the three-loop analysis. 
	According to eq. (\ref{W2kis}), we organize  the different families of two-loop diagrams in terms of three distinct classes of terms, i.e. $\cW_4^{\rm ladder}$, $\cW_4^{\rm v.m.}$ and $\cW_4^\cR$.  
	
	Let us begin with discussing the two-loop ladder-like diagrams. Expanding the Wilson loop operator (\ref{eq:1/2 BPS supersymmetric Wilson loop}) at order $g_B^4$ and employing the tree-level propagators of the adjoint scalar and gauge field (\ref{eq:tree-level prop}), we find the ladder corrections
	\begin{align}
		\label{W4ladder}
		\mathord{
			\begin{tikzpicture}[baseline=-0.65ex,scale=0.45]
				\draw [black] (0,0) circle [radius=2cm];
				\begin{feynman}
					\vertex (A) at (-0.75,1.8);
					\vertex (B) at (-0.75,-1.8);
					\vertex (C) at (0.75,1.8);
					\vertex (D) at (0.75,-1.8);
					\diagram*{
						(A) -- [photon] (B),
						(A) --[ fermion] (B),
						(C) -- [photon] (D),
						(C) --[ fermion] (D)
					};
				\end{feynman}
			\end{tikzpicture} 
		} 
		&		=\dfrac{g_B^4}{N}\oint_{\cD}\dd^4\tau\Biggl\{ \ C^{aabb}\left(\widehat{\Delta}(x_{12})\widehat{\Delta}(x_{34})+ \widehat{\Delta}(x_{14})\widehat{\Delta}(x_{23})\right) + C^{abab}\widehat{\Delta}(x_{13})\widehat{\Delta}(x_{24})\Biggr\}\notag\\
		&=	\cW_4^{\rm ladder}    \ .
	\end{align}
	In the previous expression, the domain of integration $\cD$ denotes the ordered region $\tau_1>\tau_2>\tau_3>\tau_4$, the propagator $\widehat{\Delta}(x)$ is defined in eq. (\ref{defDeltahat}) and we introduced the SU($N$) tensor 
	\begin{equation}
		\label{eq:trace of four generators}
		C^{abcd} = \tr T^a T^b T^c T^d \ .
	\end{equation}  
	Using the properties of the non-Abelian exponentiation of the Wilson loop \cite{GATHERAL198390,FRENKEL1984231}, we can reduce eq. (\ref{W4ladder}) to the following expression  
	\begin{equation}
		\label{eq:ladder with non-maximally abelian part}
		\cW_4^{\rm ladder} = 
		\dfrac{1}{2}\left(\cW^{\rm ladder}_2\right)^2 + \dfrac{\hat{g}_B^4}{2N} \tr \left(\big[T^b,T^a\big]\right)^2\oint_\cD\dd^4\tau \widehat{\Delta}(x_{13})\widehat{\Delta}(x_{24})  \ , 
	\end{equation}	
	where $\cW^{\rm ladder}_2$ is the ladder-like contribution of eq. (\ref{eq:ladder g^2 def and definition a_0}), while the second term defines the so-called \textit{maximally non-Abelian part} of the diagrams. The nested integration in this last term is treated in detail in Appendix \ref{sec:Fourier} by Fourier representations. Employing the parametrization (\ref{eq:parametrization}) and eq. (\ref{eq:integral ladder non-maximally def}), we finally find 
	\begin{align}
		\label{eq:ladderg4finalresult} 
		\cW_4^{\rm ladder} 
		&=  \hat g_B^4\dfrac{C_F (2N^2-3)}{12N} B^2_1(\epsilon) 
		-\epsilon \hat g_B^4 \dfrac{C_F N \zeta(3)}{16\pi^2} \ + \cO(\epsilon)^2~. 
	\end{align}
	Note that the term proportional to $\zeta(3)$ arises from the maximally non-Abelian part of the diagram. Further expanding the function $B_1(\epsilon)$, by employing eq. (\ref{defBn}), we can determine the complete expression of the evanescent correction $\delta\cW_4^{\rm ladder}$. For convenience, however,  we will present this calculation in Section \ref{eq:renormalization}, where we will discuss the renormalization of the Wilson loop.
	
	Secondly, we analyse the quantity $\cW_4^{\rm v.m.}$, which encodes all the two-loop diagrams uniquely characterized by internal vertices and lines associated with the vector-multiplet.  The only non-trivial contributions result from the 
	\textit{Mercedes-like} diagrams\footnote{In principle, one could also expect single-exchange diagrams dressed with the one-loop corrections to the adjoint scalar and gauge field propagators resulting from the vector-multiplet interaction. However, it follows from eq.s (\ref{eq:one-loop adjoint scalar appendix}) and (\ref{eq:one-loop correction gauge field appendix}) that these specific contributions are not present.}:
	%
	\begin{align}
		\label{W4vm}
		\cW_4^{\rm v.m.} = 
		\mathord{
			\begin{tikzpicture}[radius=2.cm, scale=0.5, baseline=-0.65ex]
				\draw [black] (0,0) circle [];
				\begin{feynman}is convenient to express  order.  
					\vertex (A) at (0,2.);
					\vertex (C) at (0,0);
					\vertex (D) at (-1.5, -1.3);
					\vertex (B) at (1.5, -1.3);
					\diagram*{
						(A) -- [photon] (C),
						(A) --[ fermion] (C),
						(C) -- [photon] (B),
						(C) --[fermion] (D),
						(C) --[ photon] (D)
					};
				\end{feynman} 
			\end{tikzpicture} 
		}~.	
	\end{align}
	This class of corrections  were originally discussed in \cite{Erickson:2000af}, where the authors studied the supersymmetric Wilson loop in $\mathcal{N}=4$ SYM and showed that 
	\begin{equation}
		\begin{split}
			\label{eq:real calculation Sigma 3}
			\mathord{
				\begin{tikzpicture}[radius=2.cm, scale=0.5, baseline=-0.65ex]
					\draw [black] (0,0) circle [];
					\begin{feynman}
						\vertex (A) at (0,2.);
						\vertex (C) at (0,0);
						\vertex (D) at (-1.5, -1.3);
						\vertex (B) at (1.5, -1.3);
						\diagram*{
							(A) -- [photon] (C),
							(A) --[ fermion] (C),
							(C) -- [photon] (B),
							(C) --[fermion] (D),
							(C) --[ photon] (D)
						};
					\end{feynman} 
				\end{tikzpicture} 
			} =  -~ 
			\mathord{
				\begin{tikzpicture}[radius=2.cm, baseline=-0.65ex,scale=0.5]
					\draw [black] (0,0) circle [];
					\filldraw[color=white!80, fill=white!15](0,0) circle (1);	
					\draw [black,thick]   (0,0) circle [radius=1.];
					\begin{feynman}
						\vertex (A) at (0,2.);
						\vertex (C) at (0, 1.);
						\vertex (D) at (0, -1.);
						\vertex (B) at (0,-2.);
						\diagram*{
							(A) -- [photon] (C),
							(A) --[ charged scalar] (C),
							(D) -- [photon] (B),
							(D) --[ charged scalar] (B)
						};
					\end{feynman},
			\end{tikzpicture} } 
			+ \delta \cW^{\rm v.m.}_4=-\cW_4^{\rm Adj}+  \delta \cW^{\rm v.m.}_4 \ .
		\end{split}
	\end{equation}
	This expression provides a concrete realization of eq.  (\ref{vmisad}) at two loops. In particular,  the bubble-like contribution denotes the one-loop correction to the adjoint scalar and gauge field propagator in $\mathcal{N}=4$ SYM, where the hypermultiplets are in the adjoint representation, while the evanescent correction $\delta\cW_4^{\rm v.m.}$ is given by 
	\begin{equation}
		\label{deltapW4}
		\delta \cW_4^{\rm v.m.} = 
		\epsilon \ \dfrac{\hat{g}_B^4 C_FN \Gamma(1-2\epsilon)  }{(2\pi)^{-2\epsilon}128\pi^4} \int_0^1 \dd F (\alpha\beta\gamma)^{-\epsilon} \ \oint\dd^3 \tau \ \varepsilon(\tau)  \ \dfrac{\sin \tau_{13}} {Q^{1-2\epsilon}}
		+\cO (\epsilon)^2 \ .
	\end{equation}
	In the previous expression, we introduced the quantities 
	\begin{align}
		\label{eq:def of Q}
		Q&=\alpha\beta(1-s\tau_{ 12})+\beta\gamma(1-s\tau_{ 23}) +\gamma\alpha(1-\cos\tau_{ 13})    \ ,  \\[0.2em] 
		\label{eq:integration over the unit cube}
		\dd F&=\dd\alpha\,\dd\beta\,\dd\gamma\,\delta(1-\alpha-\beta-\gamma) \ ,  \\[0.2em]
		\label{eq:defepsilon}
		\varepsilon(\tau)  &= \theta(\tau_{12})\theta(\tau_{23}) -\theta(\tau_{13})\theta(\tau_{32}) + \text{ permutations} \ .
	\end{align} 
	The path-ordered integral in eq. (\ref{deltapW4}) is completely regular in the limit $\epsilon\to 0 $ and is evaluated in Appendix \ref{sec:Fourier}. In particular, using eq. (\ref{eq:leading order I1}), we find that 
	\begin{equation}
		\label{eq:I1 main}
		\int_0^1 \dd F \ (\alpha\beta\gamma)^{-\epsilon} \  \oint \dd^3 \tau \ \varepsilon(\tau)\dfrac{\sin \tau_{ 13}}{Q^{1-2\epsilon}}
		=- 16 \pi^2\,\zeta(3)  +\mathcal{O}(\epsilon) \ .
	\end{equation}
	Substituting this expression in eq. (\ref{deltapW4}) and expanding the prefactor about $\epsilon\to 0 $, we finally arrive at the following result: 
	\begin{equation}
		\label{eq:interaction contributions at two-loop with evanescent term}
		\delta\cW^{\rm v.m.}_4 
		= -\epsilon \dfrac{\hat{g}_B^4 C_F N\zeta(3)  }{8\pi^{2} }   + \cO(\epsilon)^2\ .
	\end{equation}  
	
	The last quantity we have to determine is the correction $\cW_4^\cR$, which encodes all the diagrams characterized by internal lines associated with the matter hypermultiplets in the representation $\cR$. At two loops, we find 
	\begin{equation}
		\label{W4Ris}
		\cW_4^\cR = \mathord{
			\begin{tikzpicture}[radius=2.cm, baseline=-0.65ex,scale=0.5]
				\draw [black] (0,0) circle [];
				\filldraw[color=white!15, fill=white!15](0,0) circle (1);	
				\draw [black,dashed,thick]   (0,0) circle [radius=1.];
				\begin{feynman}
					\vertex (A) at (0,2.);
					\vertex (C) at (0, 1.);
					\vertex (D) at (0, -1.);
					\vertex (B) at (0,-2.);
					\diagram*{
						(A) -- [photon] (C),
						(A) --[ charged scalar] (C),
						(D) -- [photon] (B),
						(D) --[ charged scalar] (B)
					};
				\end{feynman},
		\end{tikzpicture} }  \ ,
	\end{equation}
	where the dashed virtual loop denote the one-loop corrections to the adjoint scalar and gauge field propagator resulting from matter field in the representation $\cR$. We can now combine the previous expression with eq. (\ref{eq:real calculation Sigma 3}) to construct the difference theory diagrams at two-loop, i.e.
	\begin{align}
		\label{diffW4}
		\cW_4^\prime = \cW_4^\cR - \cW_4^{\rm Adj} =
		\mathord{
			\begin{tikzpicture}[radius=2.cm, baseline=-0.65ex,scale=0.5]
				\draw [black] (0,0) circle [];
				\filldraw[color=white!15, fill=white!15](0,0) circle (1);	
				\draw [black,dashed,thick]   (0,0) circle [radius=1.];
				\begin{feynman}
					\vertex (A) at (0,2.);
					\vertex (C) at (0, 1.);
					\vertex (D) at (0, -1.);
					\vertex (B) at (0,-2.);
					\diagram*{
						(A) -- [photon] (C),
						(A) --[ charged scalar] (C),
						(D) -- [photon] (B),
						(D) --[ charged scalar] (B)
					};
				\end{feynman},
		\end{tikzpicture} } -
		\mathord{
			\begin{tikzpicture}[radius=2.cm, baseline=-0.65ex,scale=0.5]
				\draw [black] (0,0) circle [];
				\filldraw[color=white!80, fill=white!15](0,0) circle (1);	
				\draw [black,thick]   (0,0) circle [radius=1.];
				\begin{feynman}
					\vertex (A) at (0,2.);
					\vertex (C) at (0, 1.);
					\vertex (D) at (0, -1.);
					\vertex (B) at (0,-2.);
					\diagram*{
						(A) -- [photon] (C),
						(A) --[ charged scalar] (C),
						(D) -- [photon] (B),
						(D) --[ charged scalar] (B)
					};
				\end{feynman},
		\end{tikzpicture} } 
		\equiv
		\mathord{					\
			\begin{tikzpicture}[radius=2.cm, baseline=-0.65ex,scale=0.5]
				\draw [black] (0,0) circle [];
				\filldraw[color=white!80, fill=white!15](0,0) circle (1);	
				\draw [black, thick]   (0,0) circle [radius=1.];
				\draw [black, thick, dashed]   (0,0) circle [radius=0.9];
				\begin{feynman}
					\vertex (A) at (0,2.);
					\vertex (C) at (0, 1.);
					\vertex (D) at (0, -1.);
					\vertex (B) at (0,-2.);
					\diagram*{
						(A) -- [photon] (C),
						(A) --[ charged scalar] (C),
						(D) -- [photon] (B),
						(D) --[ charged scalar] (B)
					};
				\end{feynman}
		\end{tikzpicture} }  \ .
	\end{align}	
	
	Thus, we remain with a single-exchange contribution dressed with  the one-loop correction to the adjoint scalar and gauge field propagator in the difference theory approach.   	
	The expression of these propagators in configuration space are given by eq.s (\ref{eq:one-loop correction adjoint scalar main text}, \ref{eq:one-loop correction gauge-field position main text}).
	Note that the correction to the gluon propagator involves the  gauge-like term $\partial_{1,\mu}\partial_{2,\nu} 
	\Delta^{(1),{\rm g}}(x_{12})$ which,  when contracted with the tangent vectors $\dot x_1^\mu \dot x_2^\nu$, gives rise to total derivatives integrated over a closed path. These obviously  vanish and we remain with
	\begin{equation}
		\label{W4ris}
		\cW_4^\prime = \frac{g_B^2\, C_F}{2} \oint \dd^2\tau\,\widehat \Delta^{(1)}(x_{12})~.		
	\end{equation}
	In analogy to the ladder-like correction (\ref{eq:ladder g2}), we introduced an effective one-loop propagator on the Wilson loop contour
	\begin{align}
		\label{deltahat1is}
		\widehat{\Delta}^{(1)}(x_{12})& 
		= (R^2-\dot{x}_1\cdot\dot{x}_2)\Delta^{(1)}(x_{12})
		\notag\\[0.4em]
		& 
		= 
		\,
		\frac{g_B^2 f^{(1)}(\epsilon)\Gamma(1-2\epsilon)}{2^{3+2\epsilon} \pi^{2-\epsilon} \Gamma(1+\epsilon)}
		\left(4 R^2 \sin^2(\frac{\tau_{12}}{2})\right)^{2\epsilon}~,
	\end{align}		
	where, to obtain the second equality,  we used the explicit definition of the function $\Delta^{(1)}(x_{12})$, given by eq. (\ref{eq:one-loop correction adjoint scalar main text}), and the parametrization (\ref{eq:parametrization}).
	Performing the integration over the contour by eq. (\ref{a0is}) and by employing the definition of $f^{(1)}(\epsilon)$ in eq. (\ref{eq:scalar gluon pol 1-loop difference}), we produce a factor $2\pi^2\, a_0(-2\epsilon)$ and arrive at the following result:
	\begin{align}
		\label{W4Rfinal}
		\cW_4^\prime
		= \hat g_B^4 C_F\, P_2(\epsilon)\, B_2(\epsilon)~, \quad \text{where} \quad 	
		P_2(\epsilon) = -\dfrac{\beta_0}{\epsilon(1 - 2 \epsilon)}
	\end{align}
	and we recall that the function $B_2(\epsilon)$ was defined in eq. (\ref{defBn}). Combining together the relations we derived in this subsection, we find that the two-loop corrections to Wilson loop v.e.v can be written as follows:
	\begin{equation}
		\label{eq:W4 final result}
		\cW_4 = \cW^{\rm ladder}_4+\cW_4^\prime+\delta\cW^{\rm v.m.}_4   \ .
	\end{equation}
	
	\subsection{Three-loop corrections}
	\label{subsec:3lcorr}
	The calculation of the three-loop diagrams is significantly more involved and technical than the two-loop one. However, the logical steps are identical except for the fact that we do not have to calculate the evanescent corrections since, upon renormalization, they contribute  to four loops. This means that the three-loop corrections take the following form   
	\begin{equation}
		\label{W6issimple}
		\cW_6 = \left.\cW_6^{\rm ladder}\right|_{d=4} +  \cW^\prime_6+ \cO(\epsilon) \ .
	\end{equation}
	
	Let us begin with analysing the ladder diagrams. In $d= 4$ dimensions, their expression is captured by eq. (\ref{expW0}). We find that 
	\begin{equation}
		\label{eq:ladder g6}
		\cW_6^{\rm ladder}   
		= \mathord{
			\begin{tikzpicture}[baseline=-0.65ex,scale=0.5]
				\draw [black] (0,0) circle [radius=2cm];
				\begin{feynman}
					\vertex (A) at (-0.75,1.8);
					\vertex (B) at (-0.75,-1.8);
					\vertex (C) at (0.75,1.8);
					\vertex (D) at (0.75,-1.8);
					\vertex (E) at (0,2);
					\vertex (F) at (0,-2);
					\diagram*{
						(A) -- [photon] (B),
						(A) --[ fermion] (B),
						(C) -- [photon] (D),
						(C) --[ fermion] (D),
						(E) --[ fermion] (F),
						(E) --[ photon] (F),
					};
				\end{feynman}
			\end{tikzpicture} 
		} 
		= \dfrac{\hat{g}_B^6C_F(N^4-3N^2+3)}{4608N^2} + \cO(\epsilon)~.
	\end{equation}
	
	The three-loop interaction contributions are encoded in the difference-theory term $\cW^\prime_6=\cW_6^\cR-\cW_6^{\rm Adj}$. Unlike its two-loop counterpart (\ref{diffW4}), $\cW_6^\prime$ consists of three different classes of Feynman diagrams which can be organized according to the number of insertions in the Wilson loop contour. We use the notation
	\begin{equation}
		\label{W6k}
		\cW_6^\prime = \cW_{6(2)}^\prime + \cW_{6(3)}^\prime+ \cW_{6(4)}^\prime~,
	\end{equation} to distinguish each contribution which we will  discuss in turn.
	
	\subsubsection{Diagrams with two insertions}
	At order $g_B^6$, we can insert in the Wilson loop contour a single scalar/gauge-field propagator dressed with the two-loop corrections in the difference theory approach. The explicit expressions of these corrections in configuration space is computed in Appendix \ref{sec:scalar polarization operator}, see eq.s (\ref{eq:two loop corrections adjoint conf},\ref{eq:two loop corrections gluon conf}). Expanding the Wilson loop at order $g_B^2$ and employing these relations, we find, using the usual difference-theory notation, the following expression
	\begin{equation}
		\label{eq:bubble-exchange}
		\cW_{6(2)}^\prime
		= \mathord{
			\begin{tikzpicture}[radius=2.cm, baseline=-0.65ex,scale=0.7]
				\draw [black] (0,0) circle [];
				\filldraw[color=gray!80, fill=gray!15](0,0) circle (1);	
				\draw [black, thick]   (0,0) circle [radius=1.];
				\draw [black, thick, dashed]   (0,0) circle [radius=0.9];
				\begin{feynman}
					\vertex (A) at (0,2.);
					\vertex (C) at (0, 1.);
					\vertex (D) at (0, -1.);
					\vertex (d) at (0.1,0) {\text{\footnotesize 2-loop }\normalsize} ;
					\vertex (B) at (0,-2.);
					\diagram*{
						(A) -- [photon] (C),
						(A) --[ fermion] (C),
						(D) -- [photon] (B),
						(D) --[ fermion] (B)
					};
				\end{feynman}
		\end{tikzpicture} }
		= \frac{g_B^2\, C_F}{2} \oint \dd^2\tau\, \hat\Delta^{(2)}(x_{12})~.
	\end{equation}
	In analogy to the one/two-loop corrections (\ref{eq:ladder g2}) and (\ref{diffW4}), we  defined the two-loop effective propagator on the Wilson loop contour as follows 
	\begin{align}
		\label{Deltahat2is}
		\hat\Delta^{(2)}(x_{12}) &= (R^2 - \dot x_1\cdot\dot x_2)  \Delta^{(2)}(x_{12})\notag \\[0.4em]
		&= f^{(2)}(\epsilon) \, \frac{g_B^4\Gamma(1-3\epsilon)}{2^{3+4\epsilon}\pi^{2-\epsilon} \Gamma(1 + 2 \epsilon)} 
		\frac{1}{\left(4 R^2 \sin^2 \frac{\tau_{12}}{2}\right)^{-3\epsilon}}~,
	\end{align} 
	where to obtain the second equality we employed eq. (\ref{eq:two loop corrections adjoint conf}) and the parametrization (\ref{eq:parametrization}).
	
	Substituting eq. (\ref{Deltahat2is}) in eq. (\ref{eq:bubble-exchange}), we can easily integrate over the Wilson loop contour by means of eq. (\ref{a0is}). Moreover, recalling that $f^{(2)}(\epsilon)$, given by eq.s (\ref{fis},\ref{fiis}), contains four different terms, we finally find
	\begin{align}
		\label{fourF}
		\cW_{6(2)}^\prime =\sum_{i=1}^4  F^{(2)}_i \ , \quad \text{where} \quad 
		F_i^{(2)} = f_i^{(2)}(\epsilon) \frac{\hat{g}_B^6C_F\Gamma(1 - 3\epsilon)}{2^{3+4\epsilon}\pi^{-\epsilon} \Gamma(1 + 2 \epsilon)}  a_0(-3\epsilon)~. 
	\end{align}

	Using the explicit form (\ref{fiis}) of the functions $f_i^{(2)}(\epsilon)$  and simple manipulations, we can recast these contributions as follows:
	\begin{align}
		\label{Fiis}
		F^{(2)}_1 & =
		-\hat{g}_B^6 \frac{C_F \,i_\cR }{8\pi^2} \frac{P_2(\epsilon) B_3(\epsilon)}{\epsilon(1-2\epsilon)} + \cO(\epsilon)	~,
		\notag\\
		F^{(2)}_2 & = - \hat{g}_B^6 \frac{C_F\, N}{16\pi^2} \frac{P_2(\epsilon) B_3(\epsilon)}{\epsilon(1-3\epsilon)}~,
		\notag\\
		F^{(2)}_3 & = \hat{g}_B^6 \frac{C_F\, N}{32\pi^2} \frac{P_2(\epsilon) B_3(\epsilon
			)}{\epsilon(1 + \epsilon)}~,
		\notag\\
		F^{(2)}_4 & = \hat{g}_B^6 \frac{\cK^\prime_4}{N} \frac{3\zeta(3)}{(4\pi)^4} + \cO(\epsilon)~.
	\end{align} Note that only the last contribute is regular in the limit $\epsilon\to 0$, while the others exhibit single and double UV poles. Note also that the contribution $F^{(2)}_3$ arises from the gauge-like part of the gluon self-energy in the second diagram of eq. (\ref{eq:decorating 1}). By gauge invariance, we expect that it should eventually cancel against similar contributions resulting from other diagrams.  
	
	\subsubsection{Diagrams with three insertions}
	\label{sec:diagrams with three insertions}
	The three-loop diagrams  with three insertions on the Wilson loop contour fall in two distinct classes, corresponding to one-loop  reducible and irreducible corrections to the gauge-scalar and pure gauge vertex in the difference theory approach. These diagrams are computed  in Appendix \ref{sec:Mercedes diagram} and Appendix \ref{sec:lifesaver diagrams appendix}. The  complexity of the calculation lies on the  \textit{path-ordered} integration over the contour which we have to perform in arbitrary dimension $d$ due to the presence of UV singularities. Although the computations are extremely technical, the final result is quite simple and follows from eq.s (\ref{eq:Sigma_+ final}) and (\ref{eq:lifesaver final result}). We find
	\begin{equation}
		\begin{split}
			\label{W63is}
			\cW^\prime_{6(3)} & = 			
			\mathord{
				\begin{tikzpicture}[radius=2.cm, scale=0.6, baseline=-0.65ex]
					\begin{feynman}
						\vertex (A) at (0,2.);
						\vertex (C) at (0,0);
						\vertex (D) at (-1.5, -1.3);
						\vertex (B) at (1.5, -1.3);
						\vertex (B1) at (1.,-0.8);
						\vertex (B2) at (0.5,-0.5);
						\diagram*{
							(A) -- [photon] (C),
							(A) --[ fermion] (C),
							(C) -- [photon] (B2),
							(C) -- [plain] (B2),
							(B) --[photon] (B1),
							(B) --[plain] (B1),
							(C) --[ fermion] (D),
							(C) --[ photon] (D)
						};
					\end{feynman}
					\draw [black] (0,0) circle [];
					\filldraw[color=white!80, fill=white!15](0.75,-0.65) circle (0.7);	
					\draw [black, thick, dashed] (0.75,-0.65) circle [radius=0.63cm];
					\draw [black] (0.75,-0.65) circle [radius=0.7cm];
				\end{tikzpicture} 
			} +	\mathord{
				\begin{tikzpicture}[radius=2.cm, baseline=-0.65ex,scale=0.6]
					\draw [black] (0,0) circle [];
					\draw [black] (0,0) circle [radius=0.8cm];
					\draw [black, dashed,thick] (0,0) circle [radius=0.7cm];
					\begin{feynman}
						\vertex (A) at (0,2);
						\vertex (C) at (0,0.8);
						\vertex (D) at (-1.5, -1.3);
						\vertex (B) at (-0.7, -0.4);
						\vertex (B1) at (0.7,-0.4);
						\vertex (B2) at (1.5,-1.3);
						\diagram*{
							(A) -- [photon] (C),
							(A) --[ charged scalar] (C),
							(B) --[ anti fermion] (D),
							(B) --[photon] (D),
							(B1) --[fermion] (B2),
							(B1) --[photon] (B2),
						};
					\end{feynman}
				\end{tikzpicture} 
			}  	
			\\		
			& = \dfrac{N}{i_\cR} F^{(2)}_1 - F^{(2)}_2 - 2 F^{(2)}_3  
			+ \hat{g}_B^6 \dfrac{C_F N \beta_0}{4\pi^2} \zeta(3)  + \cO(\epsilon)~.
		\end{split}
	\end{equation} 
	Thus, up to a finite term proportional to $\zeta(3)$, these diagrams with internal vertices are expressible as linear combinations of the bubble-like contributions $F^{(2)}_i$ that emerge from the single-exchange corrections of the same order, see eq.s (\ref{fourF},\ref{Fiis}). As it occurred in eq. (\ref{fourF}), the $F^{(2)}_3$ contribution above results from diagrams involving the gauge-like part of the gluon self-energy at one-loop.  
	
	\subsubsection{Diagrams with four insertions}
	\label{sec:diagrams with four insertions}
	This class of corrections arises when dressing the internal lines of the two-loop ladder-like corrections (\ref{W4ladder}) with the one-loop correction to the adjoint scalar and gauge field propagator in the difference approach. The intermediate steps of the calculation are reported in Appendix \ref{sec:diagrams with four emissions appendix}. In particular, by employing eq.s (\ref{eq:sigma4prime final result}) and (\ref{eq:sigma4primeprime def}), we find that 
	\begin{equation}
		\label{W64is}
		\begin{split}
			\cW^\prime_{6(4)} 
			& = 			
			\mathord{
				\begin{tikzpicture}[baseline=-0.65ex,scale=0.6]
					\draw [black] (0,0) circle [radius=2cm];
					\draw [black] (-0.75,0) circle [radius=0.45cm];
					\begin{feynman}
						\vertex (A) at (-0.75,1.8);
						\vertex (b) at (-0.75,0.45);
						\vertex (d) at (-0.75,-0.45);
						\vertex (B) at (-0.75,-1.8);
						\vertex (C) at (0.75,1.8);
						\vertex (D) at (0.75,-1.8);
						\diagram*{
							(A) -- [photon] (b),
							(A) --[ fermion] (b),
							(d) --[fermion] (B),
							(d) --[photon] (B),
							(C) -- [photon] (D),
							(C) --[fermion] (D)
						};
					\end{feynman}
					\filldraw[color=white!80, fill=white!15](-0.75,0) circle (0.7);	
					\draw [black, thick, dashed] (-0.75,0) circle [radius=0.63cm];
					\draw [black] (-0.75,0) circle [radius=0.7cm];
				\end{tikzpicture} 
			}  \\
			& = F^{(2)}_3 + \hat{g}_B^6\dfrac{C_F(2N^2-3)}{6N}B_1(\epsilon)B_2(\epsilon)P_2(\epsilon) 
			+ \hat{g}_B^6 C_F N \beta_0  \frac{3\zeta(3)}{16\pi^2}+ \cO(\epsilon)~,
		\end{split}
	\end{equation} where we recall that $F_3^{(2)}$ is the three-loop bubble-like contributions defined in eq. (\ref{Fiis}) and, again, it results  from diagrams involving the gauge-like part of the gluon self-energy.
	
	\subsection{Summary of the three-loop results}
	\label{subsec:summary3}
	Let us summarise our findings at three-loop accuracy for the difference-theory interaction correction defined in eq. (\ref{W6k}). Using the results (\ref{fourF}, \ref{W63is}, \ref{W64is}), we obtain
	\begin{equation}
		\label{W6res}
		\cW_6^\prime 
		=\frac{i_\cR-N}{i_\cR} F^{(2)}_1 
		+ \hat{g}_B^6 C_F\left(\dfrac{ 2N^2-3}{6N}B_1(\epsilon)B_2(\epsilon)P_2(\epsilon)   
		+N\beta_0 \frac{7\zeta(3)}{16\pi^2}   
		+ \frac{\cK_{4}^\prime}{N}\dfrac{3\zeta(3) }{2^{8}\pi^4C_F}\right)
		+ \cO(\epsilon)~,
	\end{equation} 
	where we recall that the functions $B_n(\epsilon)$ and $P_2(\epsilon)$ are defined, respectively,  in eq.s (\ref{defBn}) and  (\ref{W4Rfinal}). As anticipated, the final result does not include any  $F_3^{(2)}$ contributions as a consequence due to gauge invariance. Actually, an analogous cancellation also occurs for the $F_2^{(2)}$ contributions and, as we will shortly see, this is essential to ensure the correct renormalization properties of the Wilson loop observable.  
	
	The first contribution in the previous expression can be further simplified by  using the explicit definition of the bubble-like contribution $F^{(2)}_1$ given by eq. (\ref{Fiis}). We find that it accounts for a double insertion in the single-exchange diagram (\ref{eq:ladder g2}) of the one-loop correction to the adjoint scalar and gauge field in the difference theory:      
	\begin{equation}
		\label{eq:DeltaW^2_6}
		\frac{i_\cR-N}{i_\cR} F_1^{(2)} 
		=\hat{g}_B^6 C_F P^2_2(\epsilon)B_3(\epsilon) + \cO(\epsilon) 
		= \mathord{
			\begin{tikzpicture}[baseline=-0.65ex,scale=0.6]
				\draw [black] (0,0) circle [radius=2cm];
				\begin{feynman}
					\vertex (A) at (0,2);
					\vertex (b) at (-0,0.45);
					\vertex (d) at (-0,-0.45);
					\vertex (B) at (-0,-2);
					\diagram*{
						(A) -- [photon] (B),
						(A) --[ plain] (B),
					};
				\end{feynman}
				\filldraw[color=white!80, fill=white!15](-0.,0.85) circle (0.73);	
				\draw [black, thick, dashed] (-0.,0.85) circle [radius=0.66cm];
				\draw [black] (-0,0.85) circle [radius=0.73cm];
				\filldraw[color=white!80, fill=white!15](-0.,-0.85) circle (0.73);	
				\draw [black, thick, dashed] (-0.,-0.85) circle [radius=0.66cm];
				\draw [black] (-0,-0.85) circle [radius=0.73cm];
			\end{tikzpicture} 
		} + \cO(\epsilon
		)~. 
	\end{equation}
	Let us note that the internal exchange in the previous expression \textit{does not} represent the reducible component of the internal correction associated with the contribution $\cW^\prime_{6(2)}$ (\ref{eq:bubble-exchange}) which, instead,  is given  by
	\begin{align}
		\label{redpart}
		\mathord{
			\begin{tikzpicture}[baseline=-0.65ex,scale=0.6]
				\draw [black] (0,0) circle [radius=2cm];
				\begin{feynman}
					\vertex (A) at (0,2);
					\vertex (b) at (-0,0.45);
					\vertex (d) at (-0,-0.45);
					\vertex (B) at (-0,-2);
					\diagram*{
						(A) -- [photon] (B),
						(A) --[ plain] (B),
					};
				\end{feynman}
				\filldraw[color=white!80, fill=white!15](-0.,0.85) circle (0.7);	
				\draw [black, thick, dashed] (-0.,0.85) circle [radius=0.7cm];
				\filldraw[color=white!80, fill=white!15](-0.,-0.85) circle (0.7);	
				\draw [black, thick, dashed] (-0.,-0.85) circle [radius=0.7cm];
			\end{tikzpicture} 
		} 
		-
		\mathord{
			\begin{tikzpicture}[baseline=-0.65ex,scale=0.6]
				\draw [black] (0,0) circle [radius=2cm];
				\begin{feynman}
					\vertex (A) at (0,2);
					\vertex (b) at (-0,0.45);
					\vertex (d) at (-0,-0.45);
					\vertex (B) at (-0,-2);
					\diagram*{
						(A) -- [photon] (B),
						(A) --[ plain] (B),
					};
				\end{feynman}
				\filldraw[color=white!80, fill=white!15](-0.,0.85) circle (0.7);	
				\draw [black, thick] (-0.,0.85) circle [radius=0.7cm];
				\filldraw[color=white!80, fill=white!15](-0.,-0.85) circle (0.7);	
				\draw [black, thick] (-0.,-0.85) circle [radius=0.7cm];
			\end{tikzpicture} 
		}~.
	\end{align}  
	In fact, eq. (\ref{eq:DeltaW^2_6}) arises when adding to the previous diagrams the first term in eq. (\ref{W63is}), resulting from  the diagrams with internal vertices $\cW^\prime_{6(3)}$ (\ref{eq:DeltaW^2_6}). This additional correction introduces the ``cross terms'' characterized by one of the two internal bubbles in the representation $\cR$ and the second one in the adjoint. 
	
	Altogether, taking into account all the results described above, we get the following expression of the Wilson loop v.e.v. up to three loops:
	\begin{align}
		\label{restotW}
		\cW  = 1 &+ \hat{g}_B^2 C_F B_1(\epsilon) 
		+ \hat{g}_B^4 C_F \left(\dfrac{ (2N^2-3)}{12N}B_1^2(\epsilon) + P_2(\epsilon)B_2(\epsilon)-\epsilon N \dfrac{3\zeta(3)}{16\pi^2} \right)  \notag \\[0.4em]
		& + \hat{g}_B^6
		C_F \left(\frac{N^4-3N^2+3}{46098 N^2} + \dfrac{2N^2-3}{6N}B_1(\epsilon)B_2(\epsilon)P_2(\epsilon)
		+ P_2^2(\epsilon)B_3(\epsilon)  + \beta_0   N \dfrac{7\zeta(3)}{16\pi^2}\right) \notag\\[0.4em] 
		&+ \hat{g}_B^6 \dfrac{3\zeta(3)\cK_{4}^\prime }{2^{8}\pi^4N} +\ldots~,
	\end{align} 
	where the dots stand for  $\cO(\epsilon)$ terms which only contribute at four loops.

\section{Renormalization }
\label{eq:renormalization}
The vacuum expectation value of the  1/2 BPS Wilson loop (\ref{eq:1/2 BPS supersymmetric Wilson loop}) is (UV) divergent and we have to renormalize it in order to obtain a finite result. The divergences are encoded in the function $P_2(\epsilon)$, defined in eq. (\ref{W4Rfinal}), which is singular in the limit $\epsilon\to 0$.   
Since the circular Wilson loop operator is defined over a smooth curve, the singularities are  reabsorbed by the charge renormalization \cite{Dotsenko:1979wb,Brandt:1981kf,Korchemsky:1987wg} which, in terms of $\hat{g}_B=g_B R^\epsilon$, reads  
\begin{equation}
	\label{eq:renormalized coupling}
	\hat{g}_B=g_*\,\left(R M \right)^\epsilon Z_{g_*}(\epsilon) \ .
\end{equation} 
In the previous expression,  $g_*$ is the renormalized coupling evaluated at the renormalization scale $M$, while $Z_{g_*}(\epsilon)$ encodes the so-called subtraction terms. These can be easily calculated by the explicit expression of the $\beta$-function (\ref{eq:beta0}). In particular, acting on 
eq. (\ref{eq:renormalized coupling}) with the logarithmic derivative with respect to $M$ and requiring that  $g_B$ does not depend on $M$ we find, in the MS scheme,  that
\begin{equation}
	\label{eq:Z_{g_*}}
	\begin{split}
		Z_{g_*}(\epsilon) &= \exp(-\int_0^{g_*}\dfrac{\dd{t}}{t} \dfrac{(\epsilon t+\beta(t)}{\beta(t)}) \\[0.6em]
		& = \left(1-\dfrac{\beta_0 g_*^2}{\epsilon} \right)^{-\frac{1}{2}} 
		= 1 + \dfrac{\beta_0 g_*^2}{2\epsilon}+ \dfrac{3}{8}\dfrac{(\beta_0)^2 g_*^4}{\epsilon^2}
		+\ldots  \  .
	\end{split}
\end{equation}

The renormalized Wilson loop average is obtained by replacing the bare coupling $\hat{g}_B$ with the renormalized one $g_*$ in the dimensionally regularized observable (\ref{restotW}) and taking the limit $\epsilon\to 0$, i.e.   
\begin{equation}
	\label{eq:W renormalized}
	W_* =\lim_{\epsilon\to 0} \cW(g_*) \ . 
\end{equation} 

Note that when  $\epsilon\to 0$, the overall dependence on the renormalization scale $M$ must vanish. This means that  $W_*$ satisfies a Callan-Symanzik equation \cite{Billo:2023igr} which constrains the dependence of the renormalized Wilson loop average on $M$, $g_*$ and $R$, making them to appear in the running coupling constant $g(R)$,  defined in eq. (\ref{eq:classical}).

If we consider the three-loop results in eq. (\ref{restotW}), we can verify that all the divergences cancel out upon introducing the renormalized coupling and taking the limit $\epsilon\to 0$. Moreover, the final result can be expressed in terms of the running coupling. For instance, let us examine the terms
\begin{align}
	\label{sse}
	\hat{g}_B^2C_F B_1(\epsilon) + \hat{g}_B^4C_F P_2(\epsilon)B_2(\epsilon) + \hat{g}_B^6C_F P_2^2(\epsilon) B_3(\epsilon) \ ,
\end{align}	
which correspond, respectively, to a single-exchange diagrams dressed with zero, one or two corrections to the adjoint scalar and gauge field  propagator at one-loop in the difference theory. To proceed with the computation, we use eq.s (\ref{defBn}) and (\ref{W4Rfinal}) to expand the functions $B_n(\epsilon)$ and $P_2(\epsilon)$ about $\epsilon\to 0$, i.e. 
\begin{align}
	\label{expPB}
	P_2(\epsilon) & = -\beta_0 \left( \frac{1}{\epsilon} + 2 + 4\epsilon + \cO(\epsilon^2)\right)~,
	\notag \\
	B_1(\epsilon) & = \frac 14  + \frac14 \left(\gamma + \log\pi\right) \epsilon + \frac{1}{16} \left(\pi^2 + \left(\gamma + \log\pi\right)^2\right) \epsilon^2
	+ \cO(\epsilon^3)~,
	\notag\\
	B_2(\epsilon) & = \frac 14  + \frac 12 \left(\gamma + \log\pi\right) \epsilon + \cO(\epsilon^2)~,
	\notag\\
	B_3(\epsilon) & = \frac 14  + \cO(\epsilon)~,
\end{align} and we replace the bare coupling in eq. (\ref{sse}) with the renormalized one  (\ref{eq:renormalized coupling}). By employing  the subtraction terms (\ref{eq:Z_{g_*}}) and the expansions (\ref{expPB}), it is straightforward to verify that the final result is divergence free.  Analogously, it is also straightforward to show that, up to four-loop terms, the finite term takes the following form: 
\begin{align}
	\label{3lresg}
	& \frac{g_*^2}{4} 
	\left(1 - \beta_0 g_*^2 \left(\log M^2 R^2 + 2 + \gamma + \log \pi\right)
	+ \beta_0^2 g_*^4 \left(\left(\log M^2 R^2 + 2 + \gamma + \log \pi\right)^2 + \frac{\pi^2}{3}\right)
	\right) \ .
\end{align}
Let us focus on the regime (\ref{eq:range of energies and lambda di strong couplign }) in which we derived the matrix model on $\mathbb{S}^4$. Within this range,  $\log RM \gg 0$ so that the logarithmic terms, associated with the short-distance properties  of theory, dominate over $\cO(M^0)$ ones. Thus, we can write\footnote{These (scheme-dependent) finite terms are not completely captured by the matrix model, even if we could reabsorb many of them by using as a renormalization scale the quantity $\tilde M$ such that $\log \tilde M^2 R^2 = \log M^2 R^2 + 2 + \gamma + \log \pi$.}
\begin{align}
	\label{3lresgsim}
	& \frac{g_*^2}{4} 
	\left(1 - \beta_0 g_*^2 \log M^2 R^2 
	+ \beta_0^2 g_*^4 \left(\log M^2 R^2 \right)^2\right) + \cO(g_*^8)
	= \frac{g^2}{4} + \cO(g^8)~,
\end{align}
where we recognized the expansion up to order $g_*^6$ of the running coupling constant defined in eq. (\ref{eq:classical}). It is interesting to observe that the previous expression admits a graphical description in terms of a resummation of single-exchange:
\begin{equation}
	\label{resgraph}
	\mathord{
		\begin{tikzpicture}[baseline=-0.65ex,scale=0.5]
			\draw [black] (0,0) circle [radius=2cm];
			\begin{feynman}
				\vertex (A) at (0,2);
				\vertex (B) at (0,-2);
				\diagram*{
					(A) -- [photon] (B),
					(A) --[ fermion ] (B),
				};
			\end{feynman}
		\end{tikzpicture} 
	}
	+ \mathord{
		\begin{tikzpicture}[radius=2.cm, baseline=-0.65ex,scale=0.5]
			\draw [black] (0,0) circle [];
			\filldraw[color=white!15, fill=white!15](0,0) circle (1);	
			\draw [black,dashed,thick]   (0,0) circle [radius=.66];
			\draw [black,thick]   (0,0) circle [radius=.73];
			\begin{feynman}
				\vertex (A) at (0,2.);
				\vertex (C) at (0, 0.73);
				\vertex (D) at (0, -0.73);
				\vertex (B) at (0,-2.);
				\diagram*{
					(A) -- [photon] (C),
					(A) --[ fermion] (C),
					(D) -- [photon] (B),
					(D) --[ fermion] (B)
				};
			\end{feynman},
	\end{tikzpicture} } 
	+ \mathord{
		\begin{tikzpicture}[baseline=-0.65ex,scale=0.5]
			\draw [black] (0,0) circle [radius=2cm];
			\begin{feynman}
				\vertex (A) at (0,2);
				\vertex (b) at (-0,0.45);
				\vertex (d) at (-0,-0.45);
				\vertex (B) at (-0,-2);
				\diagram*{
					(A) -- [photon] (B),
					(A) --[ plain] (B),
				};
			\end{feynman}
			\filldraw[color=white!80, fill=white!15](-0.,0.85) circle (0.73);	
			\draw [black, thick, dashed] (-0.,0.85) circle [radius=0.66cm];
			\draw [black,thick] (-0,0.85) circle [radius=0.73cm];			
			\filldraw[color=white!80, fill=white!15](-0.,-0.85) circle (0.73);	
			\draw [black, thick, dashed] (-0.,-0.85) circle [radius=0.66cm];
			\draw [black,thick] (-0,-0.85) circle [radius=0.73cm];
		\end{tikzpicture} 
	} 
	= 	\mathord{
		\begin{tikzpicture}[baseline=-0.65ex,scale=0.5]
			\draw [black] (0,0) circle [radius=2cm];
			\begin{feynman}
				\vertex (A) at (0,2);
				\vertex (B) at (0,-2);
				\vertex (d) at (0,2.5){$g$};
				\vertex (c) at (0,-2.5){$g$};
				\diagram*{
					(A) -- [photon] (B),
					(A) --[ fermion ] (B),
				};
			\end{feynman}
		\end{tikzpicture} 
	}
	+ \cO(g^8)~.
\end{equation} The right-hand side of the previous expression highlights that the final result can be obtained by the usual ladder-like contribution by replacing the bare coupling with the running one, defined in eq.  (\ref{eq:classical}).   

Going back to eq. (\ref{restotW}), we repeat the same analysis for the terms proportional to  the colour factor $(2N^2-3)$, characterizing the double-exchange diagrams. Exploiting analogous manipulations, we find, within the regime (\ref{eq:range of energies and lambda di strong couplign }), that
\begin{align}
	\label{double_exchange_1}
	\hat g_B^4 C_F \frac{2N^2 -3}{12 N}
	\left(B_1^2(\epsilon) + 2 \hat g_B^2 B_1(\epsilon) B_2(\epsilon) P_2(\epsilon)\right)
	= g^4 C_F \frac{2N^2 -3}{192 N} + \cO(g^8)~.
\end{align} 

Let us now consider the terms in (\ref{restotW}) proportional to $\zeta(3)$ and characterized by the colour factor $C_F N$, for which we have
\begin{align}
	\label{zetab0}
	\hat g_B^4 C_F N \frac{\zeta(3)}{16 \pi^2} \left(-3 \epsilon + 7 \beta_0 \hat g_B^2\right)
	= g^6 C_F N\beta_0 \frac{\zeta(3)}{16 \pi^2} + \cO(g^8)~.
\end{align}
Note that each coefficient on the l.h.s. represents the sum of two types of contributions:   $-3\epsilon \hat g_B^4 = (-2\epsilon - \epsilon) \hat g_B^4$ and $7 \beta_0 \hat g_B^6 = (4 + 3) \beta_0 \hat g_B^6$. More specifically, the $(-2\epsilon \hat g_B^4)$ term results from  the evanescent correction $ \delta\cW^{\rm v.m.}_4$ of the two-loop diagrams with internal vertices, defined in eq. (\ref{eq:real calculation Sigma 3}) and explicitly given by eq. (\ref{eq:interaction contributions at two-loop with evanescent term}). Upon renormalization, this  evanescence interferes with the UV poles of the bare coupling and \textit{precisely remove} the $4 \beta_0 \hat g_B^6$ term, resulting from the same family of diagrams at three-loop, i.e. the Mercedes and lifesaver corrections we presented in eq. (\ref{W63is}). This means that all the terms proportional to $\beta_0\zeta(3)$ only originate from the \textit{ladder-like} diagrams depicted in eq.s (\ref{eq:ladderg4finalresult}) and (\ref{W64is}), which are responsible, respectively, for the contributions $-\epsilon\hat{g}_B^4$ and $3 \beta_0 \hat g_B^6$.

In graphical terms, we can summarize the content of eq.s (\ref{double_exchange_1}) and (\ref{zetab0}) as follows%
\footnote{This is actually not precise: the right hand side includes also, as reported in eq. (\ref{W64is}), the term $F_3^{(2)}$. However, as we already pointed out, this contribution does not contribute since it is exactly removed by analogous contributions resulting from the correction $\cW^\prime_{6(3)}$, see eq. (\ref{W63is}), and $\cW^\prime_{6(2)}$, given by eq. (\ref{eq:DeltaW^2_6}).}:
\begin{align}
	\label{double-graphical}
	\mathord{
		\begin{tikzpicture}[baseline=-0.65ex,scale=0.5]
			\draw [black] (0,0) circle [radius=2cm];
			\begin{feynman}
				\vertex (A) at (-0.75,1.8);
				\vertex (B) at (-0.75,-1.8);
				\vertex (C) at (0.75,1.8);
				\vertex (D) at (0.75,-1.8);
				\diagram*{		
					(A) -- [fermion] (B),
					(A) --[ photon] (B),
					(C) -- [photon] (D),
					(C) --[fermion] (D)
				};
			\end{feynman}
		\end{tikzpicture} 
	} 
	+
	\mathord{
		\begin{tikzpicture}[baseline=-0.65ex,scale=0.5]
			\draw [black] (0,0) circle [radius=2cm];
			\draw [black] (-0.75,0) circle [radius=0.45cm];
			\begin{feynman}
				\vertex (A) at (-0.75,1.8);
				\vertex (b) at (-0.75,0.45);
				\vertex (d) at (-0.75,-0.45);
				\vertex (B) at (-0.75,-1.8);
				\vertex (C) at (0.75,1.8);
				\vertex (D) at (0.75,-1.8);
				\diagram*{
					(A) -- [photon] (b),
					(A) --[ fermion] (b),
					(d) --[fermion] (B),
					(d) --[photon] (B),
					(C) -- [photon] (D),
					(C) --[fermion] (D)
				};
			\end{feynman}
			\filldraw[color=white!80, fill=white!15](-0.75,0) circle (0.7);	
			\draw [black, thick, dashed] (-0.75,0) circle [radius=0.63cm];
			\draw [black,thick] (-0.75,0) circle [radius=0.7cm];
		\end{tikzpicture} 
	} 
	= 
	\mathord{
		\begin{tikzpicture}[baseline=-0.65ex,scale=0.5]
			\draw [black] (0,0) circle [radius=2cm];
			\begin{feynman}
				\vertex (A) at (-0.75,1.8);
				\vertex (Ap) at (-0.75,2.4) {$g$};
				\vertex (B) at (-0.75,-1.8);
				\vertex (Bp) at (-0.75,-2.4) {$g$};
				\vertex (C) at (0.75,1.8);
				\vertex (Cp) at (0.75,2.4) {$g$};
				\vertex (D) at (0.75,-1.8);
				\vertex (Dp) at (0.75,-2.4) {$g$};
				\diagram*{		
					(A) -- [photon] (B),
					(A) --[fermion] (B),
					(C) -- [photon] (D),
					(C) --[fermion] (D)
				};
			\end{feynman}
		\end{tikzpicture} 
	} 
	+ g^6 C_F N \beta_0 \frac{\zeta(3)}{16 \pi^2} 
	+ \cO(g^8)~.		
\end{align}

Since our analysis regards the three-loop correction, the renormalization of the  triple-exchange terms (\ref{eq:ladder g6}) is trivial and provides us with the following contribution
\begin{align}
	\label{tripleis}
	g^6 \dfrac{C_F(N^4-3N^2+3)}{4608N^2} + \cO(g^8)~. 
\end{align}

The last term  in eq. (\ref{restotW}), proportional to the colour factor $\cK_{4}^\prime$, results  from  the irreducible part of the internal correction in the single-exchange diagrams (\ref{eq:bubble-exchange}),  namely from the $F^{(2)}_4$ function in eq. (\ref{eq:bubble-exchange}). We find that 
\begin{align}
	\label{key}	
	\hat g_B^6 \frac{\cK^\prime_4}{N} \frac{3\zeta(3)}{2^8\pi^4} = g^6 \frac{\cK^\prime_4}{N} \frac{3\zeta(3)}{2^8\pi^4} + \cO(g^8)~.
\end{align} 

Collecting all the  results we derived in this subsection, we can write the renormalized Wilson loop vev $W_*$ in terms of the running coupling constant $g$ as follows: 
\begin{align}
	\label{Wsres}
	W_* = W_0 + g^6 \frac{\cK^\prime_4}{N} \frac{3\zeta(3)}{2^8\pi^4}  + g^6 C_F N \beta_0 \frac{\zeta(3)}{16 \pi^2}  + \cO(g^8)~,
\end{align}
where $W_0$ was introduced in eq. (\ref{expW0}) and contains the ladder diagrams computed with the running coupling constant $g$, while the two terms proportional to $\zeta(3)$ \emph{coincide exactly with the prediction of the localization matrix model}, as follows from eq.s (\ref{WexpSint}) and (\ref{eq:conneted correlator}). Let us stress that this agreement holds within the regime  (\ref{eq:range of energies and lambda di strong couplign }). From the field theory point of view, the final result, when expressed in terms of the running coupling, is purely due to ladder-like diagrams, see eq.s (\ref{resgraph},\ref{double-graphical}) and (\ref{tripleis}). Moreover, the final outcome also ties perfectly in with the matrix model diagrams (\ref{eq:matrix model diagrams}), which suggest that  the two terms proportional to $\zeta(3)$ have to be associated with single-exchange diagrams. Indeed, as we previously explained, the correction involving the coefficient $\cK_{4}^\prime$ results from the diagram (\ref{eq:bubble-exchange}), while the term $\beta_0\zeta(3)$ is proportional to the fundamental Casimir $C_F$, which is the expected colour coefficient of the as single-exchange diagrams (\ref{eq:ladder g2 pic}). 

\section{Conclusions and outlook}
\label{sec:conclusions}
In this paper,  we  investigated the relation between supersymmetric localization on $\mathbb{S}^4$ and standard perturbative techniques in flat space for a generic $\mathcal{N}=2$ SYM theory with non-vanishing $\beta$-function. The analysis has been performed by studying the vacuum expectation value of the 1/2 BPS Wilson loop, for which localization provides an explicit result in term of an interacting matrix model. Although conformal invariance is broken at quantum level, preventing a direct connection between the sphere and the Euclidean space, we found a precise agreement in  the specific regime described in eq. (\ref{eq:range of energies and lambda di strong couplign }). Within this range of validity, the contribution of instantons and power-like corrections are suppressed and we showed that the matrix model predictions match standard perturbation theory based on Feynman diagrams techniques in flat space up to order $g^6$. At this perturbative order the matrix model produces two non-trivial $\zeta(3)$-like terms, that have a different origin: one is already present in the conformal case \cite{Billo:2019fbi,Andree:2010na}, while the other is  peculiar of the models  with non-vanishing $\beta$-function. We successfully compared the effective matrix diagrams associated with these contributions with the flat-space perturbative expansion, finding crucial interference effects between evanescent terms and the UV divergences of the bare coupling constant. Our results not only provide a non-trivial test of the localization approach for generic $\mathcal{N}=2$ SYM theories, but also make manifest the subtle reorganization of the conventional Feynman diagrams into the matrix-model average. On the technical side, the perturbative computations of the three-loop contributions involved multiple ordered integrations of position-space Green functions along a circular domain. As far as our knowledge is concerned, this type of calculations have never been considered before at such precision level: we have devoted a series of appendices to illustrate the procedure and the actual emergence of the evanescent terms and finite contributions relevant for the final result.

Clearly, there are some possible improvements and extensions of our work. It would be interesting to expand our analysis to the next perturbative order and try to generalize the understanding at all loops. This would imply a more systematic approach to the calculation of Feynman diagrams for circular Wilson loops involving complicated path-ordered trigonometric integrations. In the case of cusped Wilson loops, the path-integration is performed over straight lines by techniques involving heavy quark effective theory. These have provided beautiful results for the cusp anomalous dimension \cite{Korchemsky:1987wg} at high-loop order, both in supersymmetric and non-supersymmetric theories (see \cite{Grozin:2022umo} for status review). It would be nice to develop an analogous tool to face circular contours.
Another natural investigation would be to examine correlators of local operators in this non-conformal set-up: supersymmetric localization still gives exact results for classes of two-point functions that can be compared with flat-space perturbation theory \cite{Billo:2019job}. Studying these local observables in light of the present computations could further improve our understanding of the effects associated with a non-trivial $\beta$-function. 
We plan to explore these two-point functions in the near future. A more speculative direction concerns the study the large-order behaviour of the perturbative series in presence of a running coupling constant. Exact all-orders expressions on $\mathbb{S}^4$ have been already used to explore asymptotic properties of the matrix-model perturbative expansion, in connection with resurgent techniques \cite{Aniceto:2014hoa}. The analysis has been performed for different $\mathcal{N}=2$ SYM theories, obtaining explicit results in the conformal and massive cases. It would be interesting to reconsider the non-conformal case and its relation with a flat-space set-up to shed light on the convergence properties of the perturbative series and, possibly, on some gauge-invariant resummations.

	\vskip 1cm
	\noindent {\large {\bf Acknowledgments}}
	\vskip 0.2cm
	We thank M. Frau, F. Galvagno, G. Korchemsky, A. Lerda, I. Pesando and P. Vallarino for lively exchange of ideas. A.T. is grateful to the Institut de Physique Théorique (CEA) for the kind hospitality during part of this work. This research is partially
	supported by the MIUR PRIN contract 2020KR4KN2 “String Theory as a bridge between
	Gauge Theories and Quantum Gravity” and by the INFN projects ST\&FI “String Theory
	\& Fundamental Interactions” and GAST "Gauge Theory And Strings".
	\vskip 1cm
	
	\appendix
	
	\section{Field theory set-ups and conventions}
	\label{sec:Conventions}
	Our conventions follow those of \cite{Billo:2017glv,Billo:2018oog,Billo:2019fbi}. In Euclidean space the spin group is $\mathrm{Spin}(4)\simeq \mathrm{SU}(2)_{\alpha}\otimes \mathrm{SU}(2)_{\dot{\alpha}}$.  Chiral spinors carry  undotted indices  $\alpha, \beta,\ldots$, while  anti-chiral spinors carry dotted indices $\dot{\alpha},\dot{\beta},\dots\ $, which are contracted as follows 
	\begin{equation}
		\psi\chi\equiv\psi^{\alpha}\chi_{\alpha} \ , \quad \quad \bar{\psi}\bar{\chi}\equiv\bar{\psi}_{\dot{\alpha}}\bar{\chi}^{\dot{\alpha}} \ .
	\end{equation} In the following, we raise and lower indices as follows 
	\begin{equation}
		\label{eq:Fierz identities}
		\psi^{\alpha}=\epsilon^{\alpha \beta}\psi_{\beta}, \quad \quad \bar{\psi}^{\dot{\alpha}}=\epsilon^{\dot{\alpha}\dot{\beta}}\bar{\psi}_{\dot{\beta}} \ ,
	\end{equation}
	where $\epsilon^{12}=\epsilon_{21}=\epsilon^{\dot{1}\dot{2}}=\epsilon_{\dot{2}\dot{1}}=1$. Let us note in Euclidean spacetime spinors satisfy \textit{pseudoreality} conditions, i.e. \begin{equation}
		(\psi_\alpha)^\dagger = \psi^\alpha \ .
	\end{equation} 

	The matrices $(\bar{\sigma}^{\mu})^{\dot{\alpha}\alpha}$ and $(\sigma^{\mu})_{\alpha \dot{\beta}}$ are defined as follows
	\begin{equation}
		\label{eq:sigma matrices}
		\sigma^{\mu}=(\vec{\tau},-\mathrm{i}\mathbb{I}) \ , \quad \quad \bar{\sigma}^{\mu}= (-\vec{\tau}, -\mathrm{i}\mathbb{I}) \ ,
	\end{equation}
	where $\vec{\tau}$ are the ordinary Pauli matrices. Furthermore, these matrices are such that
	\begin{equation}(\bar{\sigma}^{\mu})^{\dot{\alpha}\alpha}=\epsilon^{\dot{\alpha}\dot{\beta}}\epsilon^{\alpha \beta}(\sigma^{\mu})_{\beta \dot{\beta}}
	\end{equation}
	and satisfy the Clifford algebra 
	\begin{align}
		\label{eq:clifford}
		\sigma^{\mu}\bar{\sigma}^{\nu}+ \sigma^{\nu}\bar{\sigma}^{\mu}&=-2\delta^{\mu \nu}\mathbb{I} \ , \\[0.5em] 
		\bar{\sigma}^{\mu}\sigma^\nu + \bar{\sigma}^\nu\sigma^\mu &= -2\delta^{\mu\nu}\mathbb{I} \ .
	\end{align} The previous expressions obviously implies that \begin{equation}
		\label{eq:double trace}
		\Tr \sigma^\mu\bar{\sigma}^\nu=-2\delta^{\mu \nu} \ .
	\end{equation} It also is straightforward to show that the following set of relations hold \begin{align}
		\tr(\bar{\sigma}^\mu\sigma^\nu\bar{\sigma}^\rho\sigma^\sigma) &=2\big(\delta^{\mu\nu}\delta^{\rho\sigma}-\delta^{\mu\rho}\delta^{\nu\sigma}+\delta^{\mu\sigma}\delta^{\nu\rho}+\epsilon^{\mu\nu\rho\sigma}\big)\notag \ , \\	\tr({\sigma}^\mu\bar{\sigma}^\nu\sigma^\rho\bar{\sigma}^\sigma)
		\label{eq:trace}
		&=2\big(\delta^{\mu\nu}\delta^{\rho\sigma}-\delta^{\mu\rho}\delta^{\nu\sigma}+\delta^{\mu\sigma}\delta^{\nu\rho}-\epsilon^{\mu\nu\rho\sigma}\big) \ , \\
		\bar{\sigma}^\mu\sigma^\nu\bar{\sigma}^\rho&=-\delta^{\mu\nu}\bar{\sigma}^\rho+\delta^{\mu\rho}\bar{\sigma}^\nu-\delta^{\nu\rho}\bar{\sigma}^\mu-\epsilon^{\mu \nu \rho\alpha}\bar{\sigma}_\alpha  \ , \notag
	\end{align}  where we normalize $\epsilon^{1234}=\epsilon_{1234}=1$.

	\subsection{Euclidean actions in flat space}
	\label{sec:actions in flat space}
	We consider $\mathcal{N}=2$ super-Yang-Mills theories with gauge group $\mathrm{SU}(N)$ and with massless hypermultiplets in an arbitrary representation $\mathcal{R}$. The Lie algebra of the gauge group  is $\mathfrak{su}(n)$ and spanned by hermitian traceless generators $T^a$, with $a=1,\ldots,N^2-1$, satisfying      \begin{equation}
		[T^a, T^b]=\mathrm{i}f^{abc}T^c \ .
	\end{equation} 
	
	In the $\mathcal{N}=2$ language, the vector multiplet consists of one gauge field and one complex scalar fields, denoted as $A_\mu$ and $\phi$, along with their fermionic partners $\psi$ and $\lambda$, to which we will sometimes refer as the \textit{gauginos}. In  Euclidean space, the dynamics of this supermultiplet is described by the following  gauged-fixed action 
	\begin{equation}
		\label{eq:pure super-Yang-Mills N=2}
		\begin{split}
			S^{\mathrm{gauge}}_0= &\int \dd^4{x} \ \text{Tr} \bigg[ -\dfrac{1}{2} F_{\mu \nu}F^{\mu \nu} -2\mathrm{i} \lambda \sigma^{\mu} D_{\mu}\Bar{\lambda} -2\mathrm{i}  \psi \sigma^{\mu} D_{\mu}\Bar{\psi}
			-2 D_{\mu}\Bar{\phi}D^{\mu}\phi -2\partial_{\mu}\Bar{c}D^{\mu}c \bigg] \ , \\
			S_{\mathrm{int}}=	&\int \dd^4{x} \ \text{Tr} \bigg[2\mathrm{i}g_B \sqrt{2}\Big( \Bar{\phi}\big\{\lambda^\alpha,\psi_\alpha\big\}-\phi\big\{\Bar{\psi}_{\dot{\alpha}},\Bar{\lambda}^{\dot{\alpha}}\big\} \Big) -\xi (\partial_{\mu}A^{\mu})^2 -g_B^2\big[\phi,\bar{\phi} \big]^2\bigg] \ ,
		\end{split}
	\end{equation} where in the previous expression we denoted with $c$ the ghost field. Let us note that with these conventions the actions are negative defined and consequently, they appear as $\mathrm{e}^S$ in the path integral. The field-strength and the adjoint covariant derivatives are \begin{equation}
		\label{eq:field strength and adjoint cov}
		\begin{split}
			F_{\mu \nu}=\partial_{\mu} A_{\nu} -\partial_{\nu}A_{\mu} -\mathrm{i} g_B [A_{\mu},A_{\nu}]\ , \quad 
			D_{\mu}=A_{\mu} -\mathrm{i}g_B[A_{\mu}, \bullet ] \ .
		\end{split}
	\end{equation}

	In the $\mathcal{N}=2$ language matter sits in the hypermultiplets. Their spacetime field content consists of two complex scalars fields, i.e. $q$ and $\tilde{q}$, along with their fermionic partners $\eta$ and $\tilde{\eta}$. In particular, $q$ and $\eta$ transform in the representation $\cR$, while the $\tilde{q}$ and $\tilde{\eta}$ in the conjugated one, i.e. $\cR^*$. The dynamics is encoded in the following actions \begin{equation}
		\label{eq:actions matter in flat space}
		\begin{split}
			S_0^Q = \int \dd^4{x} \bigg[  &-D_{\mu}\Bar{q} D^{\mu}q -\mathrm{i} \Bar{\eta}\Bar{\sigma}^{\mu}D_{\mu}\eta   - D_{\mu}{\Tilde{q}} D^{\mu}\bar{\Tilde{q}} -\mathrm{i} {\tilde{\eta}}{\sigma}^{\mu}D_{\mu}\bar{\tilde{\eta}} \bigg] \\
			S_{\mathrm{int}}^Q = \int \dd^4{x} \ \bigg[ & \mathrm{i}\sqrt{2}g_B \Big( \Tilde{q}\Bar{\lambda}\Bar{\tilde{\eta}} -\tilde{\eta}\lambda \Bar{\Tilde{q}}\Big) + \mathrm{i}\sqrt{2}g_B \Big( \Bar{\eta}\Bar{\phi}\Bar{\tilde{\eta}}-\tilde{\eta}\phi\eta \Big) + \mathrm{i}\sqrt{2}g_B \Big( \bar{\eta}\bar{\psi}\bar{\tilde{q}} -\tilde{q}\psi\eta \Big)\\
			& + \mathrm{i}\sqrt{2}g_B \Big( \bar{q}\bar{\psi}\bar{\tilde{\eta}}-\tilde{\eta}\psi q \Big) +\mathrm{i}\sqrt{2}g_B \Bigl(\Bar{q}\lambda \eta -\bar{\eta} \bar{\lambda} q\Bigr)
			-g_B^2V(\phi,\tilde{q},q) \bigg] \ ,
		\end{split}
	\end{equation}  where we denoted with $V(\phi,\tilde{q},q)$ the scalar potential describing quartic interactions \begin{equation}
		\begin{split}
			\label{eq:scalar potential}
			V&=\tilde{q}\{\phi,\bar{\phi}\}\bar{\tilde{q}}+\bar{q}\{\bar{\phi},\phi\} q -\left(\tilde{q}T^a_\cR\bar{\tilde{q}}\right)\left(\bar{q}T^a_\cR q\right) +2\left(\bar{q}T^a_\mathcal{R}\bar{\tilde{q}}\right)\left(\tilde{q}T^a_\cR q\right)\\
			&+\dfrac{1}{2}\left(\bar{q}T^a_\mathcal{R}{q}\right)\left(\bar{q}T^a_\mathcal{R}{q}\right)+\dfrac{1}{2}\left(\tilde{q}T^a_\cR \tilde{\bar{q}}\right)\left(\tilde{q}T^a_\cR \tilde{\bar{q}}\right)\ .
		\end{split}
	\end{equation}
	In the previous,   $T^a_\cR$ denotes the generators of the Lie algebra $\mathfrak{su}(n)$ in the representation $\cR$ of the gauge group. The covariant derivatives for a field transforming in this representation is defined as  \begin{equation}
		\label{eq:covariant derivative rep R}
		D_{\mu}=\partial_{\mu} -\mathrm{i}g_B A^a_{\mu} T^a_{\mathcal{R}}\ .
	\end{equation}


	We conclude this section by reporting our conventions for the Feynman propagators. Let us begin with considering the vector-multiplet  fields. In the Feynman gauge, i.e. $\xi=1$, the tree-level propagator of the adjoint scalar $\phi$ and of the gauge field $A_\mu$ are identical up to spacetime indices. We have \begin{equation}
		\begin{split}
			\mathord{ \begin{tikzpicture}[baseline=-0.65ex,scale=0.8]
					\begin{feynman}
						\vertex (A) at (2,0) ;
						\vertex (A1) at (2,-0.5) {$A_\mu^a$};
						\vertex (B) at (-1.5,0);
						\vertex (B1) at (-1.5,-0.5) {$A^b_\nu$};
						\diagram*{
							(B) --[ photon] (A)
						};
					\end{feynman}
				\end{tikzpicture} 
			} =\dfrac{\delta^{ab}}{p^2} \delta_{\mu \nu} \ , \quad  \quad 
			\mathord{ \begin{tikzpicture}[baseline=-0.65ex,scale=0.8]
					\begin{feynman}
						\vertex (A) at (2,0) ;
						\vertex (A1) at (2,-0.5) {$\phi^a$};
						\vertex (B) at (-1.5,0);
						\vertex (B1) at (-1.5,-0.5) {$\bar{\phi}^b$};
						\diagram*{
							(B) --[ fermion] (A)
						};
					\end{feynman}
				\end{tikzpicture} 
			}& =\dfrac{\delta^{ab}}{p^2} \ .
		\end{split}
	\end{equation} On the other hand, the tree-level propagators of the two \textit{gauginos} $\lambda$ and $\psi$ exhibits a more complicated structure. Here we consider in detail the relevant expressions for the Weyl spinor $\lambda$ but analogous results hold for $\psi$. We have two relevant Wick contractions, i.e. \begin{equation}
		\label{eq:Wick's contraction for fermions}
		\big<\lambda^a_{\alpha}(x)\bar{\lambda}^b_{\dot{\alpha}}(y)\big>_0 \ ,\quad \quad \big<\bar{\lambda}_b^{\dot{\alpha}}(y)\lambda^{\alpha}_a(x)\big>_0 \ .
	\end{equation} In our conventions, the arrow associated with the particle flow  always goes from the dotted index to the undotted one. As a result, in momentum space we represent the first contraction as follows   \begin{equation}
		\begin{split}
			\big<\lambda^a_{\alpha}(x)\bar{\lambda}^b_{\dot{\alpha}}(y)\big>_0 \quad \leftrightarrow \quad 	\mathord{ \begin{tikzpicture}[baseline=-0.65ex,scale=0.8]
					\begin{feynman}
						\vertex (A) at (2,0) ;
						\vertex (A1) at (2,-0.5) {$\alpha, \ a $} ;
						\vertex (B) at (-1.5,0);
						\vertex (B1) at (-1.5,-0.5){$\dot{\alpha}, \ b$} ;
						\diagram*{
							(B) --[ fermion, thick,  momentum={[arrow style=black]\( p \)}] (A)
						};
					\end{feynman}
				\end{tikzpicture} 
			} =\dfrac{\delta^{ab} {\sigma}_{\alpha \dot{\alpha}}\cdot p}{p^2} \ ,
		\end{split}
	\end{equation} where $\sigma_{\alpha \dot{\alpha}}\cdot p=\sigma^\mu_{\alpha \dot{\alpha}} \  p_\mu$, with  $\sigma^\mu_{\alpha \dot{\alpha}}$ defined in eq. (\ref{eq:sigma matrices}). The tree-level propagator with raised indices in eq. (\ref{eq:Wick's contraction for fermions})  is obtained from the previous expression  by employing the   $\epsilon$-tensor as explained in eq. (\ref{eq:sigma matrices}). We find \begin{equation}
		\begin{split}
			\label{eq:spinor propagator with raised indices}
			\big<\bar{\lambda}_b^{\dot{\alpha}}(y)\lambda^{\alpha}_a(x)\big>_0 \quad \leftrightarrow \quad 	\mathord{ \begin{tikzpicture}[baseline=-0.65ex,scale=0.8]
					\begin{feynman}
						\vertex (A) at (2,0) ;
						\vertex (A1) at (2,-0.5) {$\alpha, \ a $} ;
						\vertex (B) at (-1.5,0);
						\vertex (B1) at (-1.5,-0.5){$\dot{\alpha}, \ b$} ;
						\diagram*{
							(B) --[ fermion, thick, reversed momentum={[arrow style=black]\( p \)}] (A)
						};
					\end{feynman}
				\end{tikzpicture} 
			} =\dfrac{\delta^{ab} \bar{{\sigma}}^{ \dot{\alpha} \alpha}\cdot p}{p^2} \ .
		\end{split}
	\end{equation} Finally, we consider the propagators associated with the spacetime fields of the massless hypermultiplets in the representation $\cR$. For the complex scalars $q$ and $\tilde{q}$ we have
	\begin{equation}
		\begin{split}
			\label{eq:propagator scalars q tildeq}
			\mathord{ \begin{tikzpicture}[baseline=-0.65ex,scale=0.8]
					\begin{feynman}
						\vertex (A) at (2,0) ;
						\vertex (A1) at (2.,-0.5) {$q_v$};
						\vertex (B) at (-1.5,0);
						\vertex (B1) at (-1.5,-0.5) {$\bar{q}^u$};
						\diagram*{
							(B) --[ charged scalar] (A)
						};
					\end{feynman}
				\end{tikzpicture} 
			} &=\dfrac{\delta_{v}^{\ u}}{p^2} \\[0.3em]
			\mathord{ \begin{tikzpicture}[baseline=-0.65ex,scale=0.8]
					\begin{feynman}
						\vertex (A) at (2,0) ;
						\vertex (A1) at (2.,-0.5) {$\tilde{q}^v$};
						\vertex (B) at (-1.5,0);
						\vertex (B1) at (-1.5,-0.5) {$\bar{\tilde{q}}_u$};
						\newcommand\tmpda{1.4cm}
						\newcommand\tmpdb{2.6cm}
						\diagram*{
							(B) --[ ghost,with arrow=\tmpda] (A)
						};
					\end{feynman}
				\end{tikzpicture} 
			} &=\dfrac{\delta_{u}^{\ v}}{p^2} \ ,
		\end{split}
	\end{equation} 
	where $u,v=1,\ldots,\dim\cR$. Finally, we consider the fermionic propagators associated with the fermions $\eta$ and $\tilde{\eta}$. For simplicity, we only depict the contractions with lowered indices 
	i.e. \begin{align}
		\label{eq:matter spinor propagators}
		\mathord{ \begin{tikzpicture}[baseline=-0.65ex,scale=0.8]
				\begin{feynman}
					\vertex (A) at (2,0) ;
					\vertex (A1) at (2,-0.5) {$ \eta_{\alpha,u} $} ;
					\vertex (B) at (-1.5,0);
					\vertex (B1) at (-1.5,-0.5) {$\bar{\eta}^v_{\dot{\alpha}}$} ;
					\diagram*{
						(B) --[ charged scalar, thick,  momentum={[arrow style=black]\( p \)}] (A)
					};
				\end{feynman}
			\end{tikzpicture} 
		} &=\dfrac{\delta_u^{\ v} \sigma_{\alpha \dot{\alpha} }\cdot p}{p^2} \\[0.4em] 
		\mathord{ \begin{tikzpicture}[baseline=-0.65ex,scale=0.8]
				\begin{feynman}
					\vertex (A) at (2,0) ;
					\vertex (A1) at (2,-0.5) {$ \tilde{\eta}^u_{\alpha} $} ;
					\vertex (B) at (-1.5,0);
					\vertex (B1) at (-1.5,-0.5) {$\bar{\tilde{\eta}}_{\dot{\alpha},v}$} ;
					\diagram*{
						(B) --[ ghost, thick, with arrow=1.4cm,  momentum={[arrow style=black]\( p \)}] (A)
					};
				\end{feynman}
			\end{tikzpicture} 
		} &=\dfrac{\delta_v^{\ u} \sigma_{\alpha \dot{\alpha} }\cdot p}{p^2}  \ .
	\end{align} The relevant expressions with raised indices are analogous to the propagators presented in eq. (\ref{eq:spinor propagator with raised indices}).

\section{Perturbative corrections to propagators}
\label{sec:loop corrections}
In this section, we introduce our notations and conventions for the calculation of the Feynman integrals entering the perturbative corrections to the propagators at one/two-loop accuracy. We will primarily work in  momentum space and will follow the formalism presented in \cite{Grozin:2005yg}. 
At one-loop accuracy,  we consider the basis integral \begin{equation}
	\begin{split}
		\label{eq:one-loop basis integral}
		G(n_1,n_2) &= \int \dfrac{\dd^dk}{(2\pi)^d} \dfrac{1}{(k^2)^{n_1}((k+p)^2)^{n_2}}=(p^2)^{d/2-n_1-n_2} \widetilde{G}(n_1,n_2) \ ,
	\end{split}
\end{equation} where the overall dependence on external momentum $p^2$ follows from dimensionality, while  $\widetilde{G}(n_1,n_2$) is a function of the dimension $d$ and of the integers  $n_1$ and $n_2$\footnote{Let us note that  when $n_1\leq0$ or $n_2\leq0$ eq. (\ref{eq:one-loop basis integral}) vanishes.}. Employing usual Feynman parameters for the different propagators,
it is straightforward to show that  
\begin{equation}
	\label{eq:definition of g}
	\widetilde{G}(n_1,n_2)=\dfrac{\Gamma(n_1+n_1-d/2)}{(4\pi)^{d/2}\Gamma(n_1)\Gamma(n_2)}\dfrac{\Gamma(d/2-n_1)\Gamma(d/2-n_2)}{\Gamma(d-n_1-n_2)}  \ ,
\end{equation}  
where $\Gamma(x)$ is the Euler gamma function. At two-loop accuracy, the basis integral we consider  is \cite{Andree:2010na,Grozin:2005yg} 
\begin{equation}
	\label{eq:massless two-loop integral}
	\begin{split}
		G(n_1,n_2,n_3,n_4,n_5) &= \int  \dfrac{\dd^{d}k}{(2\pi)^d}  \dfrac{\dd^dl}{(2\pi)^d}
		\dfrac{  1 }{((k+p)^2)^{n_1} ((l+p)^2)^{n_2}  (k^2)^{n_3} (l^2)^{n_4} ((l-k)^2)^{n_5}}
		\notag\\
		& = (p^2)^{d-\sum n_i} \widetilde G(n_1,n_2,n_3,n_4,n_5)
	\end{split}
\end{equation} where  $n_i$ are integers. Note that the previous expression is symmetric under the interchanges $(1\leftrightarrow2,3\leftrightarrow 4)$ and $(1\leftrightarrow3,2\leftrightarrow 4)$.
When one of the parameters $n_i$ vanishes, eq (\ref{eq:massless two-loop integral}) reduces to a product of the one-loop integrals we introduced (\ref{eq:one-loop basis integral}). In particular, we will use  the identities  
\begin{align}
	\label{eq:sqaure one-loop integral}
	\widetilde G(n_1,n_2,n_3,n_4,0)& = \widetilde G(n_1,n_3) \widetilde G(n_2,n_4) \ , \\[0.4em]
	\label{eq:G(0,1,1,1,1)}
	\widetilde G(0,n_2,n_3,n_4,n_5)& = \widetilde{G}(n_3,n_5) \widetilde{G}(n_2,n_3+n_4+n_5-d/2)\ ,
\end{align} 
which can be derived by repeated applications of eq. (\ref{eq:one-loop basis integral}). When all the indices $n_i$  in eq. (\ref{eq:massless two-loop integral}) are equal to one, it is possible to employ integration by parts (see Section 5.1 of \cite{Grozin:2005yg} for the technical details) to derive the following relation:
\begin{equation}
	\label{G(1,1,1,1,1)}
	G(d) \equiv G(1,1,1,1,1)=\dfrac{2G(1,1)}{d-4}\left(G(2,1)-\left(p^2\right)^{2-d/2}G(2,3-d/2)\right) \ .
\end{equation}

Using eq. (\ref{eq:one-loop basis integral}), it is straightforward to prove that the previous expression is regular in the limit $d \to 4$ and yields the well-known result proportional to $\zeta(3)$, i.e.  \begin{equation}
	\begin{split}
		\label{eq:zeta3 bubble}
		G(d) &= (p^2)^{d-5}\widetilde{G}(d)= \dfrac{6\zeta(3)}{(4\pi)^4p^2} +\cO(d-4)  \ . 
	\end{split}
\end{equation} 

Finally, to Fourier transform in configuration space, we will employ the formula  
\begin{equation}
	\label{eq:Fourier transform for massless propagators}
	\cD(x,s) \equiv	\int \dfrac{ \dd^d{p}}{(2\pi)^d} \dfrac{e^{\mathrm{i}p\cdot x}}{(p^2)^s} = \dfrac{\Gamma(d/2-s)} {4^s \pi^{d/2} \Gamma(s)} \dfrac{1}{(x^2)^{d/2-s}} \  .
\end{equation} 
The tree-level propagators in configuration space are proportional to $\Delta(x) = \cD(x,1)$.

\subsection{One-loop corrections}
\label{sec:relevant one-loop corrections} 
In this subsection, we examine in detail the one-loop corrections to the propagators which enter the calculation of the Wilson loop.

We begin with considering the  gauge field and the adjoint scalar propagators. By gauge invariance, we can deduce that   
\begin{align}
	\label{eq:general structure one-loop correction scalar pro}
	\mathord{
		\begin{tikzpicture}[scale=0.5, baseline=-0.65ex]
			\draw [black] (0,0) circle [radius=1cm];
			\begin{feynman}
				\vertex (A) at (-2,0);
				\vertex (C) at (-1,0);
				\vertex (B) at (1, 0);
				\vertex (D) at (2, 0);
				\diagram*{
					(A) -- [fermion] (C),
					(B) --[fermion] (D),
				};
			\end{feynman}
		\end{tikzpicture} 
	} &= \dfrac{\delta^{ab}g_B^2}{(p^2)^2}\pi_S^{(1)}(p^2) \ ,\\ 
	\label{eq:general structure one-loop correction gauge pro}
	\mathord{
		\begin{tikzpicture}[scale=0.5, baseline=-0.65ex]
			\draw [black] (0,0) circle [radius=1cm];
			\begin{feynman}
				\vertex (A) at (-2,0);
				\vertex (C) at (-1,0);
				\vertex (B) at (1, 0);
				\vertex (D) at (2, 0);
				\diagram*{
					(A) -- [photon] (C),
					(B) --[photon] (D),
				};
			\end{feynman}
		\end{tikzpicture} 
	} &= \dfrac{\delta^{ab}g_B^2}{(p^2)^2}\left(\delta_{\mu \nu}-\dfrac{p_\mu p_\nu}{p^2}\right)\pi_G^{(1)}(p^2)\ , 
\end{align}where $\pi_G^{(1)}$ and $\pi_S^{(1)}$ are the gluon and scalar polarization operator, respectively. For the theories under examination, these quantities were computed in Appendix C of \cite{Billo:2023igr}, where it is explicitly showed that  they coincide in the Feynman gauge, as expected by supersymmetry. For  future reference, we report the relevant Feynman diagrams that contribute to eq. (\ref{eq:general structure one-loop correction scalar pro}). Using the conventions of Appendix \ref{sec:actions in flat space}, we find that 
\begin{equation}
	\label{eq:one-loop diagrams}
	\begin{split}
		\mathord{
			\begin{tikzpicture}[scale=0.5, baseline=-0.65ex]
				\draw [black] (0,0) circle [radius=1cm];
				\begin{feynman}
					\vertex (A) at (-2,0);
					\vertex (C) at (-1,0);
					\vertex (B) at (1, 0);
					\vertex (D) at (2, 0);
					\diagram*{
						(A) -- [fermion] (C),
						(B) --[fermion] (D),
					};
				\end{feynman}
			\end{tikzpicture} 
		}
		&= \mathord{\begin{tikzpicture}[baseline=-0.65ex,scale=0.5]
				\begin{feynman}
					\vertex (a) at (-1,0)  ;
					\vertex (b) at (4,0)  ;
					\vertex (c) at (0.5,0) ;
					\vertex (d) at (2.5,0) ;
					\vertex (e) at (1.5, 1.5) {$A A$} ;  
					\diagram*{
						(a) -- [fermion] (c),
						(c) -- [photon, half left] (d),
						(c) -- [fermion] (d),
						(d) -- [fermion] (b),
					};
				\end{feynman}
			\end{tikzpicture}
		} + \mathord{\begin{tikzpicture}[baseline=-0.65ex,scale=0.5]
				\begin{feynman}
					\vertex (a) at (-1,0)  ;
					\vertex (b) at (4,0)  ;
					\vertex (c) at (0.5,0) ;
					\vertex (d) at (2.5,0) ;
					\vertex (e) at (1.5, 1.5) {$\psi\Bar{\psi}$} ;
					\vertex (e) at (1.5, -1.5) {$\lambda\Bar{\lambda}$};
					\diagram*{
						(a) -- [fermion] (c),
						(c) -- [fermion, half left, thick] (d),
						(c) -- [fermion, half right, thick] (d),
						(d) -- [fermion] (b),
					};
				\end{feynman}
			\end{tikzpicture}
		}
		+ 	\mathord{\begin{tikzpicture}[baseline=-0.65ex,scale=0.5]
				\begin{feynman}
					\vertex (a) at (-1,0) ;
					\vertex (b) at (4,0) ;
					\vertex (c) at (0.5,0) ;
					\vertex (d) at (2.5,0) ;
					\vertex (e) at (1.5, 1.5) {$\eta\Bar{\eta}$};
					\vertex (e) at (1.5, -1.5) {$\Tilde{\eta}\Bar{\Tilde{\eta}}$};
					\diagram*{
						(a) -- [fermion] (c),
						(c) -- [anti charged scalar, half left, thick] (d),
						(c) -- [anti charged scalar, half right, thick] (d),
						(d) -- [fermion] (b),
					};
				\end{feynman}
			\end{tikzpicture}
		} \\ 
	\end{split}
\end{equation}
where  $\psi$ and $\lambda$ denote the two gauginos of the vector multiplet, while $\eta$ and $\tilde{\eta}$ are the Weyl  fermions associated with the massless hypermutliplets in the representation $\cR$. Going through the calculation of   eq. (\ref{eq:one-loop diagrams}), it is possible to show that the first two diagrams cancel each other out and consequently, we remain with \cite{Billo:2023igr} 
\begin{equation}
	\label{eq:one-loop adjoint scalar appendix}
	\begin{split}
		\mathord{
			\begin{tikzpicture}[scale=0.5, baseline=-0.65ex]
				\draw [black] (0,0) circle [radius=1cm];
				\begin{feynman}
					\vertex (A) at (-2,0);
					\vertex (C) at (-1,0);
					\vertex (B) at (1, 0);
					\vertex (D) at (2, 0);
					\diagram*{
						(A) -- [fermion] (C),
						(B) --[fermion] (D),
					};
				\end{feynman}
			\end{tikzpicture} 
		} = \mathord{\begin{tikzpicture}[baseline=-0.65ex,scale=0.5]
				\begin{feynman}
					\vertex (a) at (-1,0) ;
					\vertex (b) at (4,0) ;
					\vertex (c) at (0.5,0) ;
					\vertex (d) at (2.5,0) ;
					\vertex (e) at (1.5, 1.5) {$\eta\Bar{\eta}$};
					\vertex (e) at (1.5, -1.5) {$\Tilde{\eta}\Bar{\Tilde{\eta}}$};
					\diagram*{
						(a) -- [fermion] (c),
						(c) -- [anti charged scalar, half left, thick] (d),
						(c) -- [anti charged scalar, half right, thick] (d),
						(d) -- [fermion] (b),
					};
				\end{feynman}
			\end{tikzpicture}
		}\equiv	\mathord{
			\begin{tikzpicture}[scale=0.5, baseline=-0.65ex]
				\draw [black, dashed, thick] (0,0) circle [radius=1cm];
				\begin{feynman}
					\vertex (A) at (-2,0);
					\vertex (C) at (-1,0);
					\vertex (B) at (1, 0);
					\vertex (D) at (2, 0);
					\diagram*{
						(A) -- [fermion] (C),
						(B) --[fermion] (D),
					};
				\end{feynman}
			\end{tikzpicture} 
		}=-2\dfrac{\delta_{ab}g_B^2}{p^2}i_\mathcal{R}G(1,1) \ ,  
	\end{split}
\end{equation} 
where $G(1,1)$ is defined in (\ref{eq:one-loop basis integral}) and we recall $i_\cR$ is the Dynkin index of the representation $\cR$. Since  $\pi_G^{(1)}(p^2)=\pi_S^{(1)}(p^2)$ in the Feynman gauge, we deduce that \cite{Billo:2023igr}
\begin{equation}
	\label{eq:one-loop correction gauge field appendix}
	\begin{split}
		\mathord{
			\begin{tikzpicture}[scale=0.5, baseline=-0.65ex]
				\draw [black] (0,0) circle [radius=1cm];
				\begin{feynman}
					\vertex (A) at (-2,0);
					\vertex (C) at (-1,0);
					\vertex (B) at (1, 0);
					\vertex (D) at (2, 0);
					\diagram*{
						(A) -- [photon] (C),
						(B) --[photon] (D),
					};
				\end{feynman}
			\end{tikzpicture} 
		}&=		\mathord{
			\begin{tikzpicture}[scale=0.5, baseline=-0.65ex]
				\draw [black, dashed, thick] (0,0) circle [radius=1cm];
				\begin{feynman}
					\vertex (A) at (-2,0);
					\vertex (C) at (-1,0);
					\vertex (B) at (1, 0);
					\vertex (D) at (2, 0);
					\diagram*{
						(A) -- [photon] (C),
						(B) --[photon] (D),
					};
				\end{feynman}
			\end{tikzpicture} 
		}
		=-2i_\cR\dfrac{\delta_{ab}g_B^2}{(p^2)}\left(\delta_{\mu \nu}-\dfrac{p_\mu p_\nu}{p^2}\right)G(1,1) \ .
	\end{split}
\end{equation} 
Using these results, we can easily derive the one-loop corrections to the propagators in the difference theory method. Subtracting off the contributions of $\mathcal{N}=4$ SYM, where the hypermultiplets transform   in the adjoint representation, we find that
\begin{align}
	\label{eq:one-loop adjoint scalar difference appendix}
	\mathord{
		\begin{tikzpicture}[scale=0.5, baseline=-0.65ex]
			\draw [black] (0,0) circle [radius=1cm];
			\draw [black, thick, dashed] (0,0) circle [radius=0.9cm];
			\begin{feynman}
				\vertex (A) at (-2,0);
				\vertex (C) at (-1,0);
				\vertex (B) at (1, 0);
				\vertex (D) at (2, 0);
				\diagram*{
					(A) -- [fermion] (C),
					(B) --[fermion] (D),
				};
			\end{feynman}
		\end{tikzpicture} 
	} &=\delta_{ab} \dfrac{g_B^2 \Pi^{(1)}(p^2)}{(p^2)^2} 
	\ , \\[0.4em]
	\label{eq:one-loop gauge field difference appendix}
	\mathord{
		\begin{tikzpicture}[scale=0.5, baseline=-0.65ex]
			\draw [black] (0,0) circle [radius=1cm];
			\draw [black, thick, dashed] (0,0) circle [radius=0.9cm];
			\begin{feynman}
				\vertex (A) at (-2,0);
				\vertex (C) at (-1,0);
				\vertex (B) at (1, 0);
				\vertex (D) at (2, 0);
				\diagram*{
					(A) -- [photon] (C),
					(B) --[photon] (D),
				};
			\end{feynman}
		\end{tikzpicture} 
	}&=\delta_{ab}\left(\delta_{\mu \nu}-\dfrac{p_\mu p_\nu}{p^2}\right) \dfrac{g_B^2 \Pi^{(1)} (p^2)}{(p^2)^2} \ , 
\end{align}
with the one-loop polarization operator in the difference theory being given by
\begin{equation}
	\label{eq:scalar gluon pol 1-loop difference}
	\Pi^{(1)}(p^2) = f^{(1)}(d)(p^2)^{d/2-1} \ , \quad \text{where} \quad f^{(1)}(d)
	= - 16\pi^2 \beta_0\widetilde G(1,1)~.
\end{equation} 
We recall that the dimensionless function $\tilde{G}(1,1)$ is given by (\ref{eq:one-loop basis integral}), while $\beta_0$ is the one-loop coefficient of the $\beta$-function (\ref{eq:beta0}).
In configuration space, using eq. (\ref{eq:Fourier transform for massless propagators}) to perform the Fourier transform, we find the following result
\begin{align}
	\label{eq:one-loop correction adjoint scalar main text}
	\mathord{
		\begin{tikzpicture}[scale=0.5, baseline=-0.65ex]
			\draw [black] (0,0) circle [radius=1cm];
			\draw [black, thick, dashed] (0,0) circle [radius=0.9cm];
			\begin{feynman}
				\vertex (A) at (-2,0);
				\vertex (a) at (-2.3,-0.5) {$x_1$};
				\vertex (C) at (-1,0);
				\vertex (B) at (1, 0);
				\vertex (D) at (2, 0);
				\vertex (d) at (2.3,-0.5) {$x_2$};
				\diagram*{
					(A) -- [fermion] (C),
					(B) --[fermion] (D),
				};
			\end{feynman}
		\end{tikzpicture} 
	} &= f^{(1)}(d) g_B^2 
	\, \cD(x_{12},3-d/2)
	\equiv \Delta^{(1)}(x_{12})~ ,  
\end{align} 
for the scalar propagator. 
Repeating the same calculation for the gluon,  we have  

\begin{equation}	
	\label{eq:one-loop correction gauge-field position main text}
	\begin{split}
		\mathord{
			\begin{tikzpicture}[scale=0.5, baseline=-0.65ex]
				\draw [black] (0,0) circle [radius=1cm];
				\draw [black, thick, dashed] (0,0) circle [radius=0.9cm];
				\begin{feynman}
					\vertex (A) at (-2,0);
					\vertex (a) at (-2.3,-0.5) {$x_1$};
					\vertex (C) at (-1,0);
					\vertex (B) at (1, 0);
					\vertex (D) at (2, 0);
					\vertex (d) at (2.3,-0.5) {$x_2$};
					\diagram*{
						(A) -- [photon] (C),
						(B) --[photon] (D),
					};
				\end{feynman}
			\end{tikzpicture}
		} 	& = g_B^2 f^{(1)}(d) \,	\left(\delta_{\mu \nu} \cD(x_{12},3-d/2) - \partial_{1,\mu}\partial_{2,\nu} \cD(x_{12},4-d/2)\right)\\
		& \equiv \delta_{\mu\nu} \Delta^{(1)}(x_{12}) - \partial_{1,\mu}\partial_{2,\nu} 
		\Delta^{(1), \mathrm{g}}(x_{12})~.
	\end{split}
\end{equation}

By gauge invariance, we expect that all the Wilson loop diagrams which involves the gauge-like term  $\partial_{1,\mu}\partial_{2,\nu}\Delta^{(1), \mathrm{g}}(x_{12})$ do not  contribute to the final results and in the following, we will verify this property  explicitly.

Finally, we consider the fermionic propagators at one-loop accuracy. These will enter the calculation of the two-loop corrections to the adjoint scalar propagator we will examine in the following section. Specifically, we begin with considering the vector multiplet fermions, i.e. the gauginos $\psi$ and $\lambda$.  For the Weyl fermion $\psi$, we find 
\begin{equation}
	\label{eq:adjoint fermions}
	\begin{split}
		\mathord{
			\begin{tikzpicture}[scale=0.5, baseline=-0.65ex]
				\draw [black] (0,0) circle [radius=1cm];
				\begin{feynman}
					\vertex (A) at (-2,0);
					\vertex (C) at (-1,0);
					\vertex (B) at (1, 0);
					\vertex (D) at (2, 0);
					\diagram*{
						(A) -- [fermion, thick] (C),
						(B) --[fermion, thick] (D),
					};
				\end{feynman}
			\end{tikzpicture} 
		} 
		&= 
		\mathord{\begin{tikzpicture}[baseline=-0.65ex,scale=0.5]
				\begin{feynman}
					\vertex (a) at (-1,0)  ;
					\vertex (b) at (4,0)  ;
					\vertex (c) at (0.5,0) ;
					\vertex (d) at (2.5,0) ;
					\vertex (e) at (1.5, 1.5) ;  
					\diagram*{
						(a) -- [fermion, thick] (c),
						(c) -- [photon, half left] (d),
						(c) --[scalar, half left] (d),
						(c) -- [fermion, thick] (d),
						(d) -- [fermion, thick] (b),
					};
				\end{feynman}
			\end{tikzpicture}
		} + \mathord{\begin{tikzpicture}[baseline=-0.65ex,scale=0.5]
				\begin{feynman}
					\vertex (a) at (-1,0)  ;
					\vertex (b) at (4,0) ;
					\vertex (c) at (0.5,0) ;
					\vertex (d) at (2.5,0) ;
					\vertex (e) at (1.5, 1.5) {$\lambda\bar{\lambda}$};
					\vertex (e) at (1.5, -1.5)
					{$ \phi\bar{\phi}$} ;
					\diagram*{
						(a) -- [fermion, thick] (c),
						(c) -- [anti fermion, half left, thick] (d),
						(c) -- [fermion, half right] (d),
						(d) -- [fermion,thick] (b),
					};
				\end{feynman}
			\end{tikzpicture}
		} 
		+  \mathord{\begin{tikzpicture}[baseline=-0.65ex,scale=0.5] \draw[arrows = {-Latex[width=5pt, length=5pt]}]   (1.55,-0.9) -- (1.45,-0.9) ;
				\begin{feynman}
					\vertex (a) at (-1,0)  ;
					\vertex (b) at (4,0) ;
					\vertex (c) at (0.5,0) ;
					\vertex (d) at (2.5,0) ;
					\vertex (e) at (1.5, 1.5) {$q\bar{q}$};
					\vertex (e) at (1.5, -1.45)  {$\tilde{\eta}\bar{\tilde{\eta}}$};
					\diagram*{
						(a) -- [fermion, thick] (c),
						(c) -- [anti charged scalar, half left] (d),
						(c) -- [ghost, half right,thick] (d),
						(d) -- [fermion, thick] (b),
					};
				\end{feynman}
			\end{tikzpicture}
		} \\
		& +  \mathord{\begin{tikzpicture}[baseline=-0.65ex,scale=0.5] \draw[arrows = {-Latex[width=5pt, length=5pt]}]   (1.55,-0.9) -- (1.45,-0.9) ;
				\begin{feynman}
					\vertex (a) at (-1,0)  ;
					\vertex (b) at (4,0) ;
					\vertex (c) at (0.5,0) ;
					\vertex (d) at (2.5,0) ;
					\vertex (e) at (1.5, 1.5) {$\eta\bar{\eta}$};
					\vertex (e) at (1.5, -1.5)  {$\tilde{q}\bar{\tilde{q}}$};
					\diagram*{
						(a) -- [fermion, thick] (c),
						(c) -- [anti charged scalar, half left] (d),
						(c) -- [ghost, half right] (d),
						(d) -- [fermion, thick] (b),
					};
				\end{feynman}
			\end{tikzpicture}
		}  
		= -2(N+i_\cR ) \delta_{ab}\dfrac{g_B^2{\slashed{p}}}{p^2}G(1,1) \ ,
	\end{split}
\end{equation} where ${\slashed{p}}\equiv p_\mu{\sigma}^\mu$, with ${\sigma}^\mu$ given by (\ref{eq:sigma matrices}). In the previous expression, $q$ and $\tilde{q}$ are the complex scalars associated with hypers in the representation $\cR$, while the  first diagram  results from the interaction of the fermion $\psi$ with the gauge field $A_\mu$ and with the  real scalars $A_i$, where $i=1,\ldots,4-d$, which emerge from dimensional reduction.  We verified that the one-loop corrections to the propagator of the gaugino $\lambda$  give us the same result, as expected from supersymmetry. 
From eq. (\ref{eq:adjoint fermions}), we can easily deduce the one-loop correction to the fermion propagator in the difference method, i.e.  \begin{equation}
	\label{eq:fermion propagator one-loop difference}
	\mathord{
		\begin{tikzpicture}[scale=0.5, baseline=-0.65ex]
			\draw [black] (0,0) circle [radius=1cm];
			\draw [black, thick, dashed] (0,0) circle [radius=0.9cm];
			\begin{feynman}
				\vertex (A) at (-2,0);
				\vertex (C) at (-1,0);
				\vertex (B) at (1, 0);
				\vertex (D) at (2, 0);
				\diagram*{
					(A) -- [fermion, thick] (C),
					(B) --[fermion, thick] (D),
				};
			\end{feynman}
		\end{tikzpicture} 
	} = 2(N-i_\cR ) \delta_{ab}\dfrac{g_B^2{\slashed{p}}}{p^2}G(1,1)  \ .
\end{equation}
Finally, we consider the corrections to the propagators of the spinors $\eta$ and $\tilde{\eta}$. For the fermion $\eta$ we find 
\begin{equation}
	\label{eq:one-loop correction to the fermion in repr R}
	\begin{split}
		\mathord{
			\begin{tikzpicture}[scale=0.5, baseline=-0.65ex]
				\draw [black] (0,0) circle [radius=1cm];
				\begin{feynman}
					\vertex (A) at (-2,0);
					\vertex (C) at (-1,0);
					\vertex (B) at (1, 0);
					\vertex (D) at (2, 0);
					\diagram*{
						(A) -- [charged scalar, thick] (C),
						(B) --[charged scalar, thick] (D),
					};
				\end{feynman}
			\end{tikzpicture} 
		} 
		&= 
		\mathord{\begin{tikzpicture}[baseline=-0.65ex,scale=0.5]
				\begin{feynman}
					\vertex (a) at (-1,0)  ;
					\vertex (b) at (4,0)  ;
					\vertex (c) at (0.5,0) ;
					\vertex (d) at (2.5,0) ;
					\vertex (e) at (1.5, 1.5) ;  
					\diagram*{
						(a) -- [charged scalar, thick] (c),
						(c) -- [photon, half left] (d),
						(c) --[ scalar, half left] (d),
						(c) -- [charged scalar, thick] (d),
						(d) -- [charged scalar, thick] (b),
					};
				\end{feynman}
			\end{tikzpicture}
		} + \mathord{\begin{tikzpicture}[baseline=-0.65ex,scale=0.5]\draw[arrows = {-Latex[width=5pt, length=5pt]}]  (1.55,-0.9) -- (1.45,-0.9) ; 
				\begin{feynman}
					\vertex (a) at (-1,0)  ;
					\vertex (b) at (4,0) ;
					\vertex (c) at (0.5,0) ;
					\vertex (d) at (2.5,0) ;
					\vertex (e) at (1.5, 1.5) {$\psi\bar{\psi}$};
					\vertex (e) at (1.5, -1.5)
					{$ q\bar{q}$} ;
					\diagram*{
						(a) -- [charged scalar, thick] (c),
						(c) -- [anti fermion, half left, thick] (d),
						(c) -- [ghost, half right] (d),
						(d) -- [charged scalar,thick] (b),
					};
				\end{feynman}
			\end{tikzpicture}
		} 
		+\mathord{
			\begin{tikzpicture}[baseline=-0.65ex,scale=0.5]
				\begin{feynman}
					\vertex (a) at (-1,0)  ;
					\vertex (b) at (4,0) ;
					\vertex (c) at (0.5,0) ;
					\vertex (d) at (2.5,0) ;
					\vertex (e) at (1.5, 1.5) {$\lambda\bar{\lambda}$};
					\vertex (e) at (1.5, -1.5) {$q\bar{q}$};
					\diagram*{
						(a) -- [charged scalar, thick] (c),
						(c) -- [anti fermion, half left, thick] (d),
						(c) -- [anti charged scalar, half right] (d),
						(d) -- [charged scalar,thick] (b),
					};
				\end{feynman}
			\end{tikzpicture}
		} \\
		& +  \mathord{\begin{tikzpicture}[baseline=-0.65ex,scale=0.5] \draw[arrows = {-Latex[width=5pt, length=5pt]}]   (1.55,-0.9) -- (1.45,-0.9) ;
				\begin{feynman}
					\vertex (a) at (-1,0)  ;
					\vertex (b) at (4,0) ;
					\vertex (c) at (0.5,0) ;
					\vertex (d) at (2.5,0) ;
					\vertex (e) at (1.5, 1.5) {$\phi\bar{\phi}$};
					\vertex (e) at (1.5, -1.5) {$\tilde{\eta}\bar{\tilde{\eta}}$};
					\diagram*{
						(a) -- [charged scalar, thick] (c),
						(c) -- [anti fermion, half left, thick] (d),
						(c) -- [ghost, half right,thick] (d),
						(d) -- [charged scalar, thick] (b),
					};
				\end{feynman}
			\end{tikzpicture}
		}
		= -4 C_\cR\delta_{uv}\dfrac{g_B^2{\slashed{p}}}{p^2}G(1,1) \ ,
	\end{split}
\end{equation} where  $u,v=1,\ldots,\dim\cR$ and we recall that  $C_\cR$ is  the quadratic Casimir\footnote{The quadratic Casimir is defined via the relation $T^a_\cR T^a_\cR = C_\cR \mathbb{I}$.} of the representation $\cR$. We find an identical result for fermion $\tilde{\eta}$ as expected from supersymmetry. 

\subsection{Two-loop corrections to the propagators}
\label{sec:scalar polarization operator}
The three-loop analysis of the 1/2 BPS Wilson loop involves diagrams characterized by the two-loop corrections to the adjoint scalar  and gauge field propagator in the \textit{difference theory approach}. In the Feynman gauge, the expectation based on supersymmetry is that these quantities coincide up to spacetime indices\footnote{An explicit test of this property at two-loop accuracy can be found in \cite{Andree:2010na}, where the authors studied the 1/2 BPS Wilson loop in superconformal $\mathcal{N}=2$ QCD.}, as it occurs at one-loop accuracy (see eq.s (\ref{eq:one-loop adjoint scalar difference appendix}) and (\ref{eq:one-loop gauge field difference appendix})). Therefore, in the following, we will 
assume that 
\begin{align}
	\label{eq:two loop corrections adjoint momentum}
	\mathord{
		\begin{tikzpicture}[scale=0.6, baseline=-0.65ex]
			\filldraw[color=gray!80, fill=gray!15](0,0) circle (1);	
			\draw [black,thick] (0,0) circle [radius=1cm];
			\draw [black, thick, dashed] (0,0) circle [radius=0.9cm];
			\begin{feynman}
				\vertex (A) at (-2,0);
				\vertex (C) at (-1,0);
				\vertex (C1) at (0.1,0) {\text{\footnotesize 2-loop }\normalsize} ;
				\vertex (B) at (1, 0);
				\vertex (D) at (2, 0);
				\diagram*{
					(A) -- [fermion] (C),
					(B) --[fermion] (D),
				};
			\end{feynman}
		\end{tikzpicture} 
	} &=
	\mathord{
		\begin{tikzpicture}[scale=0.6, baseline=-0.65ex]
			\filldraw[color=gray!15, fill=gray!15](0,0) circle (1);	
			\draw [black, dashed, thick] (0,0) circle [radius=1cm];
			\begin{feynman}
				\vertex (A) at (-2,0);
				\vertex (C) at (-1,0);
				\vertex (d) at (0.1,0.3) {\text{\footnotesize 2-loop }\normalsize} ;
				\vertex (d) at (0.1,-0.3) {\text{\footnotesize $\mathcal{R}$ }\normalsize} ;
				\vertex (B) at (1, 0);
				\vertex (D) at (2, 0);
				\diagram*{
					(A) -- [fermion] (C),
					(B) --[fermion] (D),
				};
			\end{feynman}
		\end{tikzpicture} 
	} - 	\mathord{
		\begin{tikzpicture}[scale=0.6, baseline=-0.65ex]
			\filldraw[color=gray!80, fill=gray!15](0,0) circle (1);	
			\draw [black,thick] (0,0) circle [radius=1cm];
			\begin{feynman}
				\vertex (A) at (-2,0);
				\vertex (C) at (-1,0);
				\vertex (d) at (0.1,0.3) {\text{\footnotesize 2-loop }\normalsize} ;
				\vertex (d) at (0.1,-0.3) {\text{\footnotesize $\mathrm{Adj}$ }\normalsize} ;
				\vertex (B) at (1, 0);
				\vertex (D) at (2, 0);
				\diagram*{
					(A) -- [fermion] (C),
					(B) --[fermion] (D),
				};
			\end{feynman}
		\end{tikzpicture} 
	} 
	=\dfrac{\delta_{ab}g_B^4}{(p^2)^2}\Pi^{(2)}(p^2) \ , \\[0.6em]
	\label{eq:two loop corrections gluon momentum}
	\mathord{
		\begin{tikzpicture}[scale=0.6, baseline=-0.65ex]
			\filldraw[color=gray!80, fill=gray!15](0,0) circle (1);	
			\draw [black,thick] (0,0) circle [radius=1cm];
			\draw [black, thick, dashed] (0,0) circle [radius=0.9cm];
			\begin{feynman}
				\vertex (A) at (-2,0);
				\vertex (C) at (-1,0);
				\vertex (C1) at (0.1,0) {\text{\footnotesize 2-loop }\normalsize} ;
				\vertex (B) at (1, 0);
				\vertex (D) at (2, 0);
				\diagram*{
					(A) -- [photon] (C),
					(B) --[photon] (D),
				};
			\end{feynman}
		\end{tikzpicture} 
	} 
	&=\mathord{
		\begin{tikzpicture}[scale=0.6, baseline=-0.65ex]
			\filldraw[color=gray!15, fill=gray!15](0,0) circle (1);	
			\draw [black, dashed, thick] (0,0) circle [radius=1cm];
			\begin{feynman}
				\vertex (A) at (-2,0);
				\vertex (C) at (-1,0);
				\vertex (d) at (0.1,0.3) {\text{\footnotesize 2-loop }\normalsize} ;
				\vertex (d) at (0.1,-0.3) {\text{\footnotesize $\mathcal{R}$ }\normalsize} ;
				\vertex (B) at (1, 0);
				\vertex (D) at (2, 0);
				\diagram*{
					(A) -- [photon] (C),
					(B) --[photon] (D),
				};
			\end{feynman}
		\end{tikzpicture} 
	} - 	\mathord{
		\begin{tikzpicture}[scale=0.6, baseline=-0.65ex]
			\filldraw[color=gray!80, fill=gray!15](0,0) circle (1);	
			\draw [black,thick] (0,0) circle [radius=1cm];
			\begin{feynman}
				\vertex (A) at (-2,0);
				\vertex (C) at (-1,0);
				\vertex (d) at (0.1,0.3) {\text{\footnotesize 2-loop }\normalsize} ;
				\vertex (d) at (0.1,-0.3) {\text{\footnotesize $\mathrm{Adj}$ }\normalsize} ;
				\vertex (B) at (1, 0);
				\vertex (D) at (2, 0);
				\diagram*{
					(A) -- [photon] (C),
					(B) --[photon] (D),
				};
			\end{feynman}
		\end{tikzpicture} 
	} =\dfrac{\delta_{ab}g_B^4}{(p^2)^2}\left(\delta_{\mu\nu}- \frac{p_\mu p_\nu}{p^2}\right)\Pi^{(2)}(p^2) \ ,
\end{align}
and we will calculate the two-loop polarization operator $\Pi^{(2)}(p^2)$ by considering the scalar propagator. In the previous expression,  the contribution labelled by  $\cR$ encodes all the two-loop diagrams in $\mathcal{N}=2$ SYM in which the scalar $\phi$ (or the gluon) interacts with matter fields in representation $\cR$, while the other contribution denotes the corrections resulting from $\mathcal{N}=4$ SYM, where matter transforms in the adjoint representation, i.e. $\cR=\rm Adj$. By dimensional reasons, the polarization operators can be written as 
\begin{equation}
	\label{Pi2form}
	\Pi^{(2)}(p^2) = (p^2)^{d-3}\, f^{(2)}(d)~,
\end{equation}
where $f^{(2)}(d)$ is a dimensionless function of  $d$ and includes colour factors. To avoid cumbersome expressions, we find convenient to express every diagram  by the basis integrals (\ref{eq:one-loop basis integral}) and (\ref{eq:massless two-loop integral}) and  directly provide their contributions to $f^{(2)}(d)$, 
omitting  the overall prefactor  $g_B^4\delta_{ab}/(p^2)^{5-d}$. 

On the one-hand, we find that the reducible corrections are simply given by    \begin{equation}
	\begin{split}
		\mathord{
			\begin{tikzpicture}[scale=0.5, baseline=-0.65ex]
				\draw [black,dashed,thick] (0,0) circle [radius=1cm];
				\draw [black,dashed,thick] (3,0) circle [radius=1cm];
				\begin{feynman}
					\vertex (A) at (-2,0);
					\vertex (C) at (-1,0);
					\vertex (B) at (1, 0);
					\vertex (D) at (2, 0);
					\vertex (C3) at (5,0);
					\vertex (C4) at (4,0);
					\diagram*{
						(A) -- [fermion] (C),
						(B) --[fermion] (D),
						(C4) --[fermion] (C3),
					};
				\end{feynman}
			\end{tikzpicture} 
		} -	\mathord{
			\begin{tikzpicture}[scale=0.5, baseline=-0.65ex]
				\draw [black,thick] (0,0) circle [radius=1cm];
				\draw [black,thick] (3,0) circle [radius=1cm];
				\begin{feynman}
					\vertex (A) at (-2,0);
					\vertex (C) at (-1,0);
					\vertex (B) at (1, 0);
					\vertex (D) at (2, 0);
					\vertex (C3) at (5,0);
					\vertex (C4) at (4,0);
					\diagram*{
						(A) -- [fermion] (C),
						(B) --[fermion] (D),
						(C4) --[fermion] (C3),
					};
				\end{feynman}
			\end{tikzpicture} 
		} 	=	4 (i^2_\mathcal{R}-N^2)\widetilde G(1,1)^2 \ ,
	\end{split}
\end{equation} 
as it follows from eq. (\ref{eq:one-loop adjoint scalar appendix}). On the other hand, the irreducible contributions can be  organized in two classes of diagrams. 

The first one arises when decorating the internal lines of the diagrams depicted in eq. (\ref{eq:one-loop diagrams}) with the one-loop self-energies associated with the fields of the virtual loops. In the difference theory approach, we find the following  classes of diagrams 
\begin{align}
	\label{eq:decorating 1}
	\mathord{
		\begin{tikzpicture}[scale=0.6, baseline=-0.65ex]
			\filldraw[color=white!80, fill=white!15](0,0) circle (0.8);	
			\draw [black] (0,0) circle [radius=0.8cm];
			\draw [black, thick, dashed] (0,0) circle [radius=0.7cm];
			\begin{feynman}
				\vertex (a) at (-2.8,0);
				\vertex (A) at (-1.8,0);
				\vertex (C) at (-0.8,0);
				\vertex (B) at (0.8, 0);
				\vertex (D) at (1.8, 0);
				\vertex (d) at (2.8,0);
				\diagram*{
					(a) -- [ fermion] (A),
					(A)-- [fermion] (C),
					(B) --[fermion] (D),
					(A) --[photon, half left]  (D),
					(D) --[fermion] (d)
				};
			\end{feynman}
		\end{tikzpicture} 
	} &+
	\mathord{
		\begin{tikzpicture}[scale=0.6, baseline=-0.65ex]
			\begin{feynman}
				\vertex (a) at (-2.8,0);
				\vertex (A) at (-1.8,0);
				\vertex (C) at (-0.8,0);
				\vertex (B) at (0.8, 0);
				\vertex (b1) at (-0.8,1.5);
				\vertex (b2) at (0.8,1.5);
				\vertex (D) at (1.8, 0);
				\vertex (d) at (2.8,0);
				\diagram*{
					(a) -- [fermion] (A),
					(A) -- [fermion] (D),
					(A) --[photon, half left]  (D),
					(D)-- [fermion] (d),
				};
				\filldraw[color=white, fill=white](0,1.5) circle (0.8);	
				\draw [black] (0,1.5) circle [radius=0.8cm];
				\draw [black, thick, dashed] (0,1.5) circle [radius=0.7cm];
			\end{feynman}
		\end{tikzpicture} 
	} +\mathord{
		\begin{tikzpicture}[baseline=-0.6ex,scale=0.6]
			\newcommand\tmpda{-0.15cm}
			\newcommand\tmpdb{-1.45cm}
			\begin{feynman}
				\vertex (a) at (-1,0)  ;
				\vertex (b) at (4,0)  ;
				\vertex (c) at (0.5,0) ;
				\vertex (d) at (2.5,0) ;
				\vertex (d1) at (1,1) ;
				\diagram*{
					(a) -- [fermion](c),
					(c) -- [fermion,half right,thick] (d),
					(c) -- [plain,half left,thick,with  arrow=\tmpdb] (d),
					(c) -- [plain,half left,thick,with arrow=\tmpda] (d),
					(d) -- [fermion] (b),
				};
			\end{feynman}
			\filldraw[color=white, fill=white](1.5,1) circle (0.7);
			\draw [black] (1.5,1) circle [radius=0.7cm];
			\draw [black,dashed,thick] (1.5,1) circle [radius=0.6cm];
		\end{tikzpicture}
	}\notag\\
	& =2N(N-i_\cR) \left(5 \widetilde G(0,1,1,0,1)- \widetilde G(0,1,1,2,1)\right) \ , \\
	\mathord{
		\begin{tikzpicture}[baseline=-0.6ex,scale=0.6]
			\label{eq:decorating 2}
			\draw[
			black
			]
			(1.5,0) circle (1);
			\draw[
			thick, black,dashed
			]
			(1.5,0) circle (0.9);
			\filldraw[color=white!80, fill=white!15](1.5,1) circle (0.7);
			\draw [black] (1.5,1) circle [radius=0.7cm];
			\begin{feynman}
				\vertex (a) at (-1,0)  ;
				\vertex (b) at (4,0)  ;
				\vertex (c) at (0.5,0) ;
				\vertex (d) at (2.5,0) ;
				\vertex (d1) at (1,1) ;
				\diagram*{
					(a) -- [fermion](c),
					(d) -- [fermion] (b),
				};
			\end{feynman}
		\end{tikzpicture}
	} &= 16\left( C_\mathcal{R}i_\mathcal{R}-N^2\right)\left(\widetilde G(0,1,1,1,1) - \widetilde G(0,1,1,0,1)\right) \ .
\end{align} 
In eq. (\ref{eq:decorating 1}), the internal bubbles in the double dashed/continuos line notation denote, respectively, the one-loop correction to the adjoint scalar, gauge field and gaugino propagators in the difference method (see, respectively,  eq.s (\ref{eq:one-loop adjoint scalar difference appendix}), (\ref{eq:one-loop gauge field difference appendix}) and (\ref{eq:fermion propagator one-loop difference})). Similarly,  eq. (\ref{eq:decorating 2})  arises when we decorate the matter loop in eq.  (\ref{eq:one-loop adjoint scalar appendix}) with the one-loop correction (\ref{eq:one-loop correction to the fermion in repr R}) and we subtract the contribution of $\mathcal{N}=4$ SYM. 

The second class of irreducible corrections emerges from pure two-loop diagrams, which we organize in terms of  three fermionic loops and one sunset-like correction, i.e. 
\begin{align}
	\label{glasswinediagrams}
	\mathord{
		\begin{tikzpicture}[baseline=-0.6ex,scale=0.6]
			\draw[
			]
			(1.5,0) circle (1);
			\draw[
			thick, black,dashed
			]
			(1.5,0) circle (0.9);
			\begin{feynman}
				\vertex (a) at (-1,0)  ;
				\vertex (c) at (0.5,0) ;
				\vertex (d) at (2.5,0) ;
				\vertex (d1) at (1,1) ;
				\vertex (A) at (1.5,1);
				\vertex (B) at (1.5,-1);
				\diagram*{
					(a) -- [fermion](c),
					(d) -- [fermion] (b),
					(A) -- [plain] (B),
					(A) --[photon] (B)
				};
			\end{feynman}
		\end{tikzpicture}
	}&=\dfrac{2\cK_{4}^\prime}{N C_F} \Big(\widetilde G(d)- \widetilde G(1,0,1,1,1)
	+ 2 \widetilde G(0,1,1,0,1) \ \Big) \ , \\
	\label{eq:fermionic loop with gauge emission}
	\mathord{\begin{tikzpicture}[baseline=-0.65ex,scale=0.6]  
			\draw[
			]
			(1.5,0) circle (1);
			\draw[
			thick, black,dashed
			]
			(1.5,0) circle (0.9);
			\begin{feynman}
				\vertex (a) at (-1,0)  ;
				\vertex (b) at (4,0) ;
				\vertex (c) at (0.5,0) ;
				\vertex (c1) at (1.5,-0.9);
				\vertex (c3) at (3.25,0);
				\vertex (c2) at (0.4, 0);
				\vertex (d1) at (1.5,1);
				\vertex (d4) at (1.5,-0.9);
				\vertex (d) at (2.5,0) ;
				\vertex (d2) at (2.6,0);
				\vertex (e) at (1.5, 1.5) ;
				\vertex (e) at (1.5, -1.5) ;
				\diagram*{
					(a) -- [fermion] (c),
					(d) -- [fermion] (c3),
					(d1)--[photon, half left] (c3),
					(c3) --[fermion] (b),
				};
			\end{feynman}
	\end{tikzpicture}	} 
	&= - 4 N (i_\cR - N) \left(\widetilde G(0,1,1,1,1)- \widetilde G(1,1)^2\right) \ , \\[0.4em]
	\label{eq:fermionic loop with scalar exchange}
	\mathord{
		\begin{tikzpicture}[baseline=-0.6ex,scale=0.6]
			\draw[
			]
			(1.5,0) circle (1);
			\draw[
			black, thick, dashed
			]
			(1.5,-0.9) arc[start angle=-90, end angle=90, radius=0.9cm];		
			\draw[
			black, thick
			]
			(1.5,0.9) arc[start angle=90, end angle=270, radius=0.9cm];
			\begin{feynman}
				\vertex (a) at (-1,0)  ;
				\vertex (c) at (0.5,0) ;
				\vertex (d) at (2.5,0) ;
				\vertex (d1) at (1,1) ;
				\vertex (A) at (1.475,1);
				\vertex (B) at (1.475,-1);
				\vertex (D) at (1.525,1);
				\vertex (C) at (1.525,-1);
				\diagram*{
					(a) -- [fermion](c),
					(d) -- [fermion] (b),
					(C) -- [plain,dashed] (D),
					(A) --[fermion] (B)
				};
			\end{feynman}
		\end{tikzpicture}
	}
	&= 4 N (i_\cR - N)\left(4 \widetilde G(0,1,1,0,1)-2 \widetilde G(1,1)^2 \right) \ ,  \\
	\mathord{\begin{tikzpicture}[baseline=-0.65ex,scale=0.6]
			\begin{feynman}
				\newcommand\tmpda{0.9cm}
				\newcommand\tmpdb{-0.9cm}
				\draw[ with arrow=\tmpdb] (1.5,0) circle (1);
				\draw[ with arrow=\tmpda]
				(1.5,0) circle (1);
				\draw[
				black,dashed
				]
				(1.5,0) circle (0.9);
				\vertex (a) at (-1,0); 
				\vertex (b) at (4,0) ; 
				\vertex (c) at (0.5,0) ;
				\vertex (c1) at (0.55,0);
				\vertex (d) at (2.5,0) ;
				\vertex (d1) at (2.55,0);
				\vertex (e) at (1.5, 1.5);
				\vertex (e) at (1.5, -1.5);
				\diagram*{
					(a) -- [fermion] (c),
					(c) -- [fermion] (d),
					(d) -- [fermion] (b),
				};
			\end{feynman}
		\end{tikzpicture}
	} &=
	\left(8(C_\mathcal{R}i_\mathcal{R}-N^2) - 2 N (i_\cR - N)) \widetilde G(0,1,1,0,1\right)\ .
\end{align} 
The colour factor $\cK_{4}^\prime$ in eq. (\ref{glasswinediagrams}) was defined in eq. (\ref{eq:def K}), while
the double wiggly/continuous line denotes the propagation of the gauge field $A_\mu$ and of the $4-d$ real scalars resulting from dimensional reduction inside the fermion loop. Similarly, the diagrams in eq.  (\ref{eq:fermionic loop with scalar exchange}) arise from the Yukawa-like vertices in which the adjoint scalar $\phi$ interacts with the matter fermions, with the two gauginos and with the matter scalars. Note that the external continuos line with which we depicted the internal bubble  in eq. (\ref{eq:fermionic loop with scalar exchange}) has the meaning as in eq.s (\ref{eq:two loop corrections adjoint momentum}) and (\ref{eq:two loop corrections gluon momentum}).

Combining together the results we derived in this subsection, we can express the dimensionless function $f^{(2)}(d)$, that determines the scalar polarization $\Pi^{(2)}$ through eq. (\ref{Pi2form}), as the sum of four different terms
\begin{equation}
	\label{fis}
	f^{(2)}(d) = f^{(2)}_1(d) + f^{(2)}_2(d) + f^{(2)}_3(d) + f^{(2)}_4(d)~.
\end{equation}		
Recalling the explicit definition of the coefficient $\beta_0$, given by  eq.  (\ref{eq:beta0}), we finally obtain
\begin{align}
	\label{fiis}
	f^{(2)}_1 (d)& = 32\pi^2\, \beta_0\, i_\cR \, \widetilde G(1,1)^2~,\notag\\
	f^{(2)}_2(d) & = 32\pi^2\, \beta_0\, N\, \widetilde G(0,1,1,1,1)~,\notag\\
	f^{(2)}_3(d) & = 16\pi^2\, \beta_0\, N\, \widetilde G(0,1,1,2,1)~,\notag\\
	f^{(2)}_4(d) & = \frac{2 \cK^\prime_4}{N C_F}\, \widetilde{G}(d)~.
\end{align}

Finally, it is straightforward to obtain the expressions of these propagators in configuration space. By employing eq. (\ref{eq:Fourier transform for massless propagators}),  we find 
\begin{align}
	\label{eq:two loop corrections adjoint conf}
	\mathord{
		\begin{tikzpicture}[scale=0.6, baseline=-0.65ex]
			\filldraw[color=gray!80, fill=gray!15](0,0) circle (1);	
			\draw [black,thick] (0,0) circle [radius=1cm];
			\draw [black, thick, dashed] (0,0) circle [radius=0.9cm];
			\begin{feynman}
				\vertex (A) at (-2,0);
				\vertex (A1) at (-2,-0.5) {$x_1$};
				\vertex (C) at (-1,0);
				\vertex (C1) at (0.1,0) {\text{\footnotesize 2-loop }\normalsize} ;
				\vertex (B) at (1, 0);
				\vertex (D) at (2, 0);
				\vertex (D1) at (2,-0.5) {$x_2$};
				\diagram*{
					(A) -- [fermion] (C),
					(B) --[fermion] (D),
				};
			\end{feynman}
		\end{tikzpicture} 
	}
	= g_B^4 f^{(2)}(d)\, \cD(x_{12},d-5) \equiv \Delta^{(2)}(x_{12}) ~, 
\end{align}
for the adjoint scalar field. Conversely, for the gluon propagator, we get two terms:
\begin{align}
	\label{eq:two loop corrections gluon conf}
	\mathord{
		\begin{tikzpicture}[scale=0.6, baseline=-0.65ex]
			\filldraw[color=gray!80, fill=gray!15](0,0) circle (1);	
			\draw [black,thick] (0,0) circle [radius=1cm];
			\draw [black, thick, dashed] (0,0) circle [radius=0.9cm];
			\begin{feynman}
				\vertex (A) at (-2,0);
				\vertex (A1) at (-2,-0.5) {$x_1$};
				\vertex (C) at (-1,0);
				\vertex (C1) at (0.1,0) {\text{\footnotesize 2-loop }\normalsize} ;
				\vertex (B) at (1, 0);
				\vertex (D) at (2, 0);
				\vertex (D1) at (2,-0.5) {$x_2$};
				\diagram*{
					(A) -- [photon] (C),
					(B) --[photon] (D),
				};
			\end{feynman}
		\end{tikzpicture} 
	}
	& = g_B^4 f^{(2)}(d)\, \left(\delta_{\mu\nu} \cD(x_{12},5-d) - \partial_{1,\mu}\partial_{2,\nu}  \cD(x_{12},6-d)  \right)
	\notag\\
	& \equiv \delta_{\mu\nu} \Delta^{(2)}(x_{12}) 
	- \partial_{1,\mu}\partial_{2,\nu} \Delta^{(2),\mathrm{g}}(x_{12})  
	~. 
\end{align}
Note that the gauge-like term $ \partial_{1,\mu}\partial_{2,\nu}\Delta^{(2),\mathrm{g}}(x_{12})$ is completely irrelevant when inserted in the single-exchange correction (\ref{eq:bubble-exchange}) since, when contracted with the tangent vectors $\dot{x}^\mu_1\dot{x}^\nu_2$, it provides total derivates integrated over a closed path.

\section{Mercedes-like diagrams}
\label{sec:Mercedes diagram}
In this section, we provide the technical details regarding the calculation of the  \textit{Mercedes-like} correction
\begin{equation}
	\label{eq:mercedes-like diagrams}
	\begin{split}
		\mathsf{M} \equiv \mathord{
			\begin{tikzpicture}[radius=2.cm, scale=0.6, baseline=-0.65ex]
				\begin{feynman}
					\vertex (A) at (0,2.);
					\vertex (C) at (0,0);
					\vertex (D) at (-1.5, -1.3);
					\vertex (B) at (1.5, -1.3);
					\vertex (B1) at (1.,-0.8);
					\vertex (B2) at (0.5,-0.5);
					\diagram*{
						(A) -- [photon] (C),
						(A) --[ fermion] (C),
						(C) -- [photon] (B2),
						(C) -- [plain] (B2),
						(B) --[photon] (B1),
						(B) --[plain] (B1),
						(C) --[ fermion] (D),
						(C) --[ photon] (D)
					};
				\end{feynman}
				\draw [black] (0,0) circle [];
				\filldraw[color=white, fill=white!15](0.75,-0.65) circle (0.7);	
				\draw [black, thick, dashed] (0.75,-0.65) circle [radius=0.63cm];
				\draw [black] (0.75,-0.65) circle [radius=0.7cm];
			\end{tikzpicture} 
		}  \ ,
	\end{split}
\end{equation}
where we used the notation introduced in 
eq.s (\ref{eq:one-loop correction adjoint scalar main text}) and (\ref{eq:one-loop correction gauge-field position main text}). 

Expanding the Wilson loop operator (\ref{eq:1/2 BPS supersymmetric Wilson loop}) at order $g_B^3$, we obtain the following representation for the Mercedes-like corrections \begin{equation}
	\label{eq:spider 1v}
	\mathsf{M}	= \oint \dd^3\tau  \  \Bigg(  \dfrac{(\mathrm{i}g_B)^3}{3!N}  \big\langle{\tr\mathcal{P} \mathcal{A}(\tau_1)\mathcal{A}(\tau_2)\mathcal{A}(\tau_2)\big\rangle }_\mathsf{M} + \dfrac{\mathrm{i}g_B^3R^2}{2N}   \big\langle{\tr\mathcal{P}  \mathcal{A}(\tau_1)\Phi(\tau_2)\Bar{\Phi}(\tau_3)  } \big\rangle_\mathsf{M} \Bigg) \ ,
\end{equation}
where we recall that $\cP$ denotes the path-ordering operator, we  introduced the notation $\mathcal{A}_i\equiv \dot{x}^{\mu}(\tau_i)A_{\mu}^a(x(\tau_i))T^a$ and ${\Phi}_i\equiv   \phi^a(x(\tau_i)) T^a $ and we used the subscript $\mathsf{M}$ to restrict the Wick contractions inside the correlators to the  internal diagrams depicted in eq. (\ref{eq:mercedes-like diagrams}).

Before entering the calculation of eq. (\ref{eq:spider 1v}), it is convenient to recall that the one-loop correction to the gauge-field propagator, defined in eq. (\ref{eq:one-loop correction gauge-field position main text}), involves the gauge-like term $ \partial_{1,\mu}\partial_{2,\nu}\Delta^{(2),\mathrm{g}}(x_{12})$. By gauge invariance, we expect that the sum of all three-loop  corrections to the expectation value of the Wilson loop involving this gauge-like term vanishes.
To check this fact explicitly, it is convenient to introduce the following diagrammatic notation for eq. (\ref{eq:one-loop correction gauge-field position main text}): 
\begin{equation}
	\begin{split}
		\label{eq:diagrammatic split gauge field}
		\mathord{
			\begin{tikzpicture}[scale=0.5, baseline=-0.65ex]
				\draw [black] (0,0) circle [radius=1cm];
				\draw [black, thick, dashed] (0,0) circle [radius=0.9cm];
				\begin{feynman}
					\vertex (A) at (-2,0);
					\vertex at (-2,-0.5) {$x_1$};
					\vertex (C) at (-1,0);
					\vertex (B) at (1, 0);
					\vertex (D) at (2, 0);
					\vertex at (2, -0.5) {$x_2$};
					\diagram*{
						(A) -- [photon] (C),
						(B) --[photon] (D),
					};
				\end{feynman}
			\end{tikzpicture} 
		} &= \delta_{ab}\delta_{\mu\nu} \Delta^{(1)}(x_{12}) + \delta_{ab}\partial_{1,\mu}\partial_{1,\nu} 
		\Delta^{(1), \mathrm{g}}_{\mu\nu}(x_{12})\\
		&\equiv 
		\mathord{
			\begin{tikzpicture}[scale=0.5, baseline=-0.65ex]					
				\draw [black] (0,0) circle [radius=1cm];
				\draw [black, thick, dashed] (0,0) circle [radius=0.9cm];
				\begin{feynman}
					\vertex (A) at (-2,0);
					\vertex at (-2,-0.5) {$x_1$};
					\vertex (C) at (-1,0);
					\vertex (C1) at (0,0) {$\delta$};
					\vertex (B) at (1, 0);
					\vertex (D) at (2, 0);
					\vertex at (2, -0.5) {$x_2$};
					\diagram*{
						(A) -- [photon] (C),
						(B) --[photon] (D),
					};
				\end{feynman}
			\end{tikzpicture} 
		} + \mathord{
			\begin{tikzpicture}[scale=0.5, baseline=-0.65ex]
				\draw [black] (0,0) circle [radius=1cm];
				\draw [black, thick, dashed] (0,0) circle [radius=0.9cm];
				\begin{feynman}
					\vertex (A) at (-2,0);
					\vertex at (-2,-0.5) {$x_1$};
					\vertex (C) at (-1,0);
					\vertex (C1) at (0,0) {$\partial$};
					\vertex (B) at (1, 0);
					\vertex (D) at (2, 0);
					\vertex at (2, -0.5) {$x_2$};
					\diagram*{
						(A) -- [photon] (C),
						(B) --[photon] (D),
					};
				\end{feynman}
			\end{tikzpicture} 
		} \  .
	\end{split}
\end{equation} 
We use the symbols $\delta$ and $\partial$ inside the dashed/continuos bubbles, to distinguish the tensor structures of the two terms and we recall that $\Delta^{(1)}(x)$   and $\Delta^{(1),\rm g}(x)$ are defined in eq.s (\ref{eq:one-loop correction adjoint scalar main text}) and (\ref{eq:one-loop correction gauge-field position main text}) respectively, from which it follows that \begin{align}
	\label{eq:Delta(1)}
	\Delta^{(1)}(x_{12}) &= \dfrac{  g_B^2\beta_0\Gamma^2\left(d/2-1\right)}{4\pi^{d-2}(d/2-2)(d-3)(x_{12}^2)^{d-3}} \ , \\
	\label{eq:Delta(1),g}
	\Delta^{(1),g}(x_{12}) &= \dfrac{g_B^2\beta_0\Gamma^2(d/2-1)}{2^5\pi^{d-2}(3-d/2)(d-3)(2-d/2)^2(x_{12}^2)^{d-4}}  \ .
\end{align} 
Using the notation we introduced in eq. (\ref{eq:diagrammatic split gauge field}), we can organize the correction (\ref{eq:mercedes-like diagrams}) in terms of two distinct classes of diagrams, i.e.     \begin{align}
	\label{eq:sigmaprime}
	\mathsf{M}_1 &\equiv 	\mathord{
		\begin{tikzpicture}[radius=2.cm, scale=0.6, baseline=-0.65ex]
			\begin{feynman}
				\vertex (A) at (0,2.);
				\vertex (C) at (0,0);
				\vertex (D) at (-1.5, -1.3);
				\vertex (B) at (1.5, -1.3);
				\vertex (B1) at (1.,-0.8);
				\vertex (B2) at (0.5,-0.5);
				\diagram*{
					(A) -- [photon] (C),
					(C) -- [photon] (B2),
					(B) --[photon] (B1),
					(C) --[ photon] (D)
				};
			\end{feynman}
			\draw [black] (0,0) circle [];
			\filldraw[color=white, fill=white!15](0.75,-0.65) circle (0.7);	
			\draw [black, thick, dashed] (0.75,-0.65) circle [radius=0.63cm];
			\draw [black] (0.75,-0.65) circle [radius=0.7cm];
			\draw (0.4,-0.65) node[anchor=west]{$\tiny \delta$};
		\end{tikzpicture} 
	} \ + \ 
	\mathord{
		\begin{tikzpicture}[radius=2.cm, scale=0.6, baseline=-0.65ex]
			\begin{feynman}
				\vertex (A) at (0,2.);
				\vertex (C) at (0,0);
				\vertex (D) at (-1.5, -1.3);
				\vertex (B) at (1.5, -1.3);
				\vertex (B1) at (1.,-0.8);
				\vertex (B2) at (0.5,-0.5);
				\diagram*{
					(A) --[ fermion] (C),
					(C) -- [photon] (B2),
					(B) --[photon] (B1),
					(C) --[ fermion] (D),
				};
			\end{feynman}
			\draw [black] (0,0) circle [];
			\filldraw[color=white, fill=white](0.75,-0.65) circle (0.7);	
			\draw [black, thick, dashed] (0.75,-0.65) circle [radius=0.63cm];
			\draw [black] (0.75,-0.65) circle [radius=0.7cm];
			\draw (0.4,-0.65) node[anchor=west]{$\tiny \delta$};
		\end{tikzpicture} 
	}\  + \  \mathord{
		\begin{tikzpicture}[radius=2.cm, scale=0.6, baseline=-0.65ex]
			\begin{feynman}
				\vertex (A) at (0,2.);
				\vertex (C) at (0,0);
				\vertex (D) at (-1.5, -1.3);
				\vertex (B) at (1.5, -1.3);
				\vertex (B1) at (1.,-0.8);
				\vertex (B2) at (0.5,-0.5);
				\diagram*{
					(A) --[ photon] (C),
					(C) -- [anti fermion] (B),
					(C) --[ fermion] (D),
				};
			\end{feynman}
			\draw [black] (0,0) circle [];
			\filldraw[color=white, fill=white](0.75,-0.65) circle (0.7);	
			\draw [black, thick, dashed] (0.75,-0.65) circle [radius=0.6cm];
			\draw [black] (0.75,-0.65) circle [radius=0.7cm];
		\end{tikzpicture} 
	} \ , \\[0.4em] 
	\label{eq:sigmaprimeprime}
	\mathsf{M}_2&\equiv \mathord{
		\begin{tikzpicture}[radius=2.cm, scale=0.6, baseline=-0.65ex]
			\begin{feynman}
				\vertex (A) at (0,2.);
				\vertex (C) at (0,0);
				\vertex (D) at (-1.5, -1.3);
				\vertex (B) at (1.5, -1.3);
				\vertex (B1) at (1.,-0.8);
				\vertex (B2) at (0.5,-0.5);
				\diagram*{
					(A) -- [photon] (C),
					(C) -- [photon] (B2),
					(B) --[photon] (B1),
					(C) --[ photon] (D)
				};
			\end{feynman}
			\draw [black] (0,0) circle [];
			\filldraw[color=white!80, fill=white!15](0.75,-0.65) circle (0.7);	
			\draw [black, thick, dashed] (0.75,-0.65) circle [radius=0.63cm];
			\draw [black] (0.75,-0.65) circle [radius=0.7cm];
			\draw (0.4,-0.65) node[anchor=west]{$\partial$};
		\end{tikzpicture} 
	} \ + \ 	\mathord{
		\begin{tikzpicture}[radius=2.cm, scale=0.6, baseline=-0.65ex]
			\begin{feynman}
				\vertex (A) at (0,2.);
				\vertex (C) at (0,0);
				\vertex (D) at (-1.5, -1.3);
				\vertex (B) at (1.5, -1.3);
				\vertex (B1) at (1.,-0.8);
				\vertex (B2) at (0.5,-0.5);
				\diagram*{
					(A) --[ fermion] (C),
					(C) -- [photon] (B2),
					(B) --[photon] (B1),
					(C) --[ fermion] (D),
				};
			\end{feynman}
			\draw [black] (0,0) circle [];
			\filldraw[color=white!80, fill=white!15](0.75,-0.65) circle (0.7);	
			\draw [black, thick, dashed] (0.75,-0.65) circle [radius=0.63cm];
			\draw [black] (0.75,-0.65) circle [radius=0.7cm];
			\draw (0.4,-0.65) node[anchor=west]{$\partial$};
		\end{tikzpicture} 
	} \ .
\end{align} In the following subsections, we will analyse these two classes of corrections in turn.
\subsection{Computing $\mathsf{M}_1$}
\label{sec:computing sigma+}
To deduce the expression of the different diagrams contributing to  $\mathsf{M}_1$, we begin with considering the interaction action associated with the internal (gauge-scalar/pure-gauge) triple vertices. Using the conventions of Appendix \ref{sec:actions in flat space}, we find that 
\begin{equation}
	\label{eq:vertices}
	\begin{split}
		S_{\rm gs}&= g_B\int\dd^d\omega f^{abc}\left(\partial_{\mu}\bar{\phi}_bA^\mu_a \phi_c -\bar{\phi}_b A^\mu_a\partial_\mu\phi_c\right)\left(\omega\right) \\ 
		S_{\rm gg}&= g_B\int \dd^d\omega  f^{acb}\left(\partial_{\mu}A_{\nu,c}A^\mu_aA^b_\nu\right)\left(\omega\right) \ ,
	\end{split}
\end{equation} where $f^{abc}$, defined by $[T^a,T^b]=\mathrm{i}f^{abc} T_c$, are the (antisymmetric) structure constants of $\mathfrak{su}(N)$. Inserting these actions in the correlation functions of eq. (\ref{eq:spider 1v}) and decorating the proper Wick contractions with the one-loop correction to the adjoint scalar  $\delta_{ab}\Delta^{(1)}(x)$, defined in eq. (\ref{eq:one-loop correction adjoint scalar main text}), and with the tensor $\delta_{\mu\nu}\Delta^{(1)}(x)$, we arrive at the  following expression
\begin{equation}
	\label{eq:first step}
	\mathsf{M}_1=-\dfrac{g_B^4C_FN}{2}\oint \dd^3{\tau} \varepsilon(\tau)  \ (R^2-\dot{x}_1\cdot \dot{x}_3) \left( \dot{x}_2\cdot \partial_{x_1}\right) \int \dd^d \omega \sum_{i=1}^{3}\prod_{j\ne i} \Delta^{(1)}(x_{i\omega})\Delta(x_{j\omega})\ ,
\end{equation} where  $x_{i\omega}\equiv x_i-\omega$, while  the function $\Delta(x)$ and the path-ordering symbol\footnote{$\varepsilon(\tau)$ arises since the internal diagrams are proportional to the antisymmetric structure constant $f^{abc}$.} $\varepsilon(\tau)$ are defined in eq.s  (\ref{eq:Deltad}) and  (\ref{eq:defepsilon}), respectively. By integrating over $\omega$ with usual Feynman parameters and expressing the coordinates $x_i$ via the parametrization  (\ref{eq:parametrization}), we find that  \begin{equation}
	\label{eq:intermezzo}
	\begin{split}
		\mathsf{M}_1 = A_d\int_0^1\dd{\mathcal{F}}  \oint \dd^3{\tau} \varepsilon(\tau)  \ \dfrac{(1 - \cos{(\tau_{13})})(\alpha(1-\alpha)\sin{(\tau_{12})}-\alpha\gamma\sin{(\tau_{32})})} {Q^{3d/2-4}} \ .
		%
	\end{split}
\end{equation} In the previous expression, the denominator $Q$ is defined in eq. (\ref{eq:def of Q}), while the integration measure over the Feynman parameters and the multiplicative prefactor are given by  \begin{align}
	\label{eq:integration Feynman parameter Sigmaprime}
	\dd\cF 
	&=    \dd F (\alpha\beta\gamma)^{d/2-2}\left(\alpha^{d/2-2}+\beta^{d/2-2}+\gamma^{d/2-2}\right) 	\ ,  \\[0.6em]
	A_d&= \dfrac{\hat{g}_B^6C_FN\beta_0 \ \Gamma(3d/2-4)\Gamma^2(d/2-1)}{(d/2-2)\Gamma(d-2) \ (\pi)^{3d/2-6} 2^{3d/2+2}\pi^4} \label{eq:factorized pole} \ .
\end{align} In the previous expression,   $\dd F$ is the standard measure over the unit cube (\ref{eq:integration over the unit cube}). To perform the contour integration in eq. (\ref{eq:intermezzo}), we employ the following identities\footnote{This procedure is analogous to that outlined in \cite{Erickson:2000af,Bassetto:2008yf} for the calculation of the two-loop \textit{Mercedes-like diagrams} $\cW_4^{\rm v.m.}$ we defined in eq. (\ref{eq:real calculation Sigma 3}), in the context of $\mathcal{N}=4$ theories. In our case, the model is not superconformal and consequently, the analysis is more involved.} \begin{align}
	\label{eq:initial identity}
	\int_0^1\dd\cF\oint \dd^3{\tau} \dfrac{\partial}{\partial \tau_1} \Bigg( \dfrac{\epsilon(\tau) (1 - \cos\tau_{13}) } {Q^{3d/2-5}} \Bigg)&=0\ , \\[0.4em] 
	\label{eq:derivative of epsilon}
	\partial_{\tau_1} \varepsilon(\tau) - 2 \Big(\delta(\tau_{12})-\delta(\tau_{13})\Big)  &=0  \   .   
\end{align} To proceed with the calculation, it is sufficient to insert eq. (\ref{eq:initial identity}) in (\ref{eq:intermezzo}) and observe that  the measure $\dd\cF$  is completely symmetric. This enables to relabel the variables $\tau_i$ and keep the denominator $Q$ unchanged. As a result, we find that 
\begin{align}
	\label{eq:integration over the contour}
	\mathsf{M}_1&=\dfrac{2A_d}{3d/2-5} \int_0^1\dd{\cF}  \oint \dd^2{\tau}\dfrac{(1 - \cos{\tau_{23}})^{6-3d/2}}{\left(\gamma(1-\gamma)\right)^{3d/2-5}}-
	A_d\left( \dfrac{3d-12}{3d-10} \right) \int_0^1\dd{\cF} \oint \dd^3\tau\dfrac{\varepsilon(\tau) \sin{\tau_{13}}}{Q^{3d/2-5}}\notag \\[0.3em]
	&=  -2F_2^{(2)}-\dfrac{N}{i_\cR}F_1^{(2)}+ \dfrac{\hat{g}_B^6\beta_0 C_FN 9\zeta(3)}{16\pi^2}   \ . 
\end{align}
In the previous expression, the quantities
$F_1^{(2)}$ and $F_2^{(2)}$ are the (UV divergent) bubble-like contributions we defined in eq. (\ref{Fiis}). They arise from the integration over the measure $\dd\cF$ (\ref{eq:integration Feynman parameter Sigmaprime}) in the first term of the previous expression, while the $\zeta(3)$-like contribution is obtained by applying the master integral  (\ref{eq:I1 main}) to the second term.

\subsection{Computing $\mathsf{M}_2$}
\label{sec:Sigma_- gauge part}
In this section,  we turn our attention to the corrections $\mathsf{M}_2$, depicted in  eq. (\ref{eq:sigmaprimeprime}). Let us begin with discussing  the diagrams involving three gauge fields. Inserting the pure-gauge vertex (\ref{eq:vertices}) in the first correlator of eq. (\ref{eq:spider 1v}) and decorating the Wick contractions with the tensor $ \partial_{1,\mu}\partial_{2,\nu}\Delta^{(2),\mathrm{g}}(x_{12})$, we arrive at the following representation 
\begin{equation}
	\label{eq:pure gauge spider with derivatives}
	\mathord{
		\begin{tikzpicture}[radius=2.cm, scale=0.55, baseline=-0.65ex]
			\begin{feynman}
				\vertex (A) at (0,2.);
				\vertex (C) at (0,0);
				\vertex (D) at (-1.5, -1.3);
				\vertex (B) at (1.5, -1.3);
				\vertex (B1) at (1.,-0.8);
				\vertex (B2) at (0.5,-0.5);
				\diagram*{
					(A) -- [photon] (C),
					(C) -- [photon] (B2),
					(B) --[photon] (B1),
					(C) --[ photon] (D)
				};
			\end{feynman}
			\draw [black] (0,0) circle [];
			\filldraw[color=white!80, fill=white!15](0.75,-0.65) circle (0.7);	
			\draw [black, thick, dashed] (0.75,-0.65) circle [radius=0.63cm];
			\draw [black] (0.75,-0.65) circle [radius=0.7cm];
			\draw (0.3,-0.65) node[anchor=west]{$\partial$};
		\end{tikzpicture} 
	} = \dfrac{g_B^4C_FN}{2}\oint\dd^3\tau\varepsilon(\tau)\ \int\dd^d{\omega} \dfrac{\dd}{\dd \tau_1}\left(\cO(x_j)\Delta(x_{3\omega})\Delta(x_{2\omega})\Delta^{(1),\rm g}(x_{1\omega})\right)  \ ,
\end{equation} where we recall that $\Delta(x)$ is the massless tree-level propagator defined in eq. (\ref{eq:Deltad}), the function $\Delta^{(1),g}(x)$ is given by eq. (\ref{eq:Delta(1),g}), while $\cO(x_j)$ denotes the following  operator
\begin{equation}
	\label{eq:I3}
	\begin{split}
		\cO(x_j) &= \Big[ \left(\dot{x}_3\cdot\partial_1\right) \left(\partial_1-\partial_3\right)\cdot\dot{x_2}+\left(\dot{x}_2\cdot\dot{x}_3\right) \left(\partial_3\cdot\partial_1\right)\Big] \ .
	\end{split}
\end{equation} 
Let us now consider the diagrams in eq. (\ref{eq:sigmaprimeprime}) involving the propagation of two scalars and one gauge field.  Inserting the gauge-scalar vertex (\ref{eq:vertices}) in the second correlator of eq. (\ref{eq:spider 1v}), and decorating the Wick contraction of  the gauge field with the tensor  $\partial_{1,\mu}\partial_{1,\nu}\Delta^{(1),\rm g}(x_{1\omega})$, we find 
\begin{equation}
	\label{eq:gauge scalar with derivatives}
	\mathord{
		\begin{tikzpicture}[radius=2.cm, scale=0.55, baseline=-0.65ex]
			\begin{feynman}
				\vertex (A) at (0,2.);
				\vertex (C) at (0,0);
				\vertex (D) at (-1.5, -1.3);
				\vertex (B) at (1.5, -1.3);
				\vertex (B1) at (1.,-0.8);
				\vertex (B2) at (0.5,-0.5);
				\diagram*{
					(A) --[ fermion] (C),
					(C) -- [photon] (B2),
					(B) --[photon] (B1),
					(C) --[ fermion] (D),
				};
			\end{feynman}
			\draw [black] (0,0) circle [];
			\filldraw[color=white!80, fill=white!15](0.75,-0.65) circle (0.7);	
			\draw [black, thick, dashed] (0.75,-0.65) circle [radius=0.63cm];
			\draw [black] (0.75,-0.65) circle [radius=0.7cm];
			\draw (0.3,-0.65) node[anchor=west]{$\partial$};
		\end{tikzpicture} 
	} =  -\dfrac{g_B^4C_FR^2N}{2}\oint\dd^3\tau\varepsilon(\tau)\ \dfrac{\dd}{\dd \tau_1}\partial_3\cdot\partial_1 \left(\Delta(x_{3\omega})\Delta(x_{2\omega})\Delta^{(1),\rm g}(x_{1\omega})\right)  \\.
\end{equation}
Finally, by combining together eq.s (\ref{eq:pure gauge spider with derivatives}) and (\ref{eq:gauge scalar with derivatives}) and neglecting terms which provide total derivatives, we obtain the following result  \begin{equation}
	\begin{split}
		\label{eq:sigmaprimeprime def}
		\mathsf{M}_2 &= \dfrac{g_B^4C_FN}{2} \oint\dd^3\tau\varepsilon(\tau)\left(\dot{x}_2\cdot\dot{x}_3-R^2\right) 		\dfrac{\dd}{\dd\tau_1}\partial_3\cdot\partial_1\int \dd^d \omega \Delta(x_{2\omega})\Delta(x_{3\omega})\Delta^{(1),\rm g}(x_{1\omega}) \\[0.4em]
		&=g_B^4C_FN\oint\dd^3\tau (\delta(\tau_{ 12})-\delta(\tau_{13}))\left(R^2-\dot{x}_2\cdot\dot{x}_3\right) 	\partial_3\cdot\partial_1\int \dd^d \omega \Delta(x_{2\omega})\Delta(x_{3\omega})\Delta^{(1),\rm g}(x_{1\omega}) \\[0.4em]
		&=-2F_3^{(2)} \ ,
	\end{split}
\end{equation} 
where  $F_3^{(2)}$ denotes the  bubble-like contribution  defined in eq. (\ref{Fiis}) and we obtained the second line via integration by parts and using  eq. (\ref{eq:derivative of epsilon}). Combining this result with eq.  (\ref{eq:integration over the contour}), we find that  \begin{equation}
	\label{eq:Sigma_+ final}
	\begin{split}
		\mathsf{M}=	\mathsf{M}_1+\mathsf{M}_2 =-2F_2^{(2)}-2F_3^{(2)}-\dfrac{N}{i_\cR}F_1^{(2)} + 9\dfrac{\beta_0^\mathcal{R}\hat{g}_B^6C_FN}{16\pi^2} \zeta(3)  \ .
	\end{split}
\end{equation}

\section{Lifesaver diagrams}
\label{sec:lifesaver diagrams appendix}
In this section, we  examine in detail the calculation of the \textit{lifesaver-like} diagrams\footnote{We provide an extended analysis since we did not find evidence of analogous calculations in the existing literature.}
\begin{equation}
	\label{eq:lifesaver diagrams appendix}
	\mathsf{L}=	\mathord{
		\begin{tikzpicture}[radius=2.cm, baseline=-0.65ex,scale=0.55]
			\draw [black] (0,0) circle [];
			\draw [black] (0,0) circle [radius=0.8cm];
			\draw [black, dashed,thick] (0,0) circle [radius=0.7cm];
			\begin{feynman}
				\vertex (A) at (0,2);
				\vertex (C) at (0,0.8);
				\vertex (D) at (-1.5, -1.3);
				\vertex (B) at (-0.7, -0.4);
				\vertex (B1) at (0.7,-0.4);
				\vertex (B2) at (1.5,-1.3);
				\diagram*{
					(A) -- [photon] (C),
					(B) --[ anti fermion] (D),
					(B) --[photon] (D),
					(B1) --[fermion] (B2),
					(B1) --[photon] (B2),
				};
			\end{feynman}
		\end{tikzpicture} 
	} \ ,
\end{equation} where we used again the difference theory notation. In particular, the internal bubble encodes the one-loop irreducible corrections to the (gauge-scalar/pure-gauge) triple vertex in the difference theory approach. 
\subsection{Construction of the building blocks}

Expanding the Wilson loop operator (\ref{eq:1/2 BPS supersymmetric Wilson loop}) at order $g_B^3,$ we obtain the following representation for the diagrams depicted in eq. (\ref{eq:lifesaver diagrams appendix})
\begin{equation}
	\label{eq:Sigma- appendix starting}
	\mathsf{L}=\mathsf{L}_g+\mathsf{L}_{gs} \  ,
\end{equation} 
where the quantities $\mathsf{L}_{gs}$ and $\mathsf{L}_g$ encode two correlators in the difference theory approach
\begin{align}
	\label{eq:diagrammatic representation correlators Sigma_G}
	\mathsf{L}_g &=\left(\dfrac{(\mathrm{i}g_B)^3}{3!N}  \right) \left(\oint \dd^3\tau   \big\langle{ \tr \mathcal{P} \mathcal{A}(\tau_1)\cA(\tau_2)\cA(\tau_3)  } \big\rangle_{\mathsf{L}}\right) \  ,\\[0.4em]
	\label{eq:diagrammatic representation correlators Sigma_GS}
	\mathsf{L}_{gs} &=\left(\dfrac{\mathrm{i}g_B^3R^2}{2N}  \right) \left(\oint \dd^3\tau   \big\langle{ \tr \mathcal{P} \mathcal{A}(\tau_1)\Phi(\tau_2)\Bar{\Phi}(\tau_3)  } \big\rangle_{\mathsf{L}}\right) \  .
\end{align}

We begin with discussing eq. (\ref{eq:diagrammatic representation correlators Sigma_GS}), which involves the irreducible correction to the gauge-scalar vertex in the difference method.  To construct these corrections, it is sufficient to  determine the relevant diagrams characterized by internal matter line in the representation $\cR$ and subsequently, to subtract an identical contribution in which $\cR=\rm Adj$. Introducing    $\cS_{abc}^\mu(x_i)=\big<A^\mu_a(x_1)\phi_b(x_2)\bar{\phi}_c(x_3)\big>_{\mathsf{L}}$, we have
\begin{equation}
	\label{eq:F+G}
	\begin{split}
		\cS^\mu_{abc}(x_i)	&=	\mathord{
			\begin{tikzpicture}[radius=2.cm, baseline=0.65, scale=0.55]
				\draw[arrows = {-Latex[width=5pt, length=7pt]}] (0.05,-0.3)--(-0.15,-0.3) ;
				\begin{feynman}
					\vertex (A) at (0,2) ;
					\vertex at (0,2.4) {$A_a^\mu(x_1)$} ;
					\vertex (c) at (0,0.6);
					\vertex (C) at (0,0.7);
					\vertex (D) at (-2, -1.3);
					\vertex (D1) at (-2, -1.8)
					{$\bar{\phi}^{c}(x_3)$} ;
					\vertex (S) at (2, -1.3);
					\vertex (T) at (1, -0.3);
					\vertex (B) at (-1, -0.3);
					\vertex (b) at (-0.9,-0.3);
					\vertex at (2,-1.8) {$\phi^b(x_2)$} ;
					\diagram*{
						(A) --[ photon] (C),
						(B) --[ anti fermion] (D),
						(S) --[ anti fermion] (T),
						(C) -- [charged scalar, thick ] (B),
						(T) -- [ghost, thick] (B),
						(T) --[ charged scalar, thick] (C),	
					};
				\end{feynman}
			\end{tikzpicture} 
		} + 	\mathord{
			\begin{tikzpicture}[radius=2.cm, baseline=0.65, scale=0.6]
				\draw[arrows = {-Latex[width=7pt, length=5pt]}]  (-0.3,0.4) -- (-.5,0.2) ;
				\draw[arrows = {-Latex[width=7pt, length=5pt]}] (0.5,0.2)--  (.3,0.4)  ;
				\begin{feynman}
					\vertex (A) at (0,2) ;
					\vertex (a) at (0,0.6);
					\vertex     at (0,2.4) {$A_a^\mu(x_1)$} ;
					\vertex (C) at (0,0.7);
					\vertex (D) at (-2, -1.3);
					\vertex (D1) at (-2, -1.8)
					{$\bar{\phi}^{c}(x_3)$} ;
					\vertex (S) at (2, -1.3);
					\vertex (T) at (1, -0.3);
					\vertex (B) at (-1, -0.3);
					\vertex     at (2,-1.8) {$\phi^b(x_2)$} ;
					\diagram*{
						(A) --[ photon] (C),
						(B) --[ anti fermion] (D),
						(S) --[ anti fermion] (T),
						(C) -- [ghost, thick ] (B),
						(T) -- [charged scalar, thick] (B),
						(T) --[ ghost, thick] (C),	
					};
				\end{feynman}
			\end{tikzpicture} 
		}-  \left(\cR = \mathrm{Adj}\right) \\[0.4em]
		& = f_{abc} \left(\cS^\mu_1(x_i)+\cS^\mu_2(x_i)\right)\ .
	\end{split}
\end{equation} 
where we defined the functions
\begin{align}
	\cS^{\mu}_1(x_i) &= (2\mathrm{i}g_B^3)(i_\cR-N)\int\dd P\int\dfrac{\dd^dk}{(2\pi)^d} \dfrac{p^\mu_{2}k^2-p^\mu_{3}(k-p_1)^2 } {k^2(k-p_1)^2(k+p_3)^2}\label{eq:S1 difference gauge-scalar} \ , \\[0.4em]
	\cS^{\mu}_2(x_i) &= (2\mathrm{i}g_B^3)(i_\cR-N)\int\dd P\int\dfrac{\dd^dk}{(2\pi)^d} \dfrac{ p_1^2(k+p_3)^\mu -p_3^2(k-p_1)^\mu-p_2^2 k^\mu }{k^2(k-p_1)^2(k+p_3)^2} \label{eq:S2 difference gauge-scalar} \ .
\end{align}  
In the previous expression, $\dd P$ denotes the usual integration measure over the external momenta $p_i$, i.e. 
\begin{equation}
	\label{eq:integration over momenta}
	\dd{P}=     \prod_{i=1}^{3}\dfrac{\dd^dp_i}{(2\pi)^d} \dfrac{e^{-\mathrm{i}p_i\cdot x_i}}{p_i^2}(2\pi)^d\delta^d\left(\Sigma_{j} \  p_j\right) \ .
\end{equation}Let us observe that the functions in eq.s (\ref{eq:S1 difference gauge-scalar}) and (\ref{eq:S2 difference gauge-scalar}) have a different behaviour in the limit $d\to 4$. Indeed, integrating over the large loop momentum yields a pole $1/(d-4)$ in eq. (\ref{eq:S1 difference gauge-scalar}),  while the function $\cS^\mu_2(x_i)$ is regular in four dimensions. Substituting eq. (\ref{eq:F+G}) into eq. (\ref{eq:diagrammatic representation correlators Sigma_GS}), we can naturally arrange the results in terms of two distinct contributions:  \begin{align}
	\label{eq:SIgmags 1 and 2 appendix}
	\mathsf{L}_{gs}=\mathsf{L}_{gs,1}+\mathsf{L}_{gs,2} \ .
\end{align}
Specifically, the quantity  $\mathsf{L}_{gs,1}$ is given by  
\begin{equation}
	\begin{split}
		\label{eq:SigmaGs1 appendix}
		\mathsf{L}_{gs,1}&=-\dfrac{g_B^3 R^2}{4}NC_F \oint\dd^3\tau \varepsilon(\tau)\left(\dot{x}_1\cdot\cS_1\right)\\[0.4em]
		& = \cA_1\, R^2 \oint\dd^3\tau\varepsilon(\tau)\left(\dot{x}_2\cdot\partial_1\right)\int\dd^d\omega \Delta^{(1)}(x_{1\omega})\Delta(x_{2\omega})\Delta(x_{3\omega}) \ ,
	\end{split}
\end{equation}  with  $\Delta(x)$ being the tree-level propagator (\ref{eq:Deltad}) and  $\Delta^{(1)}(x)$ the (UV) divergent one-loop correction to the adjoint scalar propagator (\ref{eq:one-loop correction adjoint scalar main text}), while   
\begin{align}
	\label{eq:SigmaGs2 appendix}
	\mathsf{L}_{gs,2}&=-\dfrac{g_B^3 R^2}{4}NC_F \oint\dd^3\tau \varepsilon(\tau)\left(\dot{x}_1\cdot\cS_2\right)\notag \\[0.4em]
	&=\mathcal{A}_2 \oint\dd^3\tau\varepsilon(\tau)\int \dd P\left(-\mathrm{i}p_2^2\right) \int\dfrac{\dd^dk}{\left(2\pi\right)^2}\dfrac{R^2\left(2k\cdot\dot{x}_1-k\cdot\dot{x}_2\right)}{k^2(k+p_1)^2(k-p_3)^2} \ .
\end{align}In the previous expressions, we introduced, for the sake of conciseness, the quantities
\begin{align}
	\label{A1def}
	\cA_1 & =  \dfrac{C_FNg_B^4}{2}~,\\
	\label{eq:C2 appendix}
	\mathcal{A}_2 & = 4\pi^2C_FN\beta_0 g_B^6 \ .
\end{align}

The analysis of the internal diagrams which enter eq. (\ref{eq:diagrammatic representation correlators Sigma_G}) goes along the same lines. In particular, the irreducible one-loop correction to the pure-gauge vertex in the difference method receives corrections from both scalar and fermionic loops and for convenience, we will consider them in turn.  Matter scalars contribute via the following diagrams  
\begin{equation}
	\begin{split}
		\label{eq:loop scalars pure gauge vertex difference theory}
		\cS^{abc}_{\mu\nu\rho}(x_i) &=		\mathord{
			\begin{tikzpicture}[radius=2.cm, baseline=-0.65ex,scale=0.55]
				\begin{feynman}
					\vertex (A) at (0,2) ;
					\vertex at (0,2.4) {$A^a_\mu(x_1)$} ;
					\vertex (C) at (0,0.7);
					\vertex (D) at (-2, -1.3);
					\vertex (D1) at (-2, -1.8)
					{$A^b_\nu(x_2)$} ;
					\vertex (S) at (2, -1.3);
					\vertex (T) at (1, -0.3);
					\vertex (B) at (-1, -0.3);
					\vertex at (2,-1.8) {$A^c_\rho(x_3)$} ;
					\vertex (c) at (0,0.6);
					\vertex (b) at (-0.85,-0.25);
					\vertex (t) at (0.85,-0.25);
					\diagram*{
						(A) --[ photon] (C),
						(B) --[ photon] (D),
						(S) --[ photon] (T),
						(C) -- [ charged scalar ] (B),
						(T) -- [ anti charged scalar] (B),
						(T) --[ charged scalar] (C),	
						(c)-- [plain,dotted,thick] (b),
						(t) -- [plain,dotted,thick] (b) ,
						(c) --[plain,dotted,thick](t),
					};
				\end{feynman}
			\end{tikzpicture} 
		} \ + 	\mathord{
			\begin{tikzpicture}[radius=2.cm, baseline=-0.65ex,scale=0.55]		
				\begin{feynman}
					\vertex (A) at (0,2) ;
					\vertex at (0,2.4) {$A^a_\mu(x_1)$} ;
					\vertex (C) at (0,0.7);
					\vertex (D) at (-2, -1.3);
					\vertex (D1) at (-2, -1.8)
					{$A^c_\rho(x_3)$}	 ;
					\vertex (S) at (2, -1.3);
					\vertex (T) at (1, -0.3);
					\vertex (B) at (-1, -0.3);
					\vertex at (2,-1.8) {$A^b_\nu(x_2)$} ;
					\vertex (c) at (0,0.6);
					\vertex (b) at (-0.85,-0.25);
					\vertex (t) at (0.85,-0.25);
					\diagram*{
						(A) --[ photon] (C),
						(B) --[ photon] (D),
						(S) --[ photon] (T),
						(C) -- [charged scalar ] (B),
						(T) -- [anti charged scalar] (B),
						(T) --[ charged scalar] (C),
						(c)-- [plain,dotted, thick] (b),
						(t) -- [plain,dotted, thick] (b) ,
						(c) --[plain,dotted, thick](t),				
					};
				\end{feynman}
			\end{tikzpicture} 
		} - \  \left(\cR = \mathrm{Adj}\right)	\\[0.8em]
		&=2\ii g_B^3 f^{abc}\left(N-i_\cR\right) \int\dd P \int\dfrac{\dd^dk}{\left(2\pi\right)^d}\dfrac{\left(2k+p_1\right)_\mu(2k+p_1-p_3)_\mu(2k-p_3)_\rho}{k^2(k+p_1)^2(k-p_3)^2} \ .
	\end{split}
\end{equation}
In the previous expression, we used a double dashed/dotted line to denote, respectively,  the contributions associated with the scalars $q$ and $\tilde{q}$, which transform in the representation $\cR$ and $\cR^*$, and we recall that
$\dd P$ is defined in eq. (\ref{eq:integration over momenta}).

The  matter fermions contribute with the following diagrams: 
\begin{equation}
	\begin{split}
		\label{eq:loops of fermion eta }
		\cF_{\mu\nu\rho}^{abc}(x_i) &= 		\mathord{
			\begin{tikzpicture}[radius=2.cm, baseline=-0.65ex,scale=0.55]
				\begin{feynman}
					\vertex (A) at (0,2) ;
					\vertex at (0,2.4) {$A^a_\mu(x_1)$} ;
					\vertex (C) at (0,0.7);
					\vertex (D) at (-2, -1.3);
					\vertex (D1) at (-2, -1.8)
					{$A^b_\nu(x_2)$} ;
					\vertex (S) at (2, -1.3);
					\vertex (T) at (1, -0.3);
					\vertex (B) at (-1, -0.3);
					\vertex at (2,-1.8) {$A^c_\rho(x_3)$} ;
					\vertex (c) at (0,0.6);
					\vertex (b) at (-0.85,-0.25);
					\vertex (t) at (0.85,-0.25);
					\diagram*{
						(A) --[ photon] (C),
						(B) --[ photon] (D),
						(S) --[ photon] (T),
						(C) -- [ charged scalar,thick ] (B),
						(T) -- [ anti charged scalar,thick] (B),
						(T) --[ charged scalar,thick] (C),	
						(c)-- [plain,dotted, thick] (b),
						(t) -- [plain,dotted, thick] (b) ,
						(c) --[plain,dotted, thick](t),	
					};
				\end{feynman}
			\end{tikzpicture} 
		} \ + 	\mathord{
			\begin{tikzpicture}[radius=2.cm, baseline=-0.65ex,scale=0.55]		
				\begin{feynman}
					\vertex (A) at (0,2) ;
					\vertex at (0,2.4) {$A^a_\mu(x_1)$} ;
					\vertex (C) at (0,0.7);
					\vertex (D) at (-2, -1.3);
					\vertex (D1) at (-2, -1.8)
					{$A^c_\rho(x_3)$}	 ;
					\vertex (S) at (2, -1.3);
					\vertex (T) at (1, -0.3);
					\vertex (B) at (-1, -0.3);
					\vertex at (2,-1.8) {$A^b_\nu(x_2)$} ;
					\vertex (c) at (0,0.6);
					\vertex (b) at (-0.85,-0.25);
					\vertex (t) at (0.85,-0.25);
					\diagram*{
						(A) --[ photon] (C),
						(B) --[ photon] (D),
						(S) --[ photon] (T),
						(C) -- [charged scalar,thick ] (B),
						(T) -- [anti charged scalar, thick] (B),
						(T) --[ charged scalar,thick] (C),	
						(c)-- [plain,dotted, thick] (b),
						(t) -- [plain,dotted, thick] (b) ,
						(c) --[plain,dotted, thick](t),				
					};
				\end{feynman}
			\end{tikzpicture} 
		} -  \  \left(\cR = \mathrm{Adj}\right)
		\\[0.8em]
		& =\ii g_B^3 f^{abc}\left(N-i_\cR\right)	\int\dd P \int \dfrac{\dd^d k}{(2\pi)^d}\dfrac{\left(\Tr\bar{\sigma}_\mu\slashed{k}\bar{\sigma}_\rho(\slashed{k}-\slashed{p}_3)\bar{\sigma}_\nu(\slashed{k}+\slashed{p}_1)\right)}{k^2(k+p_1)^2(k-p_3)^2}
		\\[0.4em]
		&+ \ii g_B^3 f^{abc}\left(N-i_\cR\right)	\int\dd P \int\dfrac{\dd^d k}{(2\pi)^d}\dfrac{\left(\Tr\bar{\sigma}_\rho\slashed{k}\bar{\sigma}_\mu(\slashed{k}+\slashed{p}_1)\bar{\sigma}_\nu(\slashed{k}-\slashed{p}_3)\right)}{k^2(k+p_1)^2(k-p_3)^2} \ ,		
	\end{split}
\end{equation} 
where we used again a double dashed/dotted line to denote, respectively, the contributions of the fermions $\eta$ and $\tilde{\eta}$, which transforms in the representation $\cR$ and $\cR^*$. 
Employing the identities in eq. (\ref{eq:trace}) and neglecting terms which provide total derivatives integrated over closed paths when inserted in eq. (\ref{eq:diagrammatic representation correlators Sigma_G}), we eventually find that 
\begin{equation}
	\begin{split}
		\label{eq:fermionic looo pure-gauge vertex def}
		\cF_{\mu\nu\rho}^{abc}(x_i)=-\cS^{abc}_{\mu\nu\rho}(x_i) +f^{abc}\left(\cG_{1,\mu\nu\rho}(x_i)+
		\cG_{2,\mu\nu\rho}(x_i)\right) \  .
	\end{split}
\end{equation} 
In the previous expression, $\cS_{\mu\nu\rho}^{abc}(x_i)$ is the contribution to the pure gauge-vertex resulting from the scalar loops (\ref{eq:loop scalars pure gauge vertex difference theory}), while the quantities $\cG_{1,\mu\nu\rho}$ and  $\cG_{2,\mu\nu\rho}$ are, respectively,  the counterparts of the functions $\cS^\mu_1$ and $\cS^\mu_2$ defined in eq.s (\ref{eq:S1 difference gauge-scalar}) and (\ref{eq:S2 difference gauge-scalar}) and have a similar behaviour for $d\to 4$. Their explicit expressions are: \begin{equation}
	\label{eq:divgauge}
	\begin{split}
		\mathcal{G}_{1,\mu\nu\rho}(x_i) &= -2\mathrm{i}g_B^3(i_\mathcal{R}-N) \int \dd P  \int\dfrac{\dd^dk}{(2\pi)^d}\Bigg[\dfrac{\delta_{\mu\nu}\left(k^2p_{2,\rho}-p_{1,\rho}(k-p_3)^2\right)}{k^2(k+p_1)^2(k-p_3)^2}\\[0.6em]
		+&	\dfrac{\delta_{\mu\rho}\left(p_{1,\nu}(k-p_3)^2-p_{3,\nu}(k+p_1)^2\right)}{k^2(k+p_1)^2(k-p_3)^2}+\dfrac{\delta_{\nu\rho}\left(p_{3,\mu}(k+p_1)^2-p_{2,\mu}k^2\right)}{k^2(k+p_1)^2(k-p_3)^2} \Bigg]\ 
	\end{split}
\end{equation} 
and 
\begin{equation}
	\label{eq:finitegauge}
	\begin{split}
		\mathcal{G}_{\mu\nu\rho}(x_i) &= 2\mathrm{i}g_B^3(i_\mathcal{R}-N) \int \dd P \int\dfrac{\dd^dk}{(2\pi)^d}\Bigg[\dfrac{\delta_{\mu\nu}\left(p_1^2(k-p_3)_\rho+  p_2^2 k_\rho-p_3^2 (k+p_1)_\rho\right)}{k^2(k+p_1)^2(k-p_3)^2} \\[0.6em]
		+&\dfrac{\delta_{\mu\rho}\left(p_3^2(k+p_1)_\nu-  p_2^2 k_\nu+p_1^2 (k-p_3)_\nu\right)}{k^2(k+p_1)^2(k-p_3)^2}  -\dfrac{\delta_{\nu\rho}\left(p_1^2(k-p_3)_\mu - p_2^2 k_\mu-p_3^2 (k+p_1)_\mu\right)}{k^2(k+p_1)^2(k-p_3)^2} \Bigg]   \ .
	\end{split}
\end{equation}
%

Combining together the contribution to the pure-gauge vertex resulting from the scalars  (\ref{eq:fermionic looo pure-gauge vertex def})  with that of the fermions (\ref{eq:loop scalars pure gauge vertex difference theory}) and inserting  the result in eq. (\ref{eq:diagrammatic representation correlators Sigma_G}), we again can organize the final result in terms of two distinct contributions:
\begin{equation}
	\label{eq:G 1 and G2 appendix}
	\mathsf{L}_{g} = \mathsf{L}_{g,1} + \mathsf{L}_{g,2}~.
\end{equation} 
In analogy to eq. (\ref{eq:SIgmags 1 and 2 appendix}), $\mathsf{L}_{g,1}$ takes the following form  
\begin{equation}
	\label{eq:Sigma1G Appendix}
	\begin{split}
		\mathsf{L}_{g,1} &=\dfrac{g_B^3}{12}NC_F\oint\dd^3\tau\varepsilon(\tau)\left(\dot{x}_1^\mu\dot{x}_2^\nu\dot{x}_3^\rho\right)\cG_{1,\mu\nu\rho } \\[0.4em]
		&=- \cA_1 \oint\dd^3\tau\varepsilon(\tau)\left(\dot{x}_1\cdot\dot{x}_3\right)\left(\dot{x}_2\cdot\partial_1\right)\int\dd^d\omega \Delta^{(1)}(x_{1\omega})\Delta(x_{2\omega})\Delta(x_{3\omega}) \ .
	\end{split}
\end{equation}
where the coefficient $\cA_1$ was introduced in eq. (\ref{A1def}) while $\Delta^{(1)}(x)$ and $\Delta(x)$ are defined in eq.s (\ref{eq:one-loop correction adjoint scalar main text}) and (\ref{eq:Deltad}), respectively. Let us note that  the previous expression has the same structure of eq. (\ref{eq:SigmaGs1 appendix}) with the replacement $R^2 \to -\dot{x}_1\cdot\dot{x}_2$, as expected by supersymmetry. On the other hand, $\mathsf{L}_{g,2}$ is given by 
\begin{align}
	\label{eq:SigmaG2 Appendix}
	\mathsf{L}_{g,2}
	&=\dfrac{g_B^3}{12}NC_F\oint\dd^3\tau\varepsilon(\tau)\left(\dot{x}_1^\mu\dot{x}_2^\nu\dot{x}_3^\rho\right)
	\cG_{2,\mu\nu\rho} \\[0.4em]	&=\cA_2\oint\dd^3\tau\varepsilon(\tau)\int\dd{P}\left(\mathrm{i}p_2^2\right)\int\dfrac{\dd^d k}{\left(2\pi\right)^d}\dfrac{\left(2k\cdot\dot{x}_1\right)\left(\dot{x}_2\cdot\dot{x}_3\right)-(k\cdot\dot{x}_2)\left(\dot{x}_1\cdot\dot{x}_3\right)}{k^2(k+p_1)^2(k-p_3)^2} \ ,
\end{align}  
where the coefficient $\cA_2$ is defined in eq. (\ref{eq:C2 appendix}).
\paragraph{Summary of the results}
Using the results we derived in this section, we can construct  the final expression for the lifesaver diagram depicted in eq. (\ref{eq:lifesaver diagrams appendix}).  Starting from eq. (\ref{eq:lifesaver diagrams appendix}) and expressing the one-loop correction to the gauge-scalar and pure-gauge vertices by eq.s (\ref{eq:SIgmags 1 and 2 appendix}) and (\ref{eq:G 1 and G2 appendix}), we  can arrange the four contributions in such a way to reconstruct the usual $R^2 -\dot x_i\cdot \dot x_j$ factor. Thus, we can write   
\begin{equation}
	\begin{split}
		\label{eq:sigma- decomposition}
		\mathord{
			\begin{tikzpicture}[radius=2.cm, baseline=-0.65ex,scale=0.55]
				\draw [black] (0,0) circle [];
				\draw [black] (0,0) circle [radius=0.8cm];
				\draw [black, dashed,thick] (0,0) circle [radius=0.7cm];
				\begin{feynman}
					\vertex (A) at (0,2);
					\vertex (C) at (0,0.8);
					\vertex (D) at (-1.5, -1.3);
					\vertex (B) at (-0.7, -0.4);
					\vertex (B1) at (0.7,-0.4);
					\vertex (B2) at (1.5,-1.3);
					\diagram*{
						(A) -- [photon] (C),
						(B) --[ anti fermion] (D),
						(B) --[photon] (D),
						(B1) --[fermion] (B2),
						(B1) --[photon] (B2),
					};
				\end{feynman}
			\end{tikzpicture} 
		} &= \mathsf{L}_g + \mathsf{L}_{gs}\\
		&= \left(\mathsf{L}_{g,1} + \mathsf{L}_{gs,1}\right)
		+ \left(\mathsf{L}_{g,2} + \mathsf{L}_{gs,2}\right)
		\equiv \mathsf{L}_{1} + \mathsf{L}_{2} \ .
	\end{split}	
\end{equation} 
Explicitly, we have
\begin{equation}
	\label{eq:sigma-1 appendix}
	\mathsf{L}_{1}
	= \cA_1
	\oint\dd^3\tau\varepsilon(\tau)\left(R^2-\dot{x}_1\cdot\dot{x}_3\right)\dot{x}_2\cdot\partial_1\int\dd^d\omega \Delta^{(1)}(x_{1\omega})\Delta(x_{2\omega})\Delta(x_{3\omega}) 
\end{equation}
and
\begin{equation}
	\label{eq:sigma-2 appendix}
	\begin{split}
		\mathsf{L}_{2}
		&=\cA_2\oint\dd^3\tau\varepsilon(\tau)\int\dd{P}\left(\mathrm{i}p_2^2\right)\int\dfrac{\dd^d k}{\left(2\pi\right)^d}\dfrac{\left(2k\cdot\dot{x}_1\right)\left(\dot{x}_2\cdot\dot{x}_3-R^2\right)-(k\cdot\dot{x}_2)\left(\dot{x}_1\cdot\dot{x}_3-R^2\right)}{k^2(k+p_1)^2(k-p_3)^2} \ .
	\end{split}
\end{equation} 
\subsection{Integration over the Wilson loop contour: calculating $\mathsf{L}_1$}
In this subsection, we examine in detail the integration over the Wilson loop contour of the contribution defined in eq.  (\ref{eq:sigma-1 appendix}):
the calculation is analogous to that we described in Section \ref{sec:computing sigma+} for the correction  $\mathsf{M}_1$, defined in eq.   (\ref{eq:first step}). To begin with, we integrate over  the bulk point $\omega$ by introducing the  usual Feynman parametrizations for the propagators $\Delta(x)$ and $\Delta^{(1)}(x)$ defined, respectively,  in eq. (\ref{eq:Deltad}) and (\ref{eq:one-loop adjoint scalar difference appendix}). Using the parametrization (\ref{eq:parametrization}) for the points $x_i$ on the Wilson loop contour,  we obtain
\begin{equation}
	\label{eq:sigma_-^1 prima dell'integrale su eps}
	\begin{split}
		\mathsf{L}_{1} &= - A_d\int_0^1 \dd F  \left(\alpha^2\beta\gamma\right)^{d/2-2} \oint\dd^3\tau \dfrac{\varepsilon(\tau) \ (1-\cos\tau_{13})\left(\alpha(1-\alpha)\sin\tau_{12}+\alpha\gamma\sin\tau_{23}\right)}{Q^{3d/2-4}}\ ,
	\end{split} 
\end{equation} where the measure $\dd F$ is given by eq. (\ref{eq:integration over the unit cube}), while $A_d$ and $Q$ are defined in eq.s  (\ref{eq:factorized pole}) and (\ref{eq:def of Q}), respectively. Integrating by parts via the identity (\ref{eq:initial identity}), we find that
\begin{align}
	\label{eq:integration over the contour sigma^1_}
	\mathsf{L}_{1} &=
	\dfrac{2A_d}{5-3d/2} \int_0^1\dd F \left(\alpha^2\beta\gamma\right)^{d/2-2} \oint \dd^2{\tau}\dfrac{(1 - \cos{\tau_{23}})^{6-3d/2}}{[\gamma(1-\gamma)]^{3d/2-5}} \notag \\[0.4em]
	+&A_d\dfrac{6-3d/2}{5-3d/2} \int_0^1\dd F \left(\alpha^2\beta\gamma\right)^{d/2-2} \oint \dd^3\tau\dfrac{\varepsilon(\tau) \sin{\tau_{13}}}{Q^{3d/2-5}}    -A_d I_1(d) \ ,
\end{align}
where
\begin{equation}
	\begin{split}
		\label{eq:I_2(d) main}
		I_1(d)&= \int_0^1\dd F \left(\alpha^2\beta\gamma\right)^{d/2-2} \oint\dd^3\tau\varepsilon(\tau) \dfrac{\beta\gamma\sin\tau_{13}(1-\cos\tau_{23})+\alpha\gamma\sin\tau_{23}(1-\cos\tau_{13})}{Q^{3d/2-4}} \\[0.4em]
		&= \int_{0}^1\dd{F}\left( \alpha^{d-4}(\beta\gamma)^{d/2-1}-\beta^{d-3}\alpha^{d/2-2}\gamma^{d/2-1}\right) \oint \dd^3{\tau} \varepsilon(\tau) \dfrac{(1-\cos\tau_{23})\sin\tau_{13}}{Q^{3d/2-4}} \ .
	\end{split}
\end{equation} 
Comparing the previous expression with eq. (\ref{eq:integration over the contour}), we note that the last term is a novelty. It arises  because the integrand in eq. (\ref{eq:sigma-1 appendix}) is not completely symmetric in the exchange of the coordinates $x_i$. On the other hand, the exstra term $I_1(d)$  does not contribute to the final result; to see this, we replace the denominator $Q$ with a two-fold Mellin-Barnes integral via (\ref{eq:two-fold MB}) obtaining   \begin{equation}
	\label{eq:1/Q^3d/2-4}
	\begin{split}
		\dfrac{1}{Q^{3d/2-4}}	&=\dfrac{2^{4-3d/2}}{\Gamma(3d/2-4)}\int\frac{\dd u \dd v}{(2\pi \mathrm{i})^2}\dfrac{\Gamma(3d/2-4+u+v)\Gamma(-u)\Gamma(-v)\left(\beta\gamma\sin^2\frac{\tau_{23}}{2}\right)^u}{\left(\alpha\beta\sin^2\frac{\tau_{12}}{2}\right)^{3d/2-4+u+v}\left(\alpha\gamma\sin^2\frac{\tau_{13}}{2}\right)^{-v}} \ .
	\end{split}
\end{equation}
Inserting the previous expression into eq. (\ref{eq:I_2(d) main}), we integrate  over the Wilson loop contour via eq. (\ref{eq:third master integral}) and we obtain  
\begin{equation}
	\label{eq:I2 second manipulation}
	\begin{split}
		I_1(d) =\int\dfrac{\dd u \dd v}{(2\pi \mathrm{i})^2}&\dfrac{\Gamma(3d/2-4+u+v)\Gamma(-u)\Gamma(-v)\Gamma(d/2+u+v)}{{2^{3d/2-5}}\Gamma(3d/2-4)\Gamma(5-d)} \mathcal{J}(4-3d/2-u-v,u+1,v)\\[0.4em]  &\times \Big(\Gamma(1-d/2-u)\Gamma(4-d-v)-\Gamma(2-d/2-v)\Gamma(3-d-u)\Big) \ ,
	\end{split}
\end{equation}
where the function $\cJ(x,y,z)$ is defined in eq. (\ref{eq:third master integral}). Expanding the previous expression about $d=4$,  we arrive at
\begin{equation}
	\begin{split}
		I_1(d) &=(d-4) 
		\int_{\delta-\mathrm{i}\infty}^{\delta+\mathrm{i}\infty}  \dfrac{\dd{v}\dd{u} }{\left(2\pi \mathrm{i}\right)^2} \frac{  \csc (\pi  u) \csc (\pi  v) \csc (\pi  (u+v)) (\psi
			^{(0)}(-u)-\psi ^{(0)}(-v))}{u+v+1}
		+\ldots \  ,
	\end{split}
\end{equation} where the dots stand for terms of order $\mathcal{O}(d-4)^2$, while  $\delta\in (-1,0)$ denotes the real part of the integration variables $u$ and $v$. 
The previous expression vanishes identically because of the antisymmetry of the integrand, meaning that $I_1(d) = \cO(d-4)^2$ and $A_d I_1(d)=\cO(d-4)$, as it can be seen by employing eq. (\ref{eq:factorized pole}).

Concerning the first two terms in eq. (\ref{eq:integration over the contour sigma^1_}), one can explicit perform the integration over the Feynman parameters and apply the master integral (\ref{eq:I1 main}) to obtain
\begin{equation}
	\label{eq:sigma_-^(1) def}
	\begin{split}
		\mathsf{L}_1
		&= F_2^{(2)}  - \hat{g}_B^6\dfrac{3 C_FN\beta_0\zeta(3)}{16\pi^2}   +\mathcal{O}(d-4) \ ,
	\end{split}
\end{equation} where we recall that $F_2^{(2)}$ is the bubble-like contribution defined in eq. (\ref{Fiis}).

\subsection{Integration over the Wilson loop contour: calculating $\mathsf{L}_2$}	
The calculation of $\mathsf{L}_2$ is more complicated than that we performed in the previous subsection. To begin with, we consider eq. (\ref{eq:sigma-2 appendix}) and  we integrate over the  internal momentum $k$ by introducing the usual Feynman parameters for the three propagators. We find that
\begin{equation}
	\label{eq:I^2 first step}
	\begin{split}	
		\mathsf{L}_2	&= 2 \cA_2
		\dfrac{\mathrm{i}\Gamma(3-d/2)}{(4\pi)^{d/2}}\! \! \oint\dd^3\tau\varepsilon(\tau) (R^2-\dot{x}_1\cdot\dot{x}_3)\! \int\! \dd P \dd X\dfrac{p_3^2\left(z\dot{x}_2\cdot p_2-x\dot{x}_2\cdot p_1\right) +zp_2^2\dot{x}_2\cdot p_3}{\left(xyp_1^2+zyp_2^2+zxp_3^2\right)^{3-d/2}} \ ,
	\end{split}
\end{equation}  with $\dd X =\dd x \dd y \dd z\delta(1-x-y-z)$. The previous expression involves the quantity $p_2\cdot\dot{x}_2$ which, upon integration over the external momenta $\dd{P}$, yields a total derivative with respect to the variable $\tau_2$. As a result, the contour integration of this contribution is technically simpler to treat.  Therefore, we find convenient to express eq. (\ref{eq:I^2 first step}) as the sum of two terms, i.e. 
$\mathsf{L}_2 = \mathsf{L}_2^\prime +  \mathsf{L}_2^{\prime\prime}$,  with 
\begin{align}
	\label{eq:I^2_A}
	\mathsf{L}_2^\prime &= 2\cA_2 \dfrac{\mathrm{i}\Gamma(3-d/2)}{(4\pi)^{d/2}} \oint\dd^3\tau\varepsilon(\tau) (R^2-\dot{x}_1\cdot\dot{x}_3) \int \dd P\int_0^1 \dd X\dfrac{p_3^2\left(z\dot{x}_2\cdot p_2\right) }{M^{3-d/2}} \ , \\[0.5em]
	\label{eq:I^2_B}
	\mathsf{L}_2^{\prime\prime} &=2   \cA_2 \dfrac{\mathrm{i}\Gamma(3-d/2)}{(4\pi)^{d/2}} \oint\dd^3\tau\varepsilon(\tau) (R^2-\dot{x}_1\cdot\dot{x}_3) \int \dd P \int_0^1\dd X\dfrac{zp_2^2\left(\dot{x}_2\cdot p_3\right) - p_3^2x(\dot{x}_2\cdot p_1) }{M^{3-d/2}}  \ ,
\end{align} which we will analyse in turn. In the previous expression, the denominator $M$ is 
\begin{equation}
	\label{eq:M}
	M=xyp_1^2+zyp_2^2+zxp_3^2 \ .
\end{equation}
\subsubsection{Computing $\mathsf{L}_2^\prime$}
As we already stressed, the computation of the function $\mathsf{L}_2^\prime$ goes through the observation that the product $p_2 \cdot \dot{x}_2$ becomes a total derivative upon integration over the momenta $p_i$ (see eq. (\ref{eq:integration over momenta})). By relabelling $\tau_1\leftrightarrow \tau_2$ and recalling that $\varepsilon(\tau)$ is  antisymmetric, we find that eq. (\ref{eq:I^2_A}) can be rewritten as follows
\begin{equation}
	\begin{split}
		\label{eq:I^2_A second step}
		\mathsf{L}_2^\prime &= 2\cA_2 \dfrac{\Gamma(3-d/2)}{(4\pi)^{d/2}} \oint\dd^3\tau\varepsilon(\tau) (R^2-\dot{x}_2\cdot\dot{x}_3)\dfrac{\dd}{\dd\tau_1} \int \dd P\int_0^1 \dd X \dfrac{x p_3^2 }{M^{3-d/2}}   \\[0.4em]
		&=4\cA_2\dfrac{\Gamma(3-d/2)}{(4\pi)^{d/2}} \oint\dd^3\tau \left(\delta(\tau_{13})-\delta(\tau_{12})\right) (R^2-\dot{x}_2\cdot\dot{x}_3)\int \dd P\int_0^1 \dd X \dfrac{x p_3^2 }{M^{3-d/2}}		\ ,
	\end{split}
\end{equation} where we obtained the second line via integration by parts and eq. (\ref{eq:derivative of epsilon}). To proceed with the computation, we employ the following identity (see eq. (\ref{eq:Mellin barnes}))  \begin{equation}
	\label{eq:contour-like rep for deno}
	\dfrac{\Gamma(3-d/2) }{M^{3-d/2}} = \int\dfrac{\dd u \dd v}{(2\pi\mathrm{i})^2}\dfrac{\Gamma(3-d/2+u+v)\Gamma(-u)\Gamma(-v)}{x^{3-d/2+u}y^{3-d/2+v}z^{-u-v}}\dfrac{(p_2^2)^u(p_3^2)^v}{(p_1^2)^{3-d/2+u+v}} \ ,
\end{equation} where the integration contour separates the increasing and decreasing poles of the $\Gamma$-functions. Substituting the previous expression in eq. (\ref{eq:I^2_A second step}) and sequentially performing the integration over the Feynman parameters and the momenta $p_i$ (see eq. (\ref{eq:integration over momenta})),  we find
\begin{equation}
	\label{eq:I_A^(2) def}
	\mathsf{L}_2^\prime = \cA_2 \dfrac{\Gamma(3d/2-5)\left(\mathcal{M}^{(ii)}(d)-\mathcal{M}^{(i)}(d)\right)}{4^4\pi^{3d/2}\Gamma(d-2)\Gamma(5-d)}\oint\dfrac{R^2-\dot{x}_2\cdot\dot{x}_3}{[x_{23}^2]^{3d/2-5}} \ .
\end{equation} with the two Mellin-like amplitudes defined as   \begin{align}
	\label{eq:M^i_A}
	\mathcal{M}^{(i)}(d)&= \int\dfrac{\dd u\dd v}{(2\pi\mathrm{i})^2}\dfrac{\Gamma(d-4-u-v)\Gamma(d/2-1+u)\Gamma(5-d+v)\Gamma(-v)\Gamma(1+u+v)}{u(d/2-3-u-v)\Gamma(3d/2-5-v)[\Gamma(d/2-2-v)\Gamma(d/2-1-u)]^{-1}} \ , \\
	\label{eq:M^ii}
	\mathcal{M}^{(ii)}(d)&=  \int\dfrac{\dd u\dd v}{(2\pi\mathrm{i})^2}\dfrac{\Gamma(-u) \Gamma(4-d+u)\Gamma(1+u+v)\Gamma(d/2-1-u)\Gamma(d/2-2-v)}{\Gamma(3d/2-4-u)(3-d/2+u+v)[\Gamma(d-4-u-v)\Gamma(d/2+v)]^{-1}} \ .
\end{align} These two functions exhibit different behaviours when $d\to 4$. On the one hand, eq. (\ref{eq:M^ii}) becomes singular in this limit due to  the product $\Gamma(-u)\Gamma(4-d+u)$, which does not enable to separate the first increasing and the first decreasing pole. On the other hand,  eq. (\ref{eq:M^i_A}) is perfectly finite in four dimensions. For future reference, we provide its expansion about $d=4$, i.e. 
\begin{equation}
	\label{eq:expanding M^i_A}
	\begin{split}
		\mathcal{M}^{(i)}(d)&=\int_{\delta-\mathrm{i}\infty}^{\delta+\mathrm{i}\infty}\dfrac{\dd u\dd v}{(2\pi\mathrm{i})^2} \frac{\pi ^3 \csc (\pi  u) \csc (\pi  v) \csc (\pi  (u+v))}{v (u+v+1)} +\ldots=2\zeta(3) + \cO(d-4)\ ,
	\end{split}
\end{equation} where $\delta\in(-1,0)$ represents the real part of the variables $u$ and $v$. As we will shortly see, similar quantities will arise from the integral $\mathsf{L}_2^{\prime\prime}$, defined in eq. (\ref{eq:I^2_B}).

\subsubsection{Computing $\mathsf{L}_2^{\prime\prime}$}		
We begin with  writing eq. (\ref{eq:I^2_B}) as 
\begin{equation}				
	\label{eq:I_B^2 part 2}
	\mathsf{L}_2^{\prime\prime} = \cA_2\oint\dd^3\tau\varepsilon(\tau) (R^2-\dot{x}_1\cdot\dot{x}_3)(\dot{x}_2\cdot\partial_1)
	\widetilde{\mathsf{L}}_2^{\prime\prime} ~,
\end{equation} 
where 
\begin{equation} \widetilde{\mathsf{L}}_2^{\prime\prime} = 2 \dfrac{\Gamma(3-d/2)}{(4\pi)^{d/2}}(\dot{x}_2\cdot\partial_1) \int \dd P \ \int_0^1 \dd X\dfrac{xp_3^2+yp_2^2}{M^{3-d/2}}~.
\end{equation} 
Let us concentrate on $\widetilde{\mathsf{L}}_2^{\prime\prime}$, which contains the integration over the Feynman parameters and over the measure $\dd{P}$. 
By expressing the denominator $M$, defined in eq. (\ref{eq:M}), as a two-fold Mellin-Barnes integral via eq. (\ref{eq:contour-like rep for deno}), the integration over the measures $\dd{X}$ and $\dd{P}$ becomes elementary. The net result can be expressed as a combination  of \textit{generalized} propagators $\cD(x,s)$ defined in eq. (\ref{eq:Fourier transform for massless propagators}):    
\begin{equation}
	\begin{split}
		\widetilde{\mathsf{L}}_2^{\prime\prime}
		&=\int\dd{\Omega} \cD(x_{1\omega},\sigma) f_d(u,v)	\Big(\cD(x_{2\omega},1-u)\cD(x_{3\omega},-v)+\cD (x_{2\omega},-v)\cD (x_{3\omega},1-u)\Big)	
	\end{split}
\end{equation} where  $\dd{\Omega}=\dd^d{\omega}\dd{u}\dd{v}/(2\pi\mathrm{i})^2(2\pi)^d$ and $\sigma=4-d/2+u+v$.  The integration over the variable $\omega$ arises from the conservation of the momenta $p_i$, while    \begin{align}
	f_d(u,v)= \dfrac{\Gamma(d/2-1-u)\Gamma(d/2-2-v)\Gamma(1+u+v)\Gamma(3-d/2+v+u)\Gamma(-u)\Gamma(-v)}{\Gamma(d-2)\pi^{-d/2}} \ . 
\end{align} To proceed with the calculation, we integrate over $\dd^d{\omega}$ by introducing three Feynman parameters for the different  propagators $\cD(x,s)$. By employing eq. (\ref{eq:Fourier transform for massless propagators}) and the parametrization of the coordinates $\tau_i$, given by eq. (\ref{eq:parametrization}), we obtain 
\begin{equation}
	\label{eq:ISigmaB prima di integra sul circle}
	\begin{split}
		\mathsf{L}_2^{\prime\prime} =   \dfrac{\cA_2\Gamma(3d/2-4)R^{12-3d} }{\Gamma(d-2)2^9\pi^{3d/2}2^{3d/2-5}} \! \int \!\dd\cM \!\! \oint\dd^3\tau \dfrac{\varepsilon(\tau) (\cos\tau_{13}-1)\left(\alpha(1-\alpha)\sin\tau_{12}+\alpha\gamma\sin\tau_{23}\right)}{Q^{3d/2-4}} \ .
	\end{split} 
\end{equation} The denominator $Q$ is defined in eq. (\ref{eq:def of Q}), while the  measure $\dd\cM$ is given by \begin{align}
	\label{eq:measure def}
	\dd{\cM}&= \dfrac{\dd{u}\dd{v}}{\left(2\pi\mathrm{i}\right)^2}\dd F \ \alpha^{d-5-u-v}\beta^{d/2-2+u}\gamma^{d/2-2+v}\Big(\tilde{f}(u,v)\gamma +\tilde{f}(v,u)\beta\Big)~,
\end{align}			
with
\begin{align}
	\label{eq:tilde f}
	\tilde{f}(u,v) &=-\dfrac{\Gamma(d/2-1-v)\Gamma(d/2-2-u)\Gamma(1+u+v)}{u(3-d/2+u+v)} \ ,
\end{align} 
while $\dd F$ was defined in eq. (\ref{eq:integration over the unit cube}). Note that the integration measure  is  symmetric under the simultaneous exchange of $\beta\leftrightarrow\gamma$ and $u\leftrightarrow v$. Finally, we integrate over the coordinates $\tau_i$ by employing the identity (\ref{eq:initial identity}) and eventually obtain
\begin{equation}
	\label{eq:almost done}
	\begin{split}
		\mathsf{L}_2^{\prime\prime} = &- \cA_2 \dfrac{\Gamma(3d/2-4)R^{12-3d} }{\Gamma(d-2)2^9\pi^{3d/2}2^{3d/2-5}}\int\dd\cM \left(\dfrac{2}{3d/2-5}\oint\dd^2\tau\dfrac{(1-\cos\tau_{12})^{6-3d/2}}{[\gamma(1-\gamma)]^{3d/2-5}} \right)\\[0.6em]
		&- \cA_2	\dfrac{\Gamma(3d/2-4)R^{12-3d} }{\Gamma(d-2)2^9\pi^{3d/2}2^{3d/2-5}}\int\dd\cM \left( \dfrac{3(2-d/2)}{3d/2-5}\oint\dd^3\tau\varepsilon(\tau)\dfrac{\sin\tau_{13}}{Q^{3d/2-5}}  + T_d(\alpha,\beta,\gamma)\right) \ , 
	\end{split}
\end{equation} where 

\begin{equation}
	\label{eq:Td}
	T_d(\alpha,\beta,\gamma)= \oint\dd^3\tau\varepsilon(\tau) \dfrac{\beta\gamma\sin\tau_{13}(1-\cos\tau_{23})+\alpha\gamma\sin\tau_{23}(1-\cos\tau_{13})}{Q^{3d/2-4}}.
\end{equation} 	The first term in the second line is proportional to $(d-4)$: it vanishes in the limit $d\to 4 $ since the path-ordered integral is regular and consequently, it can be neglected for the three-loop analysis. Actually, as we will show in the following subsection, also the term involving the function $T_d(\alpha,\beta,\gamma)$ is of order $(d-4)$. Thus, we can write 
\begin{equation}
	\label{eq:I_B}
	\begin{split}
		\mathsf{L}_2^{\prime\prime} & = -\cA_2 \dfrac{\Gamma(3d/2-4)R^{12-3d} }{\Gamma(d-2)2^9\pi^{3d/2}2^{3d/2-5}}\int\dd\cM \left(\dfrac{2}{3d/2-5}\oint\dd^2\tau\dfrac{(1-\cos\tau_{12})^{6-3d/2}}{[\gamma(1-\gamma)]^{3d/2-5}} \right)+\cO(d-4)\\[0.6em]
		&=- \cA_2 \dfrac{\Gamma(3d/2-5)\left(\mathcal{M}^{(ii)}(d)+\mathcal{M}^{(i)}(d)\right)}{4^4\pi^{3d/2}\Gamma(d-2)\Gamma(5-d)}\oint\dfrac{R^2-\dot{x}_2\cdot\dot{x}_3}{(x_{23}^2)^{3d/2-5}} +\cO(d-4) \ ,
	\end{split}
\end{equation} 
where we employed the explicit form of the measure $\dd\cM$ (\ref{eq:measure def}) to integrate over the Feynman parameters and used the definitions of the amplitudes $\mathcal{M}^{(i)}(d)$ and  $\mathcal{M}^{(ii)}(d)$ given in eq.s (\ref{eq:M^i_A}) and (\ref{eq:M^ii}). Finally, combining this result with eq. (\ref{eq:I_A^(2) def}), we find that the function $\mathsf{L}_2$ (\ref{eq:I^2 first step}) can be expanded as \begin{equation}
	\label{eq:finale result for T}
	\begin{split}
		\mathsf{L}_2 &= \mathsf{L}_2^\prime+ \mathsf{L}^{\prime\prime}_2 \\[0.2em]
		&=-\dfrac{\cA_2\Gamma(3d/2-5)\mathcal{M}^{(i)}(d)}{2^7\pi^{3d/2}\Gamma(d-2)\Gamma(5-d)}\oint\dfrac{R^2-\dot{x}_2\cdot\dot{x}_3}{(x_{23}^2)^{3d/2-5}} \ +\mathcal{O}(d-4)\\
		&=-\dfrac{C_F N \hat{g}_B^6\zeta(3)}{8\pi^2}+\cO(d-4) \ .
	\end{split}
\end{equation} The last equality follows from the definition of the coefficient $\cA_2$, given by eq. (\ref{eq:C2 appendix}), from the expansion of the amplitude $\cM^{(i)}(d)$ about $d=4$ (\ref{eq:expanding M^i_A}) and from the integration over the contour. 		
\subsubsection{Evanescent integrals }
Let us conclude this section by explicitly showing that the last contribution in the second line  of eq. (\ref{eq:almost done}) is of order $(d-4)$. We consider  \begin{equation}
	\label{eq:evanescent lifesaver starting point}
	\begin{split}
		\mathsf{E}(d)&= \int\dd{\cM}T_d(\alpha,\beta,\gamma)\\[0.4em]
		&=\int_0^1 \dd F\int\dfrac{\dd u\dd v}{\left(2\pi\mathrm{i}\right)^2}\alpha^{d-5-u-v}\beta^{d/2-2+u}\gamma^{d/2-2+v}\left(\gamma\tilde{f}(u,v)+\beta\tilde{f}(v,u)\right)T_d(\alpha,\beta,\gamma) \ , 
	\end{split}
\end{equation} which  the second line follows from the definition of $\dd{\mathcal{M}}$, given by eq.  (\ref{eq:measure def}). The first term can be written as
\begin{equation}
	\mathsf{E}_1(d) =  
	2 \! \! \int\dfrac{\dd F\dd u\dd v}{\left(2\pi\mathrm{i}\right)^2}\left(\!\dfrac{\gamma^{d/2+v}\tilde{f}(u,v)}{\alpha^{5-d+u+v}\beta^{1-d/2-u}}-\dfrac{\gamma^{d/2+v}\tilde{f}(u,v)}{\beta^{4-d+u+v}\alpha^{2-d/2-u}}\!\right)\!\! \oint\dd^3\tau\varepsilon(\tau)\dfrac{\sin\tau_{13} \sin^2\frac{\tau_{23}}{2} }{Q^{3d/2-4}} \ ,
\end{equation} where we used the integral representation for the function $T_d(\alpha,\beta,\gamma)$ (\ref{eq:I_2(d) main}) and the antisymmetry of the $\varepsilon$-symbols (\ref{eq:defepsilon}). Changing variable according to $u^\prime=d/2-3-u-v$ in the second term, we find that $\mathsf{E}_1(d)=0$ for any $d$. 

The calculation of the second contribution in eq. (\ref{eq:evanescent lifesaver starting point}) is more subtle. We find that
\begin{equation}
	\label{eq:T_2 evanescent}
	\mathsf{E}_2(d) =\! 
	2\int\dfrac{\dd F\dd u\dd v}{\left(2\pi\mathrm{i}\right)^2}\!\left(\dfrac{\gamma^{d/2-1+v}\tilde{f}(v,u)}{\alpha^{5-d+u+v}\beta^{-d/2-u}}-\dfrac{\gamma^{d/2-1+v}\tilde{f}(v,u)}{\beta^{4-d+u+v}\alpha^{1-d/2-u}}\right)\!\! \oint\dd^3\tau\varepsilon(\tau)\dfrac{\sin\tau_{13} \sin^2\frac{\tau_{23}}{2} }{Q^{3d/2-4}} \ ,
\end{equation} where we employed again the integral representation of the function $T_d(\alpha,\beta,\gamma)$. To continue the calculation,  we consider separately the quantities 
\begin{equation}
	\begin{split}
		\mathsf{E}^{\prime}_2(d)& =\int\dfrac{\dd F\dd u\dd v}{\left(2\pi\mathrm{i}\right)^2} \dfrac{\gamma^{d/2-1+v}\tilde{f}(v,u)}{\alpha^{5-d+u+v}\beta^{-d/2-u}} \oint\dd^3\tau\varepsilon(\tau)\dfrac{\sin\tau_{13}\left(1-\cos\tau_{ 23}\right) }{Q^{3d/2-4}} \ , \\
		\mathsf{E}^{\prime\prime}_2(d) &=\int\dfrac{\dd F\dd u\dd v}{\left(2\pi\mathrm{i}\right)^2} \dfrac{\gamma^{d/2-1+v}\tilde{f}(v,u)}{\beta^{4-d+u+v}\alpha^{1-d/2-u}} \oint\dd^3\tau\varepsilon(\tau)\dfrac{\sin\tau_{13}\left(1-\cos\tau_{ 23}\right) }{Q^{3d/2-4}}\ . 
	\end{split}
\end{equation} Firstly focussing on  $\mathsf{E}^\prime_2(d)$, we replace the denominator $Q$ with its Mellin-Barnes image (\ref{eq:1/Q^3d/2-4}). This enables to integrate over the contour  by employing eq. (\ref{eq:third master integral}) and the result can be written as a  four-fold Mellin-Barnes integral  \begin{equation}
	\begin{split}
		\mathsf{E}^\prime_2(d) =  \int\dfrac{\dd u\dd v\dd s \dd t}{\left(2\pi\mathrm{i}\right)^4}&\dfrac{\Gamma(3d/2-4-s-t)\Gamma(-s)\Gamma(-t)\Gamma(-d/2-u-v-s)\Gamma(5-d-t+u)}{2^{3d/2-5}\Gamma(5-d)\Gamma(3d/2-4)}\\
		&\times\Gamma(d/2+v+t+s)\cJ(3d/2-4-s-t,s+1,t)\ , 
	\end{split}
\end{equation}where the function $\cJ(x,y,z)$ is defined in eq. (\ref{eq:third master integral}). Expanding the previous expression about $d\to4$ enables to integrate over $s$ and $t$ by a repeated application of Barnes's first  lemma. We eventually find that  \begin{equation}
	\mathsf{E}^\prime_2(d) = 8 \pi ^2 \int_{\delta-\mathrm{i}\infty}^{\delta+\mathrm{i}\infty}\dfrac{\dd u\dd v}{\left(2\pi\mathrm{i}\right)^2}\frac{ \Gamma (1-u) \Gamma (u+2) \Gamma (-v) \Gamma (v) \Gamma (-u-v) \Gamma (u+v+1)}{v\left(1+u+v\right) } +\cO(d-4)\ ,
\end{equation} where $\delta\in\left(-1,0\right)$ is the real part of the variables $u$ and $v$. However, it is not necessary to perform the integration over the last two variables since, repeating the same analysis for the quantity $\mathsf{E}^{\prime\prime}_2(d)$, it is possible to show its double Mellin-Barnes representation coincides with the previous expression. Exploiting this fact in eq. (\ref{eq:T_2 evanescent}), we have, expanding about $d\to 4$,  an identical cancellation. This explicitly showed that the function  $\mathsf{E}(d)$ (\ref{eq:evanescent lifesaver starting point}) is of order $\cO(d-4)$. 
\subsection{Summary of the results}
Let us briefly summarize the results for the calculation of the lifesaver diagrams (\ref{eq:lifesaver diagrams appendix}). Starting from eq. (\ref{eq:sigma- decomposition}), we finally find that 
\begin{equation}
	\label{eq:lifesaver final result}
	\begin{split}
		\mathord{
			\begin{tikzpicture}[radius=2.cm, baseline=-0.65ex,scale=0.55]
				\draw [black] (0,0) circle [];
				\draw [black] (0,0) circle [radius=0.8cm];
				\draw [black, dashed,thick] (0,0) circle [radius=0.7cm];
				\begin{feynman}
					\vertex (A) at (0,2);
					\vertex (C) at (0,0.8);
					\vertex (D) at (-1.5, -1.3);
					\vertex (B) at (-0.7, -0.4);
					\vertex (B1) at (0.7,-0.4);
					\vertex (B2) at (1.5,-1.3);
					\diagram*{
						(A) -- [photon] (C),
						(B) --[ anti fermion] (D),
						(B) --[photon] (D),
						(B1) --[fermion] (B2),
						(B1) --[photon] (B2),
					};
				\end{feynman}
			\end{tikzpicture} 
		}		 = \mathsf{L}_1+\mathsf{L}_2
		=  F_2^{(2)} - \hat{g}_B^6\dfrac{5 C_FN\beta_0^\mathcal{R}\zeta(3)}{16\pi^2}  + \mathcal{O}(d-4) \ ,
	\end{split}
\end{equation} where the second equality follows from eq.s (\ref{eq:sigma_-^(1) def}) and (\ref{eq:finale result for T}) and we recall that $F_2^{(2)}$ is the bubble-like contribution defined in eq. (\ref{Fiis}).
\section{Diagrams with four emissions}
\label{sec:diagrams with four emissions appendix}
In this section, we provide calculation details of the  following class of  diagrams 
\begin{equation}
	\label{eq:sigma4 appendix}
	\cW^\prime_{6(4)}  = 	\mathord{
		\begin{tikzpicture}[baseline=-0.65ex,scale=0.5]
			\draw [black] (0,0) circle [radius=2cm];
			\draw [black] (-0.75,0) circle [radius=0.45cm];
			\begin{feynman}
				\vertex (A) at (-0.75,1.8);
				\vertex (b) at (-0.75,0.45);
				\vertex (d) at (-0.75,-0.45);
				\vertex (B) at (-0.75,-1.8);
				\vertex (C) at (0.75,1.8);
				\vertex (D) at (0.75,-1.8);
				\diagram*{
					(A) -- [photon] (b),
					(A) --[ fermion] (b),
					(d) --[fermion] (B),
					(d) --[photon] (B),
					(C) -- [photon] (D),
					(C) --[fermion] (D)
				};
			\end{feynman}
			\filldraw[color=white!80, fill=white!15](-0.75,0) circle (0.7);	
			\draw [black, thick, dashed] (-0.75,0) circle [radius=0.63cm];
			\draw [black] (-0.75,0) circle [radius=0.7cm];
		\end{tikzpicture} 
	}  \ ,
\end{equation}
where we recall that the double dashed/continuos internal bubble denotes the one-loop correction to the adjoint scalar and gauge field in the difference approach, see
eq.s (\ref{eq:one-loop correction adjoint scalar main text}) and (\ref{eq:one-loop correction gauge-field position main text}).

Following the approach outlined  in  Section \ref{sec:Mercedes diagram} and employing eq. (\ref{eq:diagrammatic split gauge field}), we can organize the diagrams depicted in eq. (\ref{eq:sigma4 appendix}) as follows
\begin{align}
	\label{eq:sigma4prime}
	\Sigma_{4}^\prime  &= 	\mathord{
		\begin{tikzpicture}[baseline=-0.65ex,scale=0.6]
			\draw [black] (0,0) circle [radius=2cm];
			\draw [black] (-0.75,0) circle [radius=0.45cm];
			\begin{feynman}
				\vertex (A) at (-0.75,1.8);
				\vertex (b) at (-0.75,0.45);
				\vertex (d) at (-0.75,-0.45);
				\vertex (B) at (-0.75,-1.8);
				\vertex (C) at (0.75,1.8);
				\vertex (D) at (0.75,-1.8);
				\diagram*{
					(A) -- [photon] (b),
					(d) --[photon] (B),
					(C) -- [photon] (D),
				};
			\end{feynman}
			\filldraw[color=white!80, fill=white!15](-0.75,0) circle (0.7);	
			\draw [black, thick, dashed] (-0.75,0) circle [radius=0.63cm];
			\draw [black] (-0.75,0) circle [radius=0.7cm];
			\draw (-1.2,0) node[anchor=west]{$\delta$};
		\end{tikzpicture} 
	}  \  + \	\mathord{
		\begin{tikzpicture}[baseline=-0.65ex,scale=0.6]
			\draw [black] (0,0) circle [radius=2cm];
			\draw [black] (-0.75,0) circle [radius=0.45cm];
			\begin{feynman}
				\vertex (A) at (-0.75,1.8);
				\vertex (b) at (-0.75,0.45);
				\vertex (d) at (-0.75,-0.45);
				\vertex (B) at (-0.75,-1.8);
				\vertex (C) at (0.75,1.8);
				\vertex (D) at (0.75,-1.8);
				\diagram*{
					(A) -- [photon] (b),
					(d) --[photon] (B),,
					(C) --[fermion] (D)
				};
			\end{feynman}
			\filldraw[color=white!80, fill=white!15](-0.75,0) circle (0.7);	
			\draw [black, thick, dashed] (-0.75,0) circle [radius=0.63cm];
			\draw [black] (-0.75,0) circle [radius=0.7cm];
			\draw (-1.2,0) node[anchor=west]{$\delta$};
		\end{tikzpicture} 
	} +  \	\mathord{
		\begin{tikzpicture}[baseline=-0.65ex,scale=0.6]
			\draw [black] (0,0) circle [radius=2cm];
			\begin{feynman}
				\vertex (A) at (0.75,1.8);
				\vertex (b) at (0.75,0.45);
				\vertex (d) at (0.75,-0.45);
				\vertex (B) at (0.75,-1.8);
				\vertex (C) at (-0.75,1.8);
				\vertex (D) at (-0.75,-1.8);
				\diagram*{
					(A) --[ fermion] (b),
					(d) --[fermion] (B),
					(C) -- [photon] (D),
				};
			\end{feynman}
			\filldraw[color=white!80, fill=white!15](0.75,0) circle (0.7);	
			\draw [black, thick, dashed] (0.75,0) circle [radius=0.63cm];
			\draw [black] (0.75,0) circle [radius=0.7cm];
		\end{tikzpicture} 
	}\ + \	\mathord{
		\begin{tikzpicture}[baseline=-0.65ex,scale=0.6]
			\draw [black] (0,0) circle [radius=2cm];
			\draw [black] (-0.75,0) circle [radius=0.45cm];
			\begin{feynman}
				\vertex (A) at (-0.75,1.8);
				\vertex (b) at (-0.75,0.45);
				\vertex (d) at (-0.75,-0.45);
				\vertex (B) at (-0.75,-1.8);
				\vertex (C) at (0.75,1.8);
				\vertex (D) at (0.75,-1.8);
				\diagram*{
					(A) --[ fermion] (b),
					(d) --[fermion] (B),
					(C) --[fermion] (D)
				};
			\end{feynman}
			\filldraw[color=white!80, fill=white!15](-0.75,0) circle (0.7);	
			\draw [black, thick, dashed] (-0.75,0) circle [radius=0.63cm];
			\draw [black] (-0.75,0) circle [radius=0.7cm];
		\end{tikzpicture} 
	} \ , \\[0.4em]
	\label{eq:sigma4primeprime}
	\Sigma^{\prime\prime}_{4}  &= 	\mathord{
		\begin{tikzpicture}[baseline=-0.65ex,scale=0.6]
			\draw [black] (0,0) circle [radius=2cm];
			\draw [black] (-0.75,0) circle [radius=0.45cm];
			\begin{feynman}
				\vertex (A) at (-0.75,1.8);
				\vertex (b) at (-0.75,0.45);
				\vertex (d) at (-0.75,-0.45);
				\vertex (B) at (-0.75,-1.8);
				\vertex (C) at (0.75,1.8);
				\vertex (D) at (0.75,-1.8);
				\diagram*{
					(A) -- [photon] (b),
					(d) --[photon] (B),
					(C) -- [photon] (D),
				};
			\end{feynman}
			\filldraw[color=white!80, fill=white!15](-0.75,0) circle (0.7);	
			\draw [black, thick, dashed] (-0.75,0) circle [radius=0.63cm];
			\draw [black] (-0.75,0) circle [radius=0.7cm];
			\draw (-1.2,0) node[anchor=west]{$\partial$};
		\end{tikzpicture} 
	}  \ + 	\ \mathord{
		\begin{tikzpicture}[baseline=-0.65ex,scale=0.6]
			\draw [black] (0,0) circle [radius=2cm];
			\draw [black] (-0.75,0) circle [radius=0.45cm];
			\begin{feynman}
				\vertex (A) at (-0.75,1.8);
				\vertex (b) at (-0.75,0.45);
				\vertex (d) at (-0.75,-0.45);
				\vertex (B) at (-0.75,-1.8);
				\vertex (C) at (0.75,1.8);
				\vertex (D) at (0.75,-1.8);
				\diagram*{
					(A) -- [photon] (b),
					(d) --[photon] (B),
					(C) --[fermion] (D)
				};
			\end{feynman}
			\filldraw[color=white!80, fill=white!15](-0.75,0) circle (0.7);	
			\draw [black, thick, dashed] (-0.75,0) circle [radius=0.63cm];
			\draw [black] (-0.75,0) circle [radius=0.7cm];
			\draw (-1.2,-0) node[anchor=west]{$\partial$};
		\end{tikzpicture} 
	}  	\ .
\end{align} 
\subsection{Computing $\Sigma_{4}^\prime$}
Expanding the Wilson loop operator at order $g_B^4$ and decorating the Wick contractions with the one-loop correction to the adjoint scalar propagator $\Delta^{(1)}(x)$, defined in eq. (\ref{eq:one-loop correction adjoint scalar main text}), and with the tensor $\delta_{\mu\nu}\Delta^{(1)}(x)$, we find that 
\begin{equation}
	\label{Sigma_4^i starting point}
	\begin{split}
		\Sigma_{4}^\prime = \dfrac{g_B^4}{N}\int_{\tau_1>\tau_2>\tau_3>\tau_4}\dd^4\tau \  &C^{aabb}  \left(\widehat{\Delta}(x_{12})\widehat{\Delta}^{(1)}(x_{34})+ \widehat{\Delta}(x_{34})\widehat{\Delta}^{(1)}(x_{12}) \right)  +  \\[0.4em] 
		&C^{aabb} \left(\widehat{\Delta} (x_{14})\widehat{\Delta}^{(1)}(x_{23})+ \widehat{\Delta}(x_{23})\widehat{\Delta}^{(1)}(x_{14})\right) + \\[0.4em]
		& C^{abab}  \left(\widehat{\Delta}(x_{13})\widehat{\Delta}^{(1)}(x_{14})+ \widehat{\Delta}(x_{34})\widehat{\Delta}^{(1)}(x_{12})\right)
		\ , 
	\end{split}
\end{equation} where we recall  that the tensor $C^{abcd}$ is defined in eq. (\ref{eq:trace of four generators}), while  $\widehat{\Delta}(x)$ and $\widehat{\Delta}^{(1)}(x)$ are, respectively,  given by eq.s (\ref{defDeltahat}) and (\ref{deltahat1is}).  Using the non-Abelian  exponentiation rules for the Wilson loop, we rewrite the previous expression as follows \begin{equation}
	\label{Sigma_4^i second step }
	\Sigma_{4}^\prime= \cW_2\cW^\prime_4\ + \dfrac{g_B^4}{2N}\tr\left(\big[T^b,T^a\big]\right)^2\int_\cD\dd^4\tau \left(\widehat{\Delta}(x_{13})\widehat{\Delta}^{(1)}(x_{24})+ \widehat{\Delta}(x_{24})\widehat{\Delta}^{(1)}(x_{13})\right) \ ,
\end{equation} where $\cD$ denotes the ordered region $\tau_1>\tau_2\tau_3\tau_4$, the functions $\cW_2$ and $\cW^\prime_4$ are defined in eq.s (\ref{eq:ladder g^2 def and definition a_0}) and (\ref{W4Rfinal}), while the second term in the previous expression denotes the maximally non-Abelian part of the diagram. 
Going through the calculation of eq. (\ref{Sigma_4^i second step }) we encounter, by employing the parametrization eq. (\ref{eq:parametrization}), the following   integral \begin{equation}
	\int_{\tau_1>\tau_2>\tau_3>\tau_4}\dd^4\tau \dfrac{1}{\left(4\sin^2\frac{\tau_{ 13}}{2}\right)^{d/2-2} \left(4\sin^2\frac{\tau_{ 24}}{2}\right)^{d-4}} +  \dfrac{1}{\left(4\sin^2\frac{\tau_{ 24}}{2}\right)^{d/2-2} \left(4\sin^2\frac{\tau_{ 12}}{2}\right)^{d-4}}  \ .
\end{equation} Using Fourier expansion methods outlined in Appendix \ref{sec:Fourier}, the previous expressions can be evaluated  in terms of generalized hypergeometric functions (see eq. (\ref{eq:integral ladder non-maximally def})). We find that  \begin{equation}
	\label{eq:sigma4prime final result}
	\Sigma_{4}^\prime= \hat{g}_B^6 C_F \dfrac{2N^2-3}{6N}P_2(d)B_1(d)B_2(d)
	+ \dfrac{\hat{g}_B^6\beta_0 C_FN 3\zeta(3)}{2^4\pi^2} \ +\ldots \ ,
\end{equation} where the dots stand for terms proportional to $(d-4)^2$, while the function $B_n(x)$ and $P_2(x)$ are defined in eq.s (\ref{eq:ladder g^2 def and definition a_0}) and (\ref{W4Rfinal}), respectively. Note that the $\zeta(3)$-like term in the previous expression  is analogous to that we generated from the maximally non-Abelian part of the two-loop ladder-like diagram (\ref{eq:ladderg4finalresult}). In particular, the result of eq. (\ref{eq:ladderg4finalresult}) is proportional to the evanescent factor $\epsilon=2-d/2$ resulting from the integration over the contour. This factor also arises in eq. (\ref{eq:sigma4prime final result}) but it cancels against the UV pole of the one-loop correction  $\Delta^{(1)}(x)$ (\ref{eq:one-loop correction adjoint scalar main text}) and leaves a finite result. 
\subsection{Computing $\Sigma_{4}^{\prime\prime}$}
In this section, we turn our attention to the correction  $\Sigma_{4}^{\prime\prime}$, represented in eq. (\ref{eq:sigma4primeprime}). Let us begin with  considering in detail the first diagram which only involves gauge fields. We expand the Wilson loop operator at order $g_B^4$, and we decorate the relevant Wick contractions by the tensor $\Delta_{\mu\nu}^{(1),\rm g}(x) \equiv\partial_{1,\mu}\partial_{1,\nu}\Delta^{(1),g}(x)$.  We have \begin{align}
	\label{eq:sigma4primeprimepuregauge}
	\mathord{
		\begin{tikzpicture}[baseline=-0.65ex,scale=0.55]
			\draw [black] (0,0) circle [radius=2cm];
			\draw [black] (-0.75,0) circle [radius=0.45cm];
			\begin{feynman}
				\vertex (A) at (-0.75,1.8);
				\vertex (b) at (-0.75,0.45);
				\vertex (d) at (-0.75,-0.45);
				\vertex (B) at (-0.75,-1.8);
				\vertex (C) at (0.75,1.8);
				\vertex (D) at (0.75,-1.8);
				\diagram*{
					(A) -- [photon] (b),
					(d) --[photon] (B),
					(C) -- [photon] (D),
				};
			\end{feynman}
			\filldraw[color=white!80, fill=white!15](-0.75,0) circle (0.75);	
			\draw [black, thick, dashed] (-0.75,0) circle [radius=0.68cm];
			\draw [black] (-0.75,0) circle [radius=0.75cm];
			\draw (-1.2,0) node[anchor=west]{$\partial$};
		\end{tikzpicture} 
	} = g_B^4  \oint_{\cD} \dd^4\tau \ &C^{aabb} 
	\left(\dot{x}_1^\mu\dot{x}_2^\nu \Delta_{\mu\nu}^{(1),\rm g}(x_{12}) 
	\Delta(x_{34})(\dot{x}_{3}\cdot\dot{x}_4)+\binom{1\leftrightarrow3}{2\leftrightarrow4}\right) + \notag \\
	g_B^4\oint_{\cD} \dd^4\tau  \ 	&C^{abab}	\left(\dot{x}_1^\mu\dot{x}_3^\nu \Delta_{\mu\nu}^{(1),\rm g}(x_{13})
	\Delta(x_{24})(\dot{x}_{2}\cdot\dot{x}_4)+\binom{1\leftrightarrow2}{3\leftrightarrow4}\right) +   \notag \\
	&\\ \notag
	g_B^4\oint_{\cD} \dd^4\tau \  &C^{aabb}	\left(\dot{x}_1^\mu\dot{x}_4^\nu \Delta_{\mu\nu}^{(1),\rm g}(x_{14})\Delta(x_{23})(\dot{x}_{3}\cdot\dot{x}_2)+\binom{1\leftrightarrow2}{4\leftrightarrow3}\right) \ ,
\end{align} where we denoted with  $\cD$ the ordered region $\tau_1>\tau_2>\tau_3>\tau_4$ and we recall that  $\Delta(x_{12})$ is massless tree level propagator defined in eq. (\ref{eq:Deltad}), while the tensor $C^{abcd}$ is given by eq. (\ref{eq:trace of four generators}). The calculation of these diagrams can be further simplified by employing again the non-Abelian exponentiation properties of the Wilson loop. Going through the calculation, we arrive at\footnote{To obtain eq. (\ref{eq:sigma4 pure gauge def}), we neglected terms which yield total derivatives integrated over a closed path.}  
\begin{align}
	\label{eq:sigma4 pure gauge def}
	\mathord{
		\begin{tikzpicture}[baseline=-0.65ex,scale=0.55]
			\draw [black] (0,0) circle [radius=2cm];
			\draw [black] (-0.75,0) circle [radius=0.45cm];
			\begin{feynman}
				\vertex (A) at (-0.75,1.8);
				\vertex (b) at (-0.75,0.45);
				\vertex (d) at (-0.75,-0.45);
				\vertex (B) at (-0.75,-1.8);
				\vertex (C) at (0.75,1.8);
				\vertex (D) at (0.75,-1.8);
				\diagram*{
					(A) -- [photon] (b),
					(d) --[photon] (B),
					(C) -- [photon] (D),
				};
			\end{feynman}
			\filldraw[color=white!80, fill=white!15](-0.75,0) circle (0.75);	
			\draw [black, thick, dashed] (-0.75,0) circle [radius=0.68cm];
			\draw [black] (-0.75,0) circle [radius=0.75cm];
			\draw (-1.2,0) node[anchor=west]{$\partial$};
		\end{tikzpicture} 
	} 
	&=\dfrac{g_B^4}{2N}\tr\left(\big[T^a,T^b\big]\right)^2\int_\cD\dd^4\tau
	\left(\dot{x}_1^\mu\dot{x}_3^\nu \Delta^{(1),\rm g}_{\mu \nu}(x_{13})\Delta(x_{24})(\dot{x}_{2}\cdot\dot{x}_4)+\binom{1\leftrightarrow2}{3\leftrightarrow4}\right)\notag\\
	&= g_B^4\dfrac{C_F N}{2}\oint\dd^2\tau\left(\dot{x}_1\cdot\dot{x}_2\right)\Delta^{(1),\rm g}(x_{12})\Delta(x_{12}) \ ,
\end{align} where we recall that $C_F=(N^2-1)/2N$. To obtain the last equality, we integrated by parts twice. 
Repeating the same analysis for  the second diagram in eq. (\ref{eq:sigma4primeprime}), we find that   \begin{equation}
	\label{eq:sigma4primeprime scalar + gauge}
	\mathord{
		\begin{tikzpicture}[baseline=-0.65ex,scale=0.55]
			\draw [black] (0,0) circle [radius=2cm];
			\draw [black] (-0.75,0) circle [radius=0.45cm];
			\begin{feynman}
				\vertex (A) at (-0.75,1.8);
				\vertex (b) at (-0.75,0.45);
				\vertex (d) at (-0.75,-0.45);
				\vertex (B) at (-0.75,-1.8);
				\vertex (C) at (0.75,1.8);
				\vertex (D) at (0.75,-1.8);
				\diagram*{
					(A) -- [photon] (b),
					(d) --[photon] (B),
					(C) --[fermion] (D)
				};
			\end{feynman}
			\filldraw[color=white!80, fill=white!15](-0.75,0) circle (0.7);	
			\draw [black, thick, dashed] (-0.75,0) circle [radius=0.63cm];
			\draw [black] (-0.75,0) circle [radius=0.7cm];
			\draw (-1.2,0) node[anchor=west]{$\partial$};
		\end{tikzpicture} 
	}  = g_B^4\dfrac{C_F N}{2}\oint\dd^2\tau\left(-R^2\right)\Delta^{(1),g}(x_{12})\Delta(x_{12}) \ .
\end{equation} Combining together the relations we derived in this subsection, we finally arrive at the following representation for the correction $\Sigma_{4}^{\prime\prime}$, defined in (\ref{eq:sigma4primeprime}), i.e. \begin{align}
	\label{eq:sigma4primeprime def}
	\Sigma_{4}^{\prime\prime} &= -g_B^4\dfrac{C_F N}{2}\oint\dd^2\tau\left(R^2-\dot{x}_1\cdot\dot{x}_2\right)\Delta^{(1),\rm g}(x_{12})\Delta(x_{12})= F_3^{(2)} \ .
\end{align} The last equality can be explicitly proved by recalling that the functions $\Delta(x)$ and $\Delta^{(1),\rm g}(x)$ are, respectively, given by eq. (\ref{eq:Deltad}) and (\ref{eq:Delta(1),g}), and using the explicit expression bubble-like correction  $F_3^{(2)}$, given by eq. (\ref{Fiis}). Combining together the previous expression and eq. (\ref{eq:sigma4prime final result}), we reproduce eq. (\ref{W64is}). 
\section{Trigonometric integrals}
\label{sec:Fourier}
In this section, we evaluate the trigonometric integrals appearing in the calculation of the circular Wilson loop. It is convenient to firstly outline some useful relations. We will make extensively use of the following identity \cite{Beccaria:2017rbe}
\begin{equation}
	\label{eq:master integral}
	\begin{split}
		\mathcal{M}(a,b,c) &= \int_{0}^{2\pi} \dd^3\mathbb{\tau} \Big( \sin^2{\dfrac{\tau_{12}}{2}}\Big)^a \Big( \sin^2{\dfrac{\tau_{13}}{2}}\Big)^b\Big( \sin^2{\dfrac{\tau_{23}}{2}}\Big)^c\\[0.4em]
		&=8 \pi^{3/2} \dfrac{\Gamma(a+1/2)\Gamma(b+1/2)\Gamma(c+1/2)\Gamma(1+a+b+c)}{\Gamma(1+a+c)\Gamma(1+b+c)\Gamma(1+a+b)} \ .
	\end{split}
\end{equation} We can use this identity to derive other useful results. For instance, as explained in Appendix G of \cite{Bianchi:2016vvm}, the nested integral \begin{equation}
	\begin{split}
		\mathcal{I}[\alpha, \beta, \gamma]=\int_{\tau_1>\tau_2>\tau_3}\dd^3\tau & {\left[\left(\sin ^2 \frac{\tau_{12}}{2}\right)^\alpha\left(\sin ^2 \frac{\tau_{13}}{2}\right)^\beta\left(\sin ^2 \frac{\tau_{23}}{2}\right)^\gamma \cos \frac{\tau_{23}}{2}\right.} \\
		& -\left(\sin ^2 \frac{\tau_{23}}{2}\right)^\alpha\left(\sin ^2 \frac{\tau_{12}}{2}\right)^\beta\left(\sin ^2 \frac{\tau_{13}}{2}\right)^\gamma \cos \frac{\tau_{13}}{2} \\
		& \left.\left(\sin ^2 \frac{\tau_{13}}{2}\right)^\alpha\left(\sin ^2 \frac{\tau_{23}}{2}\right)^\beta\left(\sin ^2 \frac{\tau_{12}}{2}\right)^\gamma \cos \frac{\tau_{12}}{2}\right] \ ,
	\end{split}
\end{equation}  can be reduced to a linear combination of functions we introduced in eq. (\ref{eq:master integral}). The net result can be written as follows \begin{equation}
	\label{eq:second master integral}
	\mathcal{I}[\alpha, \beta, \gamma]=4 \pi^{3 / 2} \frac{\Gamma(1+\alpha+\beta+\gamma) \Gamma(1+\alpha) \Gamma(1+\beta) \Gamma\left(1/2+\gamma\right)}{\Gamma\left(3/2+\alpha+\gamma\right) \Gamma\left(3/2+\beta+\gamma\right) \Gamma(1+\alpha+\beta)}  \ .
\end{equation} Finally, by employing this useful relation, we can derive a general expression for the following path-ordered integral
\begin{equation}
	\label{eq:third trigonometric integral}
	\mathcal{J}(\alpha,\beta,\gamma) =\oint \dd^3\tau \varepsilon(\tau)\sin\tau_{13} \left(\sin ^2 \frac{\tau_{12}}{2}\right)^\alpha\left(\sin ^2 \frac{\tau_{13}}{2}\right)^\gamma\left(\sin ^2 \frac{\tau_{23}}{2}\right)^\beta  \ ,
\end{equation} where we recall that $\varepsilon(\tau)\equiv\varepsilon(\tau_1,\tau_2,\tau_3)$ is defined in terms of the Heaviside $\theta$-function in eq. (\ref{eq:defepsilon}). Employing this definition for the $\varepsilon$-symbol and relabelling the integration variables, we find that \begin{equation}
	\begin{split}
		\mathcal{J}(\alpha,\beta,\gamma) = -2 \mathcal{I}(\beta,\alpha,\gamma+1/2) -2 \mathcal{I}(\alpha,\beta,\gamma+1/2)=-4 \mathcal{I}(\alpha,\beta,\gamma+1/2) \ .
	\end{split} 
\end{equation} To obtain the last line we noted that $\mathcal{I}(\alpha,\beta,\gamma)$ is symmetric in the exchange of the first two arguments. Therefore, the final result reads \begin{equation}
	\label{eq:third master integral}
	\mathcal{J}(\alpha,\beta,\gamma)=-16\pi^{3/2}\dfrac{\Gamma(3/2+\alpha+\beta+\gamma)\Gamma(1+\alpha)\Gamma(1+\beta)\Gamma(1+\gamma)}{\Gamma(2+\alpha+\gamma)\Gamma(2+\beta+\gamma)\Gamma(1+\alpha+\beta)} \ .
\end{equation} Finally, in the calculation of the Wilson loop, we will extensively use the following identity
\begin{equation}
	\label{eq:two-fold MB}
	\frac{1}{(A+B+C)^\sigma}= \frac{1}{\Gamma(\sigma)} \int_{-\mathrm{i} \infty}^{+\mathrm{i} \infty} \dfrac{\dd{u}\dd{v}}
	{(2\pi \mathrm{i})^2}  \ \dfrac{B^u C^v}{A^{\sigma+u+v}} \Gamma(\sigma+u+v) \Gamma(-u) \Gamma(-v) \ ,
\end{equation} where the integration contour runs parallelly to imaginary axis in such a way that the increasing and decreasing poles of the $\Gamma$-functions are separated. 
\subsection{Path-ordered integrals}
\label{sec:evanescent from spider}
In this subsection, we employ some of the identities we presented  in the previous section to evaluate the path-ordered integral we introduced in eq. (\ref{eq:I1 main}), i.e.
\begin{equation}
	\label{eq:trigonometric integrals evanescent}
	\begin{split}
		E(d) &=  \int_0^1 \dd F \left(\alpha\beta\gamma\right)^{d/2-2} \  \oint \dd^3 \tau \ \varepsilon(\tau)\dfrac{\sin \tau_{ 13}}{Q^{d-3}}\\
		&=-8   \int_0^1 \dd F \left(\alpha\beta\gamma\right)^{d/2-2}\ \int_{\tau_1>\tau_2>\tau_3} \dd^3 \tau \ \dfrac{\sin \frac{\tau_{ 13}}{2}\  \sin \frac{\tau_{ 12}}{2} \ \sin \frac{\tau_{ 23}}{2}}{Q^{d-3}} \ ,
	\end{split}
\end{equation} where $\alpha,\beta$ and $\gamma$ are Feynman parameters integrated over the unit cube via the measure $\dd F$ (\ref{eq:integration over the unit cube}), while the denominator $Q$ is defined in eq. (\ref{eq:def of Q}). To obtain the second line, we employed the explicit definition of the $\epsilon$-symbol in terms of the Heavise $\theta$-function (\ref{eq:defepsilon}).
		To integrate over the Feynman parameters, we replace the denominator  $Q$ with a two-fold Mellin-Barnes representation, i.e. (see eq.s (\ref{eq:def of Q}) and (\ref{eq:two-fold MB}))  
		\begin{equation}
			\label{eq:rewriting Q}
			\begin{split}
				\dfrac{1}{Q^{\sigma}} &= \dfrac{2^{-\sigma}}{\Gamma(\sigma)} \int_{-\mathrm{i} \infty}^{+\mathrm{i} \infty} \dfrac{\dd{u}\dd{v}}
				{(2\pi \mathrm{i})^2}  \  \dfrac{\Gamma(\sigma+u+v) \Gamma(-u) \Gamma(-v)}{\left(\beta\alpha\sin^2\frac{\tau_{12}}{2}\right)^{\sigma+u+v} \ \left(\beta\gamma\sin ^2 \frac{\tau_{23}}{2} \right)^{-u}\left(\gamma\alpha\sin^2 \frac{\tau_{31}}{2} \right)^{-v}} \ ,
			\end{split}
		\end{equation} where the integration path runs parallelly to the imaginary axes and separates the increasing and the decreasing poles of the $\Gamma$-function. Substituting this identity in eq. (\ref{eq:trigonometric integrals evanescent}) and performing the integration over the Feynman parameters, we arrive at the following result  
		
		\begin{equation}
			\begin{split}
				E(d) =  \  & \int \dfrac{\dd{u}\dd{v}}
				{(2\pi \mathrm{i})^2} \ \ \dfrac{ \Gamma(d-3+u+v) \Gamma(-u) \Gamma(-v) \Gamma(2-d/2-u)}{ \left(\Gamma(2-d/2-v)\Gamma(d/2-1+u+v)\right)^{-1}} \mathcal{E}(u,v,d) \  ,
			\end{split}
		\end{equation} where in the previous expression we denoted the integral over the coordinates $\tau_i$ as  follows \begin{equation}
			\begin{split}
				\mathcal{E}(u,v,d) &=- \dfrac{2^{6-d}}{\Gamma(3-d/2)\Gamma(d-3)} \ \int_{\tau_1>\tau_2>\tau_3} 	\dd^3\tau \ \dfrac{\left(\sin^2\frac{\tau_{23}}{2}\right)^{u+1/2}\left(\sin^2\frac{\tau_{31}}{2}\right)^{v+1/2}}{ \left(\sin^2\frac{\tau_{12}}{2}\right)^{d-3+u+v-1/2}} \\[0.4em]
				&=-\dfrac{2^{6-d}}{\Gamma(3-d/2)\Gamma(d-3)3!} \oint  \ \dd^3 \tau \  \frac{\left(\sin^2\frac{\tau_{23}}{2}\right)^{u+1/2}\left(\sin^2\frac{\tau_{31}}{2}\right)^{v+1/2}}{ \left(\sin^2\frac{\tau_{12}}{2}\right)^{d-3+u+v-1/2}}\\[0.4em]
				&=- \dfrac{2^{9-d}\pi^{3/2}\Gamma(11/2-d)}{\Gamma(3-d/2)\Gamma(d-3)3!} \  \dfrac{\Gamma(u+1)\Gamma(v+1)\Gamma(4-d-u-v)}{\Gamma(2+u+v)\Gamma(5-d-u)\Gamma(5-d-v)} \ .
			\end{split}
		\end{equation}
		In the previous expression, we obtained the second line by observing that the integrand is completely symmetric. This can be proved by properly shifting the Mellin-Barnes variables and enables us to replace the nested integration with an integral  over the complete circle.
			Employing eq. (\ref{eq:master integral}), we finally find    \begin{equation}
				\label{eq:I1 prima delle mellin}
				E(d) =- \dfrac{2^{9-d}\pi^{3/2}\Gamma(11/2-d)}{\Gamma(3-d/2)\Gamma(d-3)3!}  \ M(d) \ . 
			\end{equation} In the previous expression, the amplitude $M(d)$ is a meromorphic function of the dimension $d$ which is defined in terms of the following two-fold Mellin-Barnes integral \begin{equation}
				\label{eq:Mellin barnes}
				\begin{split}
					M(d) =  \int\dfrac{\dd{u}\dd{v}}{\left(2\pi \mathrm{i}\right)^2} &\dfrac{\Gamma(v+1)\Gamma(-v)\Gamma(2-d/2-v)\Gamma(-u) \Gamma(u+1)\Gamma(2-d/2-u)}{\Gamma(5-d-v)} \  \times \\[0.4em]
					\times & \dfrac{\Gamma(d-3+u+v)  \Gamma(4-d-u-v)\Gamma(d/2-1+u+v)}{ \Gamma(2+u+v)\Gamma(5-d-u)} \  .
				\end{split}
			\end{equation}
			Since the the function $E(d)$ appears in the calculation of the Wilson loop with an evanescent coefficient (see eq. (\ref{eq:real calculation Sigma 3})), we only have to determine its behaviour for $d\to 4$. We find     \begin{equation}
				\begin{split}
					M(d)\Big |_{d=4} 
					&=		\int_{-\delta^\prime-\mathrm{i} \infty}^{+\delta^\prime+\mathrm{i} \infty} \dfrac{\dd{v\dd{u} }}{\left(2\pi \mathrm{i}\right)^2} \dfrac{-\pi^3\csc(\pi u)\csc(\pi  v)\csc( \pi   (u+v))}{(1+u+v)uv} \\[0.4em]
					&=6\zeta(3)\  , 
				\end{split}
			\end{equation}  where $\delta^\prime=\mathfrak{Re}(u)=\mathfrak{Re}(v)\in(-1,0)$, in such a way that the increasing poles are to the right of the integration contour, while the decreasing ones are to the left.
					Substituting the previous expression in eq.  (\ref{eq:I1 prima delle mellin}), we finally arrive at  \begin{equation}
						\label{eq:leading order I1}
						E(d) = -16\pi^2\zeta(3) +\mathcal{O}(d-4) \ .
					\end{equation}
					
					\subsection{Fourier expansions methods and the ladder-like diagrams}
					In this section, we will go through the calculation of the trigonometric integrals which enter the maximally non-Abelian part of the multiple-exchange diagrams (\ref{eq:ladder with non-maximally abelian part}) and (\ref{Sigma_4^i second step }). 
					
					The starting point is the Fourier expansion of the real even function $1/\sin[2\alpha](\frac{x}{2})$
					\begin{equation}
						\label{eq:Foureir expnasion sine}
						\dfrac{1}{\left(4\sin^2{\frac{x}{2}}\right)^\alpha} = \dfrac{1}{2}a_0(\alpha) +\sum_{n=1}^{\infty}a_n(\alpha) \cos{nx} \ ,
					\end{equation} where the Fourier coefficients are given by \cite{Beccaria:2017rbe} \begin{equation}
						\begin{split}
							\label{eq:Fourier coefficients}
							a_n(\alpha) &= \dfrac{1}{\pi} \int_0^{2\pi}  \dd{x} \dfrac{\cos{nx}}{\left(4\sin^2{\frac{x}{2}}\right)^\alpha}  = \dfrac{\text{sec}(\pi \alpha)\Gamma(n+\alpha)}{\Gamma(2\alpha)\Gamma(1-\alpha+n)}.
						\end{split}
					\end{equation}

					Expressing the coordinates $x_i$ in terms of trigonometric functions via eq. (\ref{eq:parametrization}), we find that the integrals appearing in eq.s (\ref{Sigma_4^i second step }) and  (\ref{eq:ladder with non-maximally abelian part}) take the following  form \begin{equation}
						\begin{split}
							\label{eq:integral ladder non-maximally}
							L \left(\alpha,\beta\right)= \int_\cD  \dfrac{\dd^4\tau}{\left(4\sin^2{\frac{\tau_{13}}{2}}\right)^\alpha\left(4\sin^2{\frac{\tau_{24}}{2}}\right)^\beta} \ ,
						\end{split}
					\end{equation}where the integration domain $\cD$ is defined by the ordered region $\tau_1>\tau_2>\tau_3>\tau_4$. Replacing the trigonometric functions via their Fourier expansions (\ref{eq:Foureir expnasion sine}) and performing the integration over the coordinates $\tau_i$, we finally arrive at the following representation \begin{equation}
						L(\alpha,\beta) = \dfrac{\pi^4}{6}a_0(\alpha)a_0(\beta)-\sum_{n=1}^{\infty} \dfrac{\pi^2}{n^2}\Big(a_0(\alpha)a_n(\beta)+a_0(\beta)a_n(
						\alpha)-a_n(\beta)a_n(\alpha)\Big) \ .
					\end{equation} The infinite sums in the previous expression can be easily performed in terms of usual generalized hypergeometric functions. After a straightforward calculation, we find that
					\begin{equation}
						\label{eq:integral ladder non-maximally def}
						\begin{split}
							\dfrac{	L(\alpha,\beta)}{\pi^2 a_0(\alpha)a_0(\beta)}=\zeta(2)-\dfrac{\alpha \prescript{}{4}{F}_3\left(\mathsf{x}_\alpha,\mathsf{y}_\alpha,1\right)}{(1-\alpha)}-\dfrac{\beta \prescript{}{4}{F}_3\left(\mathsf{x}_\beta,\mathsf{y}_\beta ,1\right)}{(1-\beta)}+ \dfrac{\alpha\beta\prescript{}{5}{F}_4\left(\mathsf{w}_{\alpha,\beta},\mathsf{z}_{\alpha,\beta},1\right)}{(1-\alpha)(1-\beta)} \ ,
						\end{split}
					\end{equation}
					where the parameters of the two generalized hypergeometric functions are encoded in the following quantities  $\mathsf{x}_\alpha=(1,1,1,1+\alpha)$, $\mathsf{y}_\alpha=(2,2,2-\alpha)$, $\mathsf{w}_{\alpha,\beta}=(1,1,1,1+\alpha,1+\beta)$ and $\mathsf{z}_{\alpha,\beta}=(2,2,2-\alpha,2-\beta)$.

	\bibliographystyle{JHEP}
	\bibliography{biblio}
\end{document}